\documentclass[preprint,10pt]{aastex}
\usepackage{amsmath}
\usepackage{amssymb}
\usepackage{graphics}
\usepackage{graphicx}
\usepackage{hyperref}
\usepackage{natbib}
\usepackage[utf8]{inputenc}

\begin{document}

\title{PHYSICAL PROPERTIES, STAR FORMATION, AND ACTIVE GALACTIC NUCLEUS ACTIVITY IN BALMER BREAK GALAXIES AT 0 $<$ $z$ $<$ 1}
\author{J. Díaz Tello\altaffilmark{1}, C. Donzelli\altaffilmark{1}, N. Padilla\altaffilmark{2}, N. Fujishiro\altaffilmark{3}, H. Hanami\altaffilmark{4}, T. Yoshikawa\altaffilmark{3}}
\and 
\author{B. Hatsukade\altaffilmark{5}.}
\email{jdiazt@oac.uncor.edu}

\altaffiltext{1}{IATE, Observatorio Astronómico de Córdoba, Universidad Nacional de Córdoba, Argentina.}
\altaffiltext{2}{Departamento de Astronomía y Astrofísica, Pontificia Universidad Católica de Chile, Chile.}
\altaffiltext{3}{Koyama Astronomical Observatory, Kyoto Sangyo University, Japan.}
\altaffiltext{4}{Physics Section, Iwate University, Japan.}
\altaffiltext{5}{Department of Astronomy, Kyoto University, Japan.}

\begin{abstract}

We present a spectroscopic study with the derivation of the physical properties of 37 Balmer break galaxies, which have the necessary lines to locate them in star-forming$-$active galactic nuclei (AGN) diagnostic diagrams. These galaxies span a redshift range from 0.045 to 0.93 and are somewhat less massive than similar samples of previous works. The studied sample has multiwavelength photometric data coverage from the ultraviolet to mid-infrared (MIR) \textit{Spitzer} bands. We investigate the connection between star formation and AGN activity via optical, mass-excitation (MEx) and MIR diagnostic diagrams. Through optical diagrams, 31 (84$\%$) star-forming galaxies, 2 (5$\%$) composite galaxies and 3 (8$\%$) AGNs were classified, whereas from the MEx diagram only one galaxy was classified as AGN. A total of 19 galaxies have photometry available in all the IRAC/\textit{Spitzer} bands. Of these, 3 AGN candidates were not classified as AGN in the optical diagrams, suggesting they are dusty/obscured AGNs, or that nuclear star formation has diluted their contributions. By fitting the spectral energy distribution (SED) of the galaxies, we derived the stellar masses, dust reddening $E(B-V)$, ages and UV star formation rates (SFRs). Furthermore, the relationship between SFR surface density ($\Sigma_{\text{SFR}}$) and stellar mass surface density per time unit ($\Sigma_{M_{\ast}/\tau}$) as a function of redshift was investigated using the [OII] $\lambda$3727, 3729, H$\alpha$ $\lambda$6563 luminosities, which revealed that both quantities are larger for higher redshift galaxies. We also studied the SFR and specific SFR (SSFR) versus stellar mass and color relations, with the more massive galaxies having higher SFR values but lower SSFR values than less massive galaxies. These results are consistent with previous ones showing that, at a given mass, high-redshift galaxies have on average larger SFR and SSFR values than low-redshift galaxies. Finally, bluer galaxies have larger SSFR values than redder galaxies and for a given color the SSFR is larger for higher redshift galaxies. 
\end{abstract}

\section{INTRODUCTION}

The redshift interval from $z$ $\sim$ 0 to 1 accounts for roughly half of the universe's age, and provides a valuable range to study the final stages of galaxy assembly. It is well known that the star formation rate (SFR) in galaxies has diminished since z $\sim$ 1 (\citeauthor{Cowie,Madau,Lilly}), with Cowie et al. coining the term ``downsizing" to describe this behaviour. This suggests that the most massive galaxies finished their star formation earlier than less massive systems, a trend observed in broad-band colors (\citeauthor{Bundy}), and radio (\citeauthor{Hopkins2}) and infrared (\citeauthor{Perez}) data. However, the reasons why star formation in the universe has been decreasing and why it stops for any given galaxy are still unresolved.

Feedback processes may play an important role in regulating the mass growth of galaxies and/or producing downsizing, with their physical origin having received much attention. Some processes implemented in numerical and semi-analytic models include regulation through feedback by supernovae (\citeauthor{Cole,Nagashima}), and active galactic nuclei (AGNs) by reduction of gas cooling and outflows of reheated gas (\citeauthor{Bower,Lagos}). The most accepted cosmological model, however, predicts that the formation of galaxies is hierarchical, with massive ellipticals being the result of both a series of mergers between smaller galaxies and by gas accretion. In this scenario without taking into account feedback processes, elliptical galaxies continue accreting gas and forming stars. As this is not observed in present day galaxies, it is important to understand the AGN activity at $z$ $<$ 1 as a feedback source. Indeed, although AGN feedback is expected to occur more often in more massive galaxies (\citeauthor{Kauffmann,Kewley}), it is also interesting to test its occurrence in low-mass galaxies. 

Galaxies show a wide variety of physical and observational properties, with their morphologies correlating with their colors, and also with characteristics of their stellar populations. The properties such as optical colors (\citeauthor{Strateva}), morphological parameters (\citeauthor{Driver}), and spectral indices (\citeauthor{Kauffmann2}), come from a bimodal distribution which defines two classes of galaxies: a red population of passive galaxies that have formed their stellar masses mostly at high redshift, and a blue population that is actively forming stars. This bimodal distribution is present at least up to $z$ $\sim$ 1-2 (\citeauthor{Giallongo,Kriek}), but its origin is still not clear. The best scenario to explain these properties may be evolutive, as galaxies in different phases of their evolution reveal different colors, star formation rates, spectral indexes and morphologies. Nevertheless, it is still unresolved how these different parameters are connected.

There is some evidence for a possible role of AGN in the transition area between the ``blue cloud'' and the ``red sequence'' on the color$-$magnitude and color$-$mass diagrams (\citeauthor{Kauffmann,Sanchez,Georgakakis}), a region also known as the ``green valley''. AGN hosts tend to have redder colors than the bulk of star-forming galaxies (\citeauthor{Yan}), suggesting that they might cause or maintain the quenching of star formation in early-type star-forming galaxies (\citeauthor{Schawinski,Georgakakis,Mendez}, for more discussion). However, in the local universe, type 2 AGNs (Seyferts) are mainly hosted in young massive star-forming galaxies (\citeauthor{Kauffmann,Kewley}), thus implying a close link between the growth of black holes (BHs) and bulges. 

About half of the sources with moderate X-ray luminosities (moderately luminous AGN) detected in deep X-ray surveys do not show broad lines or high excitation lines characteristic of AGN in their optical spectra, indicating that the optical line selection of AGN is incomplete (\citeauthor{Lacy2,Szokoly,Caccianiga}). Furthermore, AGNs with high column densities of absorbing gas and dust clouds (combined with those located in the host galaxy) may be missed in X-rays (\citeauthor{Polletta,Donley,Maiolino}). A possible explanation for this is that current X-ray and optical surveys are biased against the early phase of BH evolution. At its initial stage of formation, AGN might be deeply buried under star-forming clouds and/or show low luminosity, because although the BH mass is growing rapidly it is small. Therefore, it might be possible to use mid-infrared (MIR) wavelengths to locate these dusty AGNs by probing emission from AGN-heated dust. Moreover, this wavelength range is much less affected by extinction and is able to penetrate through the dusty envelope surrounding the AGN (\citeauthor{Lacy2,Stern,Donley}). 

Defining a galaxy population that contributes to a decrease in the SFR over cosmic time has strong implications for theories of galaxy evolution. In this investigation we studied the physical properties of a sample of 37 emission line galaxies at 0 $<$ $z$ $<$ 1, which were detected in MIR (\textit{Spitzer}) for which the multiwavelength photometry and optical spectra were available. This sample formed part of a spectroscopic survey pilot project in the Subaru XMM deep field (SXDF) out to $z$ $\sim$ 2.5, with the purpose of investigating the end of star formation in massive galaxies. This paper is organized as follows: in $\S$\ref{sample} we describe the sample selection; $\S$\ref{obs} gives details on the acquisition of the spectrophotometric data; $\S$\ref{Reduc} explains the reduction process applied to the data; $\S$\ref{results} summarizes the final sample used in this paper; in $\S$\ref{AGN} we investigate star formation and the AGN activity of our galaxies using optical, mass-excitation (MEx) and MIR diagnostic diagrams; in $\S$\ref{prop} the physical properties of our galaxy sample are derived using the spectral energy distribution (SED) fitting technique, and the relationships between SFR, specific SFR (SSFR), and SFR surface density ($\Sigma_{\text{SFR}}$) are investigated as a function of redshift, stellar mass and color. Finally, in $\S$\ref{Sum} we summarize our results and analysis.

Throughout this paper we have assumed a flat $\Lambda$-dominated cosmology with $\Omega_{m}$ = 0.28, $\Omega_{\Lambda}$ = 0.72, and 
$H_{0}$ = 70 km s$^{-1}$Mpc.

\section{SAMPLE SELECTION}\label{sample}

The galaxies presented in this paper are a subset of a sample constructed to study star formation and the AGN activity of massive galaxies in the redshift range $z$ = 0.1$-$2.5, with the parent sample being chosen from the SXDF (\citeauthor{Furusawa}). The advantage of the SXDF is that it has been observed in many photometric bands: $B, V, R, i, z$ (Subaru); $J, H, K$ (UKIRT, \citeauthor[as reference project]{Lawrence}); and 3.6$\mu$m, 4.5$\mu$m, 5.8$\mu$m, 8.0$\mu$m, 24$\mu$m, 70$\mu$m and 160$\mu$m (\textit{Spitzer}). This field, centered at R.A. = 02:18:00, Dec = $-$05:00:00, covers an area of $\sim$1.22 deg$^{2}$ and reaches depths of $\sim$27 AB magnitudes at optical wavelengths and $\sim$24 AB mag at near-infrared (NIR) and MIR wavelengths. Therefore, one can apply a number of photometric criteria in order to select galaxies with different characteristics. In our particular case, the parent sample was selected using the $\lambda$3646 Balmer and $\lambda$4000 break features as tracers of redshift, as described by Daddi et al. \citeyearpar{Daddi}, using the BzK color$-$color diagram to select star-forming galaxies in a specific redshift range, independently of the dust reddening\footnote{Details about these color criteria can also be seen in Hanami et al. \citeyearpar{Hanami}}. The limiting magnitude of the survey was set to K $<$ 23 AB mag, with an additional requirement being that the source had a counterpart in all UV$-$optical$-$NIR bands.

A total of 4 different fields were observed using 4 masks, which produced 417 spectra of the selected targets. However, only 132 of these spectra had the necessary signal-to-noise (S/N) ratio for our purposes (see Sections \ref{IMACS} and \ref{GMOS}). Additionally, out of these 132 spectra, we selected only those galaxies that showed the necessary emission lines in order to locate them in at least one of the Baldwin, Phillips and Terlevich (\citeyear{Baldwin}, hereafter BPT) diagnostic diagrams (i.e. H$\beta$4861 $\AA{}$, [OIII]$\lambda$5007, H$\alpha$6563 $\AA{}$ and [NII]$\lambda$6548, 6583 emission lines), or those presented by Lamareille \citeyearpar{Lamareille} and Juneau et al. \citeyearpar{Juneau} to identify star formation and AGN activity in galaxies at high redshifts. These latter two diagrams are particularly useful for galaxies at $z$ $>$ 0.5, since the emission lines H$\alpha$ and [NII] are beyond the spectral coverage for an optical spectrograph. Lamareille \citeyearpar{Lamareille} used the flux ratio [OIII]($\lambda$5007)/H$\beta$($\lambda$4861) versus the equivalent width ratio [OII]($\lambda$3727)/H$\beta$($\lambda$4861), while Juneau et al. \citeyearpar{Juneau} employed the flux ratio [OIII]($\lambda$5007)/H$\beta$($\lambda$4861) versus the stellar mass. Both of these approaches can be applied out to $z$ $<$ 1.0, because the emission lines H$\beta$ and [OIII] are beyond the spectral coverage at higher redshifts.

These constraints allowed us to explore the interplay between AGN and star formation in galaxies at 0 $<$ $z$ $<$ 1, independent of the dust reddening, and to comparing different optical selection methods normally employed to study these populations. Note that the presence of the aforementioned emission lines implies a galaxy sample with evidence of recent star formation activity, thus favoring the selection of late-type galaxies, where star formation and AGN activity is usually found.

\section{Observations}\label{obs}

In addition to the photometry available for the SXDF, we obtained $u$-band data along with optical spectroscopy for a broader and more accurate study of the physical properties of the galaxies. 

\subsection{\textit{$u$-Band Photometry}}

The $u$-band photometry was obtained in order to provide better constraints over the UV SFR values derived from SED fitting, and also to select Lyman break galaxies (LBGs) at redshifts $z$ $>$ 2.0. The latter will be used in future research.

The $u$-band images were obtained under photometric conditions during 2006 September 16$-$17, and 19$-$20 (PI: N. Padilla) using the MosaicII camera attached to the prime focus of the Blanco telescope at CTIO. The MosaicII is a wide field camera of 8912$\times$8192 pix$^{2}$, with a pixel scale of 0.267 arcsec/pixel. and covering a field of view of 37'$\times$37'.

We observed the region of the SXDF centered at R.A.: 02:18:00, Dec.: $-$05:00:00. During this observation, the usual calibration frames were taken including ZERO, SKYFLATS, and standard stars. A small offset was applied during the acquisition of images in order to remove the gaps between the CCDs. Table \ref{uphot} details the observation log for the whole run.

\subsection{\textit{Spectroscopic IMACS/Magellan data}}

Spectroscopic data were obtained during 2007 December 11$-$12 (PI: N. Padilla) using the Inamori-Magellan Areal Camera and Spectrograph (IMACS) attached to the Magellan telescope at LCO. Spectra were taken using the short camera mode $f$/2 configuration with a 200 lines/mm grating. In this mode, the total field of view was 27.2'$\times$27.2' and the dispersion 2.04 $\AA{}$/pixel, which gave a wavelength coverage from 450nm to 900nm.

Observations were carried out using a multi-slit mask with 227 slits. The slit size adopted was of 1" aperture, consistent with the average seeing during the observations. The typical angular sizes of our targets were of 4", and during the run 10$\times$1800 s frames were obtained, which gave a total integration time of 18000 s. The targets selected in the SXDF were centered at R.A.: 02:18:00, Dec.: $-$05:00:00, and during the observations the ZERO, FLAT, arc and flux standard calibration frames were taken. 

\subsection{\textit{Spectroscopic GMOS/Gemini South data}}

A second set of observations was made in service mode (PI: Nelson Padilla) during the second semester of 2008 with GMOS/Gemini South. In this case, the telescope detector used was a Mosaic CCD camera, which had a field of view of 5.5'$\times$5.5'. Spectra were taken in NOD$\&$SHUFFLE (N$\&$S) mode to improve the sky subtraction, using a grating of 400 lines/mm, which gave a dispersion of 3.59 $\AA{}$/pix and a wavelength coverage from 500nm to 1000nm.

Three multi-slit masks were used with 74, 61 and 55 slits, thereby covering the central region of the SXDF. The slit size adopted of a 1" aperture was the same used as that for IMACS. The centers of these regions were defined as follows: SXDF1 centered at R.A.: 02:18:04.5, Dec.: $-$05:02:05; SXDF2 at R.A.: 02:18:30.7, Dec.: $-$05:02:12; and SXDF3 at R.A.: 02:17:31.8, Dec.: $-$05:01:37. Seeing during the run was variable with a mean of 0.8". Table \ref{gmos} shows relevant details of the observations carried out for this program.

\section{REDUCTIONS}\label{Reduc}

\subsection{\textit{$u$-Band Photometry}}

The reduction process for the Mosaic $u$-band data was performed using the IRAF package \textit{mscred}, mainly following the NOAO Deep Wide Field Survey notes\footnote{http://www.noao.edu/noao/noaodeep/ReductionOpt/frames.html}. This process includes bias, flat field and World Coordinate System (WCS) astrometric corrections, together with a projection onto a standard coordinate system and image scale on single images. Reduced images were finally stacked into a single image that was complete up to a total magnitude of $u$=24.5 AB mag. 

\subsection{\textit{Spectroscopy: IMACS/Magellan data}}\label{IMACS}

The IMACS/Magellan spectra were reduced using the Car\-ne\-gie Observatories System for Multi Object Spectroscopy (COSMOS)\footnote{http://obs.carnegiescience.edu/Code/cosmos} routines, which are based on an accurate optical model of IMACS that predicts (after alignment and calibration) a precise description of the spectral features (position, angle, scale). In this way, the process of visual line searching is improved. The reduction process includes corrections by ZERO and FLAT frames, subtraction of the sky lines and background, and spectral extraction.

The COSMOS routines produce a 3D image formed by layers that contain the 2D spectrum corresponding to the mask slits, so the 2D spectra for each of the exposures were combined into a single 3D image. During this final process, the cosmic rays were removed. The 1D spectrum extraction process was carried out using the \textit{apall} routine, while the flux calibration was performed using the \textit{calibrate} routine, all belonging to IRAF. However, due to parasite light problems, only 44 spectra had a good enough S/N ratio to measure a secure redshift. Therefore, considering that the original mask had 227 slits, the success rate was 19$\%$.
 
\subsection{\textit{Spectroscopy: GMOS/Gemini South data}}\label{GMOS}

The reduction process was carried out using the \textit{gemini-gmos} routines within IRAF, adopting the usual factors including bias and flat-field corrections, N$\&$S sky substraction, and wavelength and flux calibrations. The reduced 2D spectra were processed to recover the positive and negative spectra in order to join them in a single 2D spectrum using standard IRAF routines. Finally, the 1D extraction process and flux calibration were performed using the \textit{apall} and \textit{calibrate} routines within IRAF. Similar problems were encountered as those described in the previous section. From 190 slits arranged in 3 masks, a secure redshift was only obtained for 89 spectra, representing a success rate of 46$\%$. 

\subsection{\textit{Emission Line Fluxes}}\label{fluxes}

Fluxes of the emission lines were measured using the \textit{splot} routine within IRAF, adopting the Gaussian mode as it provides an adequate representation of the observed line profiles. Prior to taking the emission line measurements, the spectrum of the underlying stellar population was subtracted in order to produce a pure emission line spectrum. This is particularly important since both the H$\beta$ and H$\alpha$ emission lines are affected by Balmer stellar absorption. Furthermore, prior to the emission line measurements, the aperture photometry was used to correct errors in the spectra due to flux loss. Stellar template spectra were obtained from stellar population synthesis models as described in section \ref{SED}.

Based on the technique applied to Sloan Digital Sky Survey (SDSS) galaxies by Tremonti et al. \citeyearpar{Tremonti}, the calculated $B$-, $V$-, $R$-, $i$-, $z$-band fluxes were compared with the fluxes from the spectrum. The correction factor for each spectrum was estimated as the average of the flux ratio between the aperture photometry and the continuum calculated for each filter, with typical values of such corrections being about 30$\%$. The flux limit in the emission lines was 4$\times$10$^{-18}$ erg s$^{-1}$cm$^{-2}$ at the 3$\sigma$ level. In H$\alpha$, this value corresponded to a limit in the SFR of $\sim$0.0001 $M_{\odot}$/yr at $z$=0.045, and of $\sim$0.1 $M_{\odot}$/yr, for the most distant galaxy of our sample at $z$=0.93.

\subsection{\textit{Spectral Energy Distribution Fitting}}\label{SED}

The GALAXEV code (\citeauthor{Bruzual}) was used to generate a database of SED templates spanning the ages of [0.001Gyr; 13.5Gyr] and the metallicities Z of [0.004; 0.05]. We assumed a Salpeter initial mass function (IMF, Salpeter \citeyear{Salpeter}) and two types of star formation histories (SFHs): passive evolution of stars (single burst; i.e. simple stellar population, SSP) and an exponentially declining one (SFR($t$) $\sim$ $e^{-t/\tau}$, with $\tau$ spanning [0.25Gyr; 15Gyr]). A dust reddening law (\citeauthor{Calzetti2}) was applied to each template using $E(B-V)$ values spanning [0; 0.5] (see section \ref{dust} for more details). Then, a least-squares $\chi^{2}$ fitting method was used to obtain the best template that matched the dominant stellar population, with emission lines being masked prior to the fitting. By using these templates the stellar mass of the sample galaxies was estimated, which revealed values of total stellar masses in the range 10$^{9}-$10$^{10}M_{\odot}$. To avoid the age$-$metallicity degeneracy of the SED fits, a constant metallicity of Z=0.004 was used. This value was taken as the mean metallicity value found in galaxies that had the same stellar mass range as our galaxies (\citeauthor{Gallazzi}), ignoring possible mass$-$metallicity evolution.

Both total and aperture photometry of the galaxies were used to perform SED fitting. The $u, B, V, R, i, z$, $J, H, K$ aperture magnitudes were calculated in a diaphragm with the same aperture of the slit as that used for spectroscopy (1"), with SExtractor\footnote{http://www.astromatic.net/software/sextractor} code being used to obtain the magnitudes. The photometric data roughly covered the range 3550$-$22.000 $\AA{}$, which allowed us to be able to estimate the stellar continuum beyond the spectral coverage of the individual spectra (4500$-$10000 $\AA{}$). Therefore, the set of possible solutions for each of our galaxies obtained by GALAXEV was substantially reduced. Throughout this paper, all the physical properties of the galaxies are relative to the spectroscopy aperture except for stellar masses. Aperture photometry parameters were used to compare them directly to the properties derived from the spectra (reddening, SFR) and to compute color index and age, whereas total magnitudes were used to compute total stellar masses (section 7.3). A 90$\%$ confidence interval was used to determine the uncertainties of the derived physical properties.

\begin{figure}
\begin{center}
\includegraphics[width=0.32\textwidth]{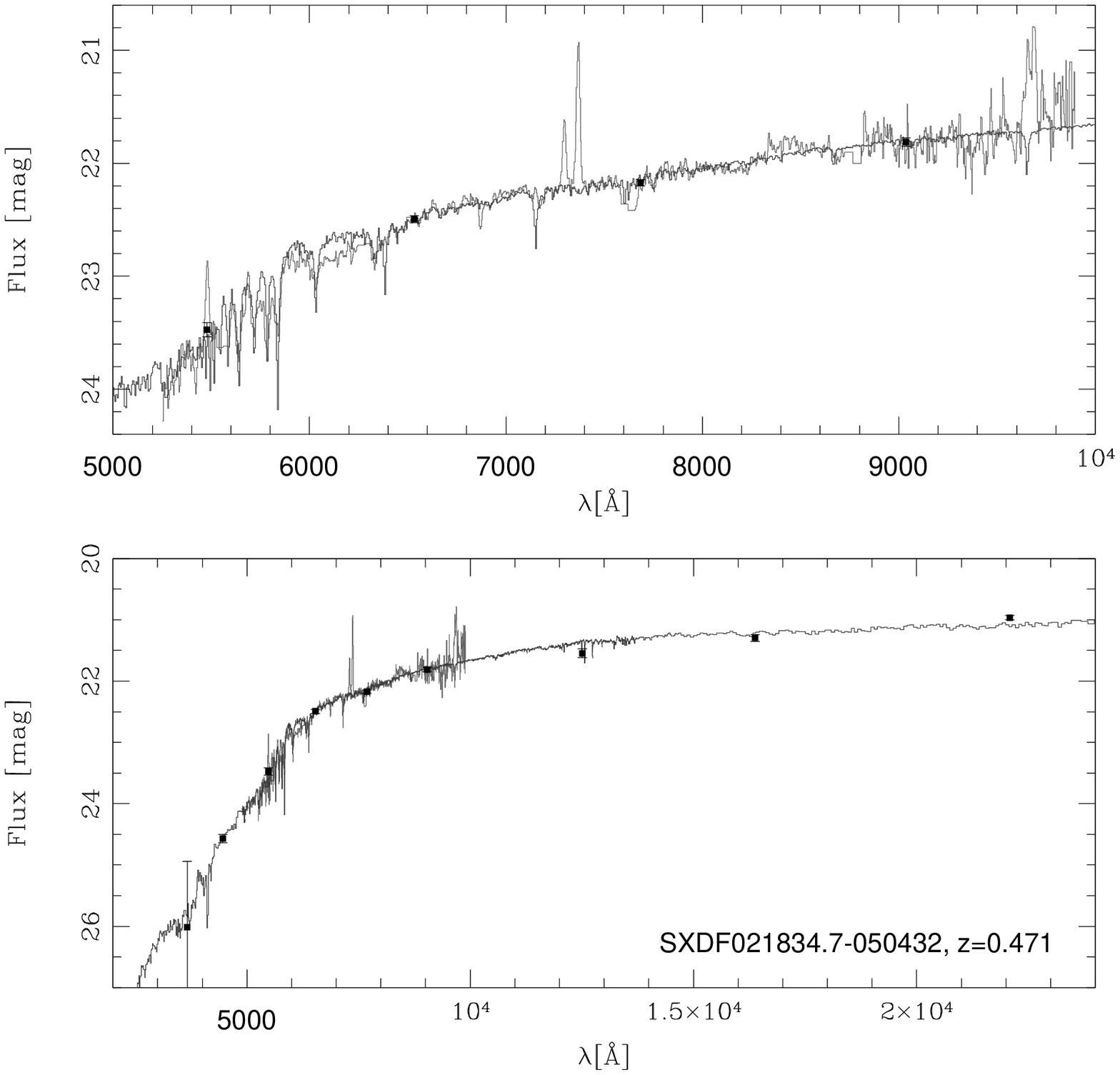}
\includegraphics[width=0.32\textwidth]{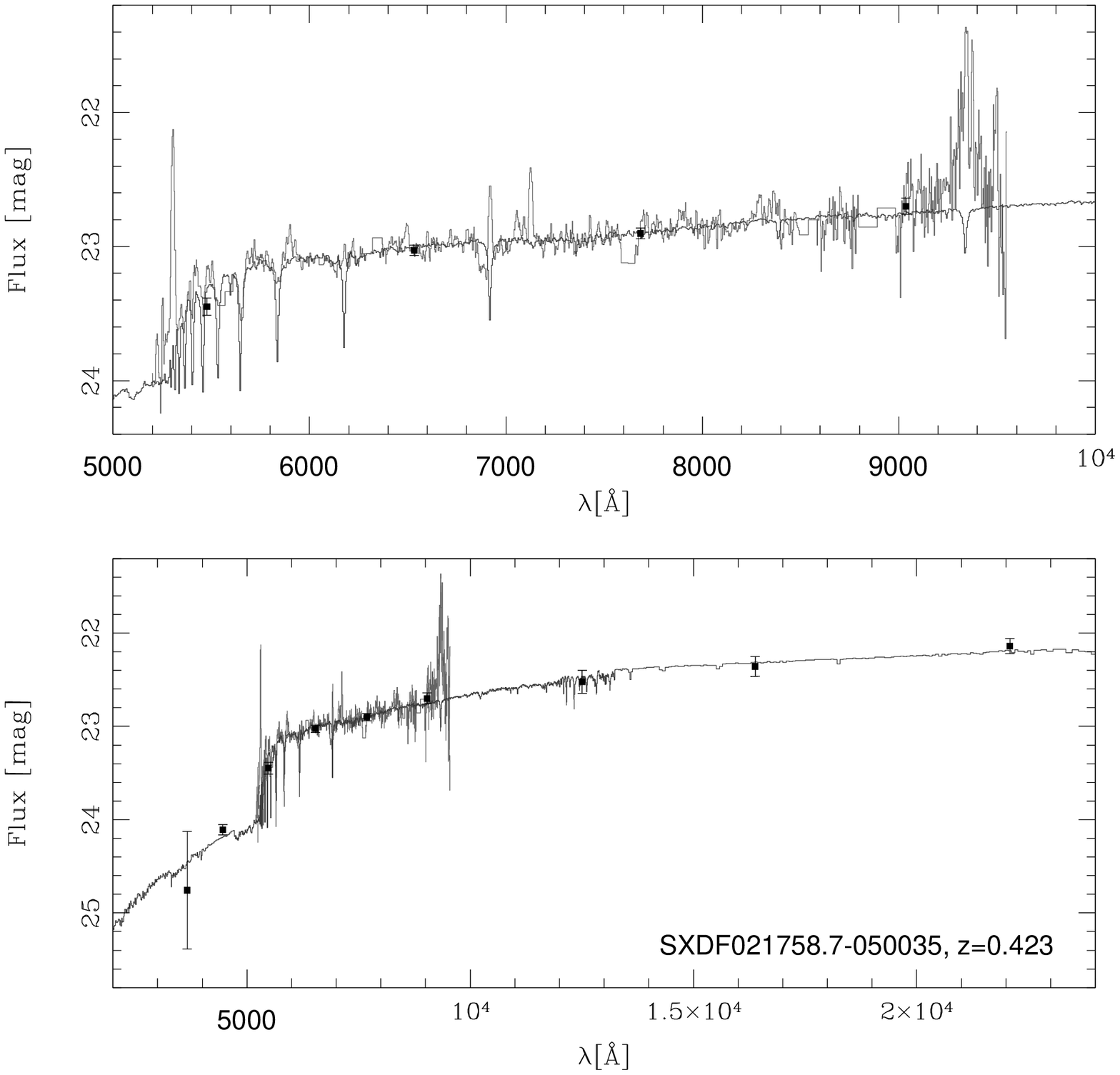}
\includegraphics[width=0.32\textwidth]{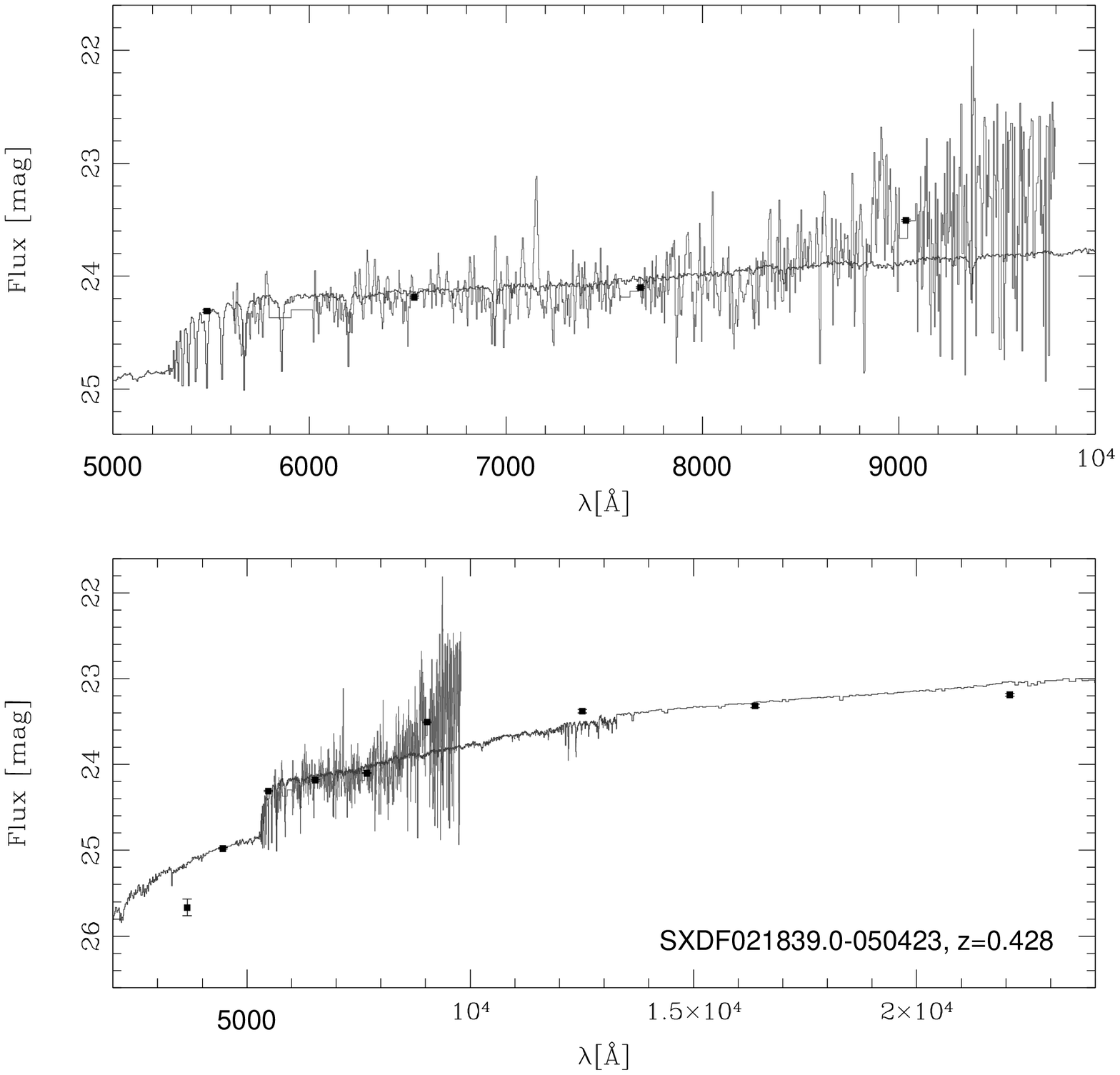}
\includegraphics[width=0.32\textwidth]{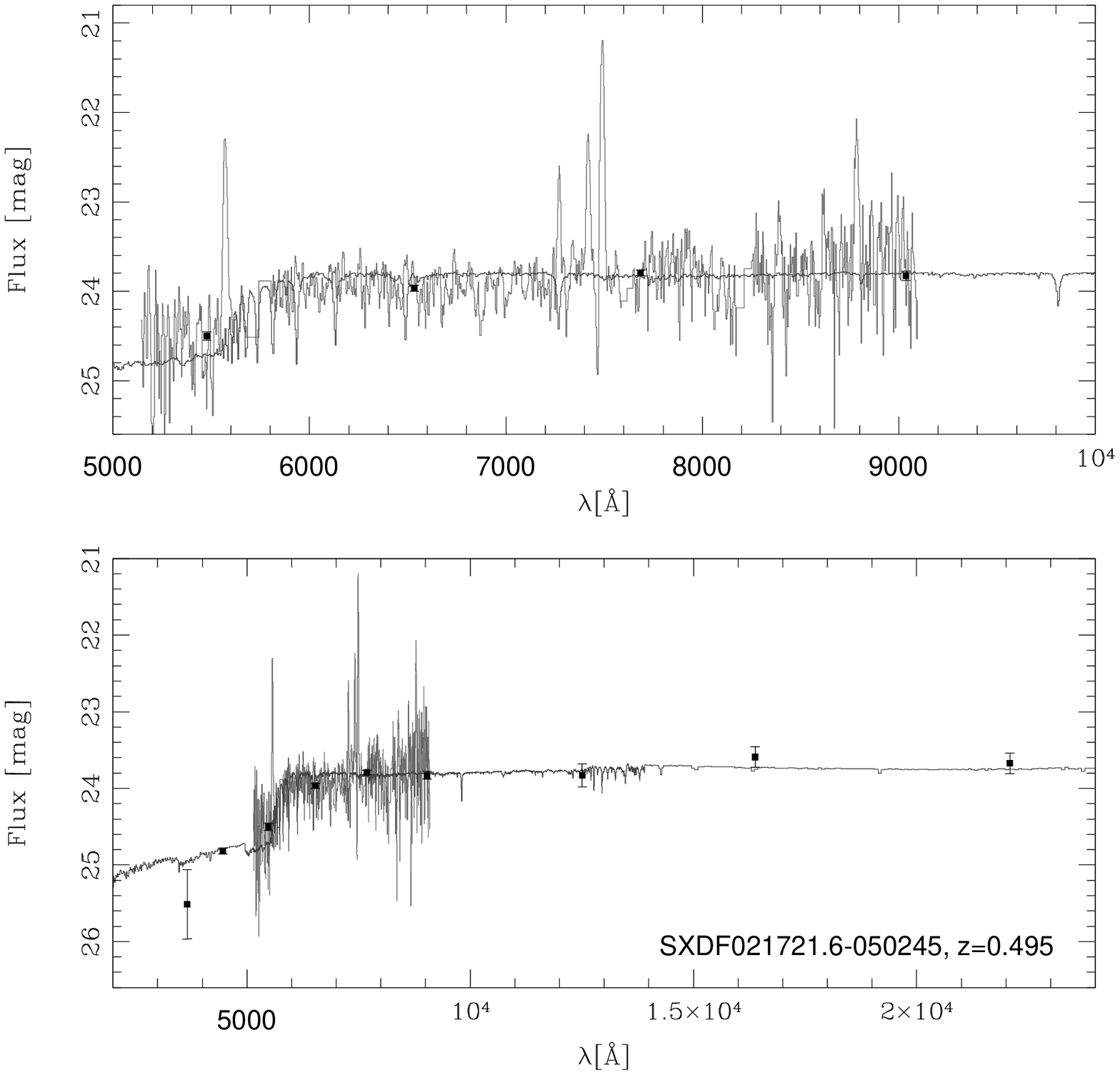}
\includegraphics[width=0.32\textwidth]{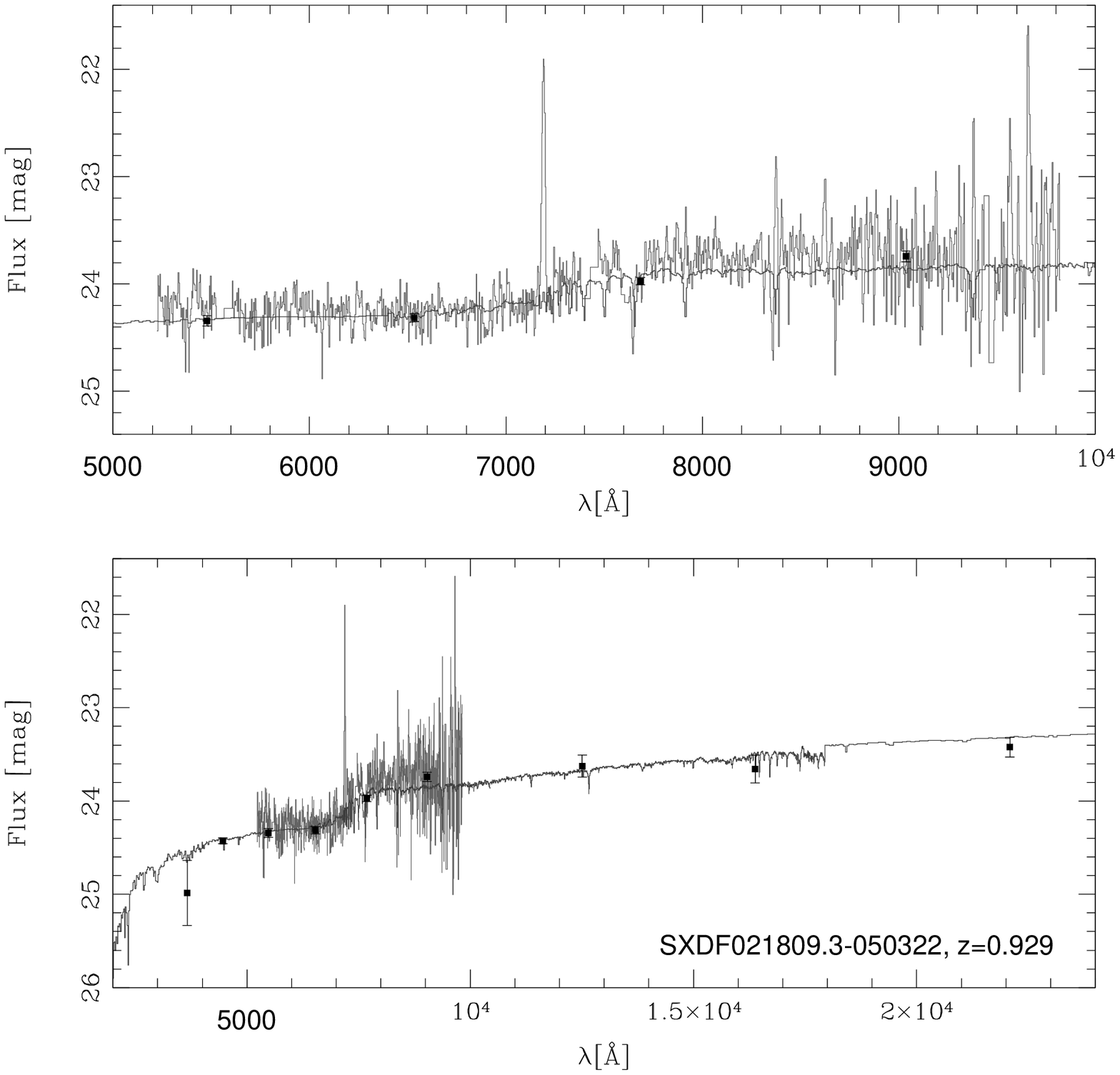}
\includegraphics[width=0.32\textwidth]{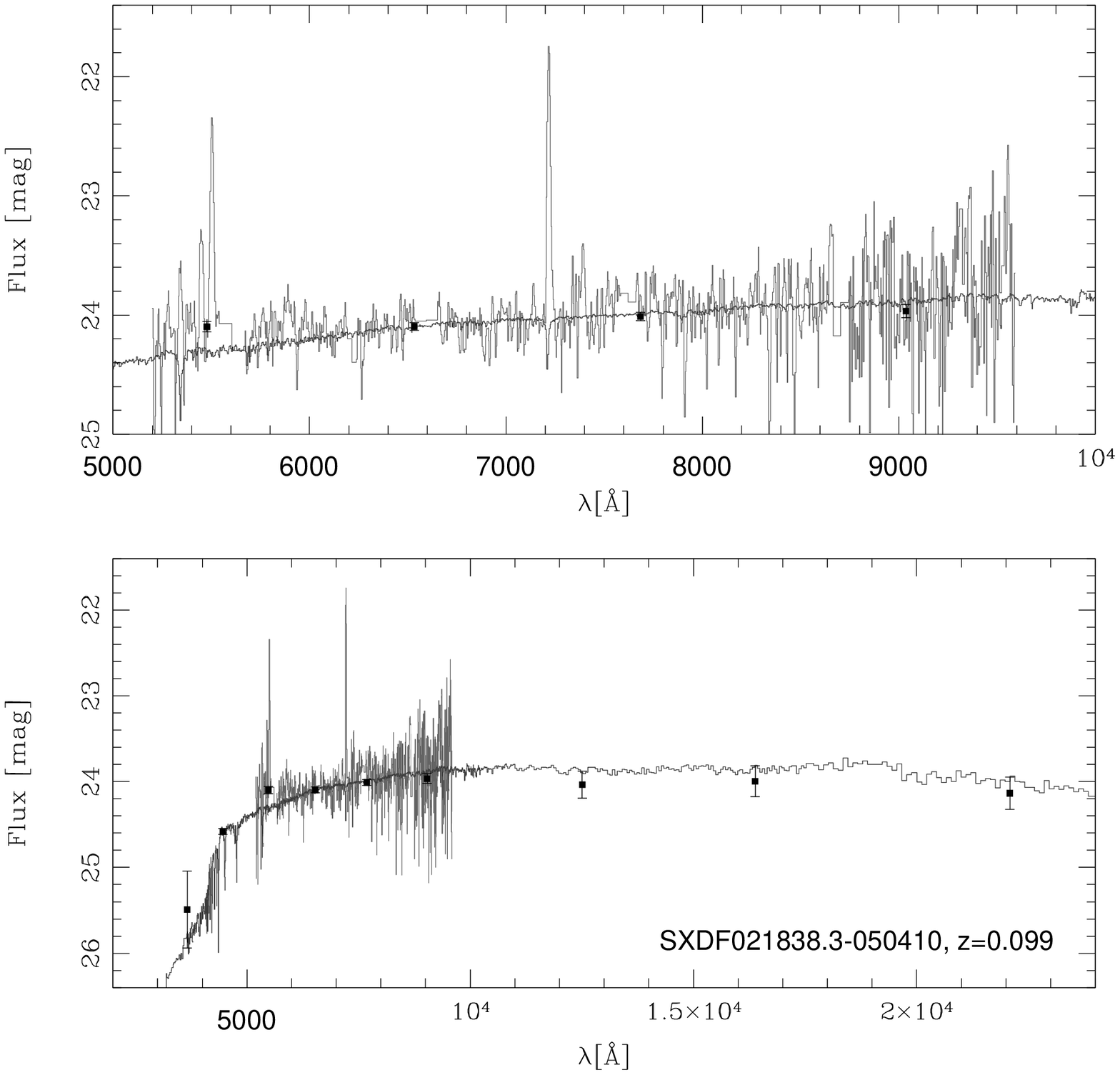}
\includegraphics[width=0.32\textwidth]{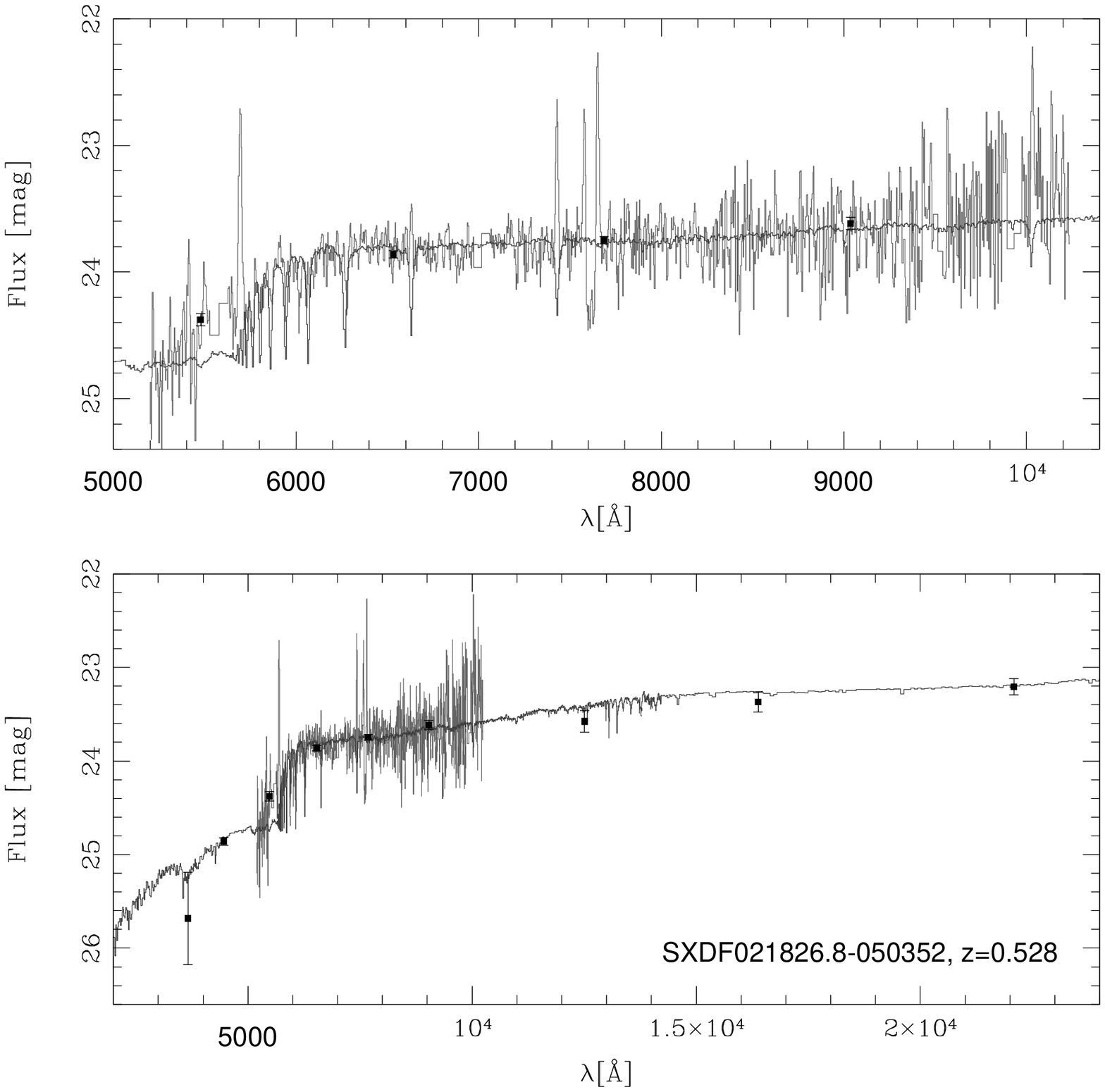}
\includegraphics[width=0.32\textwidth]{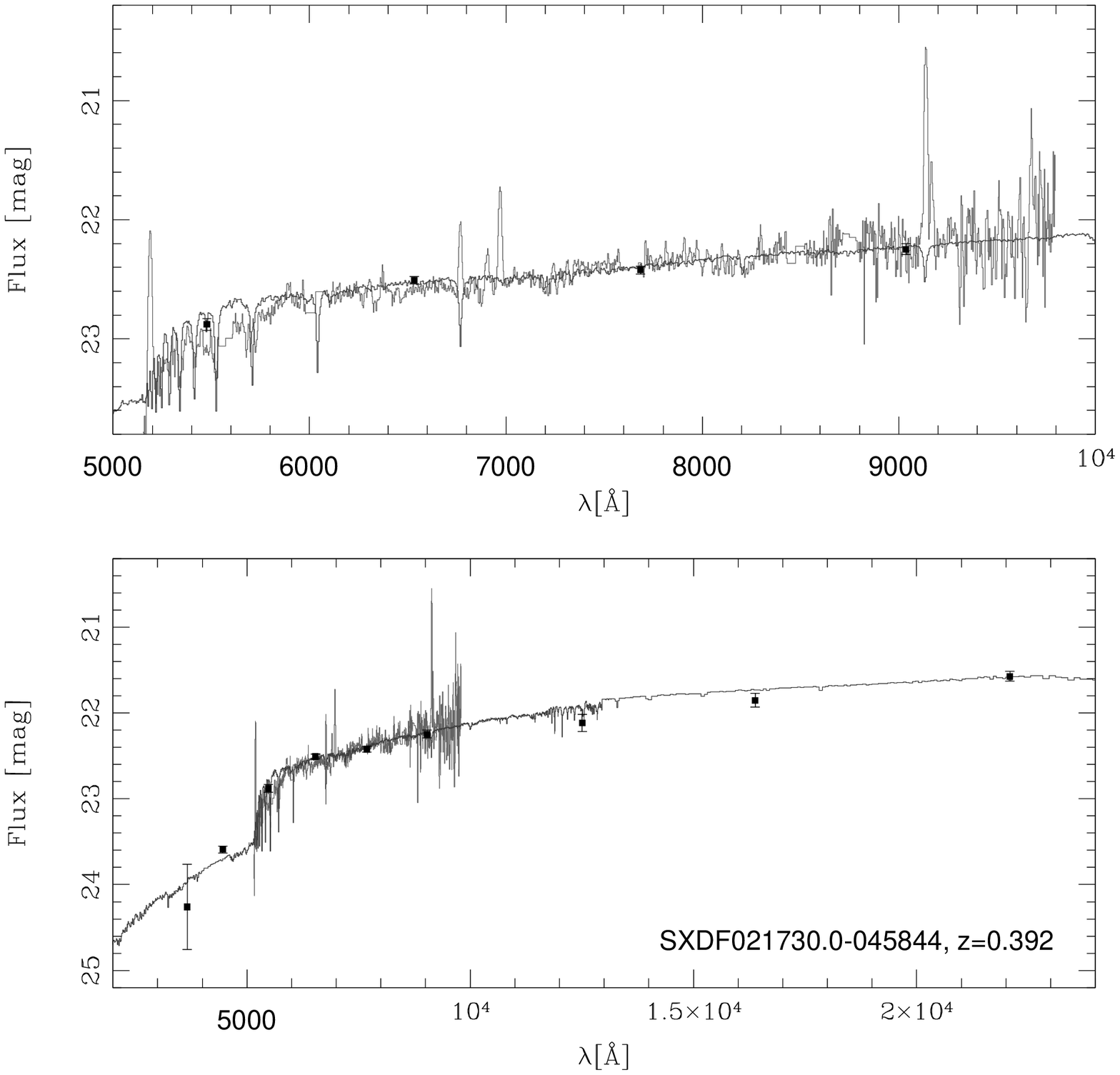}
\includegraphics[width=0.32\textwidth]{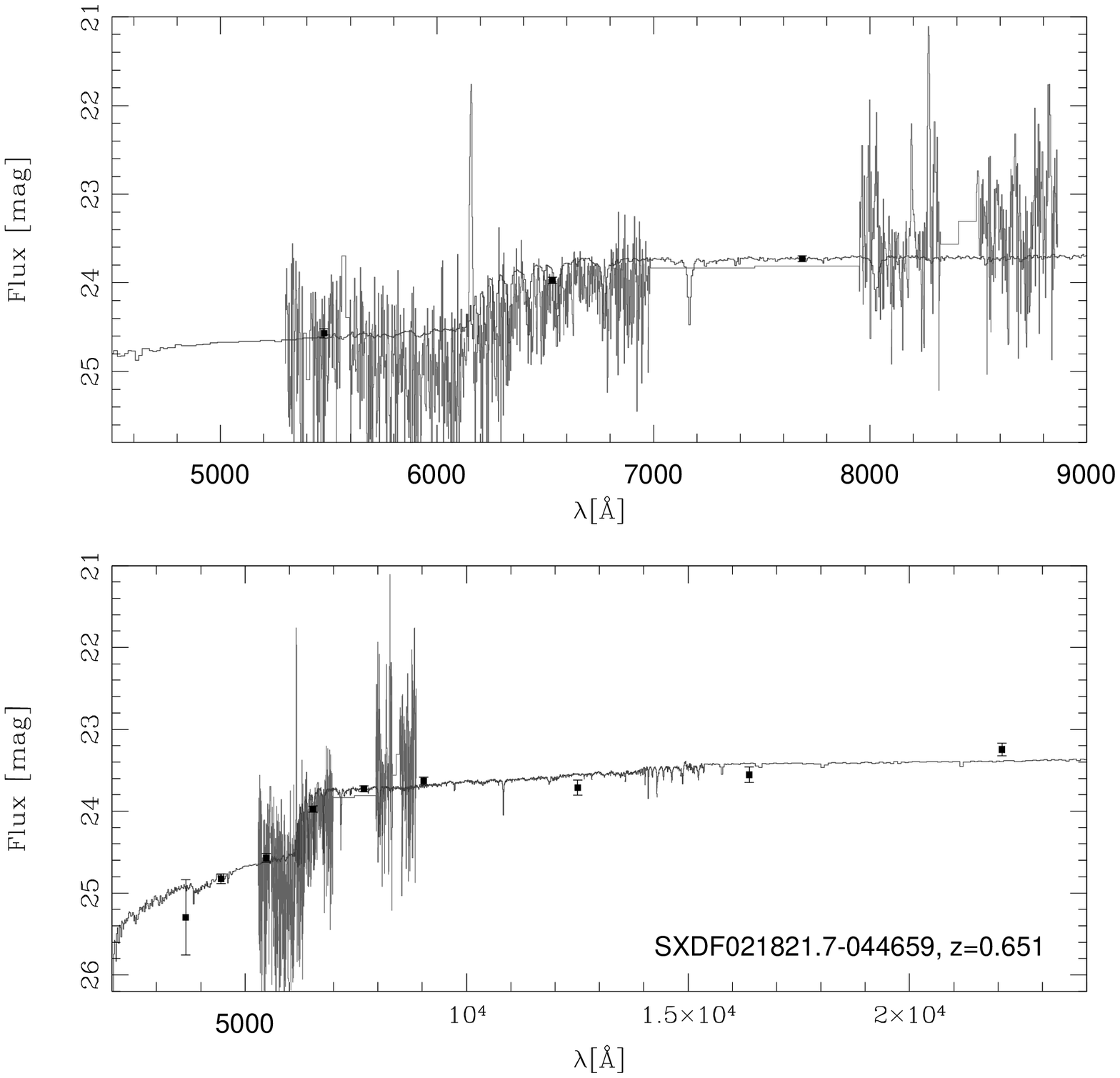}
\caption{\footnotesize{Spectral Energy Distribution (SED) fitting to the spectra and aperture photometry. The figure shows two panels for each galaxy. The lower panel shows the 1" aperture optical, NIR photometry (black squares) and the observed spectrum (in light gray), together with the best fitting template (in dark gray). The upper panel shows in more detail how the template fits the observed optical spectrum.}}\label{SEDs}
\end{center}
\end{figure}

Figure \ref{SEDs} shows the SED fitting to both the spectrum and aperture photometry for our sample galaxies. For each galaxy, the lower panel shows the whole range of SED fitting, while upper panel shows only the wavelength range of the optical spectrum. Black squares represent the photometric data.

\section{THE FINAL SAMPLE}\label{results}

As we remarked in Section \ref{sample}, we only selected those galaxies that revealed the necessary emission lines in order to classify them according to the BPT, Lamareille (\citeyear{Lamareille}, hereafter L10) or Juneau et al. (\citeyear{Juneau}, hereafter J11) diagnostic diagrams, with 37 galaxies being found which fulfilled the required conditions. GMOS observations provided twenty seven of these spectra, while the remaining ten spectra came from the IMACS observations. The parent sample composed of 132 galaxies with valid spectroscopic redshifts formed part of a spectroscopic survey pilot project in the Subaru XMM deep field (SXDF) out to $z$ $\sim$ 2.5, with the purpose of investigating the end of star formation in massive galaxies. Table \ref{spectra} summarizes the adopted names for the objects and origin of the spectrum, together with their coordinates, spectroscopic redshifts, and computed observed frame u$_{AB}$ magnitudes. Figure \ref{redshift} shows the redshift distribution of our galaxy sample. As can be observed, the galaxies were mainly located at 0.3 $<$ $z$ $<$ 0.7, with this range containing 70$\%$ of the sample.

\section{AGN$-$SF ACTIVITY DIAGNOSTIC DIAGRAMS}\label{AGN}

\defcitealias{Lamareille}{L10}
\defcitealias{Juneau}{J11}

AGN emission lines are due to non-thermal ionizing sources and/or shocks, while in star forming (or HII) galaxies they originate from hot young stars. However, AGN and starbursts can also coexist in galaxies (composite galaxies). Throughout this paper, we will refer to AGN as Seyfert 2 or LINER types since there were no Seyfert 1 galaxies in our sample.

In this section we report on the star formation and AGN activity of our galaxies. Different diagnostic diagrams were utilized. As noted in section \ref{sample}, for galaxies at $z$ $<$ 0.5, the BPT diagnostic diagrams were useful to separate star-forming activity from AGN, whereas for galaxies at higher redshifts, the \citetalias{Lamareille} and \citetalias{Juneau} diagnostic diagrams allowed us to separate them well up to $z$ $\sim$ 1. Finally, we used the AGN color$-$color diagrams proposed by Stern et al. \citeyearpar{Stern} and Lacy et al. \citeyearpar{Lacy} to investigate the existence of dusty AGN in the sample. Using the former diagram, galaxies were classified through the (5.8$-$8.0$\mu$m) versus (3.6$-$4.5$\mu$m) colors, while using the latter one the classification was made using the $S$5.8/$S$3.6 versus $S$8.0/$S$4.5 flux ratios. These color indices are responsive to the power law nature of the AGN continuum, which shows redder MIR colors than star-forming galaxies. 

\subsection{\textit{Optical Star-forming$-$AGN Diagrams}}\label{opt}

Kewley et al. \citeyearpar{Kewley} used the SDSS DR4 to show that Seyferts and LINERs formed separate branches on the standard optical BPT diagnostic diagrams, and thereby improved the empirical separation between star-forming, Seyferts, LINER and composite galaxies. Figure \ref{bpts} shows the [OIII]($\lambda$5007)/H$\beta$($\lambda$4861) (hereafter [OIII]/H$\beta$) versus [NII]($\lambda$6548, 6583)/H$\alpha$($\lambda$6563) (hereafter [NII]/H$\alpha$, left panel) and the [OIII]/H$\beta$ versus [SII]($\lambda$6717, 6731)/H$\alpha$($\lambda$6563) (hereafter [SII]/H$\alpha$, right panel) diagnostic diagrams for the sample galaxies at $z$ $<$ 0.5 and $z$ $<$ 0.48, respectively. On the left panel, three galaxies (SXDF021834.7-050432, SXDF021758.7-050035, and SXDF021839.0-050423) that could host an AGN can be observed, while the remaining ones are clearly of the star-forming type. It is interesting to note that two of these galaxies (SXDF021834.7-050432, SXDF021758.7-050035) were also classified as AGN from their MIR colors (see section \ref{mir}). On the other hand, the right panel does not show any AGN candidates. Unfortunately, it was not possible to measure the [SII]$\lambda$6717, 6731 lines in the previous AGN candidates due to their redshifts ($z$ $>$ 0.48), which placed the emission lines beyond the observed spectral range. Neither was it possible to detect the [OI]$\lambda$6300 emission line in any sample galaxy, since this line was relatively weak, and our spectra had low S/Ns.

\defcitealias{Kewley}{K06}

According to the \citetalias{Kewley} diagrams, there were 1 Seyfert 2, 2 composite galaxies and 15 star-forming ones out of the sample of 18 classified galaxies, corresponding to percentages of 5$\%$, 11$\%$ and 83$\%$, respectively. This result is in reasonable agreement with that presented by \citetalias{Kewley}, who found 3$\%$ of Seyfert 2, 7$\%$ of LINERs and 7$\%$ of composite galaxies for a sample with 0.04 $<$ $z$ $<$ 0.1. Although our sample had almost twice the percentage of Seyfert and composite galaxies than that of \citetalias{Kewley}, care must be taken about this result since our sample is small and incomplete. 

\subsubsection{The Blue Star-forming-AGN Diagram}

In our study, the \citetalias{Lamareille} diagram proved useful for galaxies at $z$ $>$ 0.5, since H$\alpha$6563$\AA{}$ was well beyond the spectral range available in our data. Figure \ref{hzbpt} shows [OIII]/H$\beta$ versus the equivalent width ratio W$_{[OII]}$($\lambda$3727)/W$_{H\beta}$($\lambda$4861) diagram and reveals three galaxies classified as AGN: two are Seyfert 2 galaxies (SXDF021834.7-050432 and SXDF021721.6-050245), and one is LINER (SXDF021821.7-044659). It is interesting to note that these galaxies were also classified as AGN by their MIR color diagrams (see section \ref{mir}). From these latter diagrams two additional galaxies were AGN candidates, with one being a star-forming$-$Seyfert 2 (SXDF021758.7-050035), and the other a star-forming$-$composite (SXDF021836.9-045046).

Out of 23 galaxies that were in the \citetalias{Lamareille} diagram, there were 2 (8$\%$) Seyfert 2, 1 (4$\%$) LINER, 10 (43$\%$) star-forming$-$Seyfert 2 galaxies, 3 (13$\%$) star-forming$-$composite galaxies and 7 (30$\%$) star-forming galaxies. Assuming that the star-forming$-$Seyfert 2 and star-forming$-$composite regions are the only star-forming types, we obtained an upper limit of 20 (87$\%$) for star-forming. In a much larger sample of 1213 galaxies with 0.5 $<$ $z$ $<$ 0.9, Lamareille et al. \citeyearpar{Lamareille2} found 3$\%$ of Seyfert 2's, no LINERs, 19$\%$ of composites galaxies and 78$\%$ of star-forming galaxies. However, it should be noted that the classification scheme used by Lamareille et al. \citeyearpar{Lamareille2} was slightly different from the one we used. In our case, the new classification proposed by \citetalias{Lamareille} was used.

\subsubsection{Mass-excitation Star-forming-AGN Diagram}\label{MEx}

\citetalias{Juneau} proposed the Mass-Excitation (MEx) diagnostic diagram to identify star-forming galaxies and AGNs at intermediate redshift, using the SDSS DR4 to show that combining the [OIII]/H$\beta$ line ratio with the total stellar mass successfully distinguished AGN from star formation emission. This diagram relies on the fact that AGNs are mainly hosted in massive galaxies with stellar masses of M$_{\ast}\gtrsim10^{10}M_{\odot}$ (\citeauthor{Kauffmann,Kewley}) and that the [OIII]/H$\beta$ line ratio is a well known feature for recognizing AGNs in classical BPT diagrams. 

The total stellar masses were estimated by SED fitting (see section \ref{mass} for more details), with figure \ref{maex} showing the MEx diagram for our sample of galaxies. Note that this MEx diagram was calibrated using the Chabrier IMF (\citeauthor{Chabrier}), and therefore in this figure our stellar mass values were converted from the Salpeter IMF to the Chabrier IMF. According to \citetalias{Juneau}, this diagram shows 3 regions. Of these, the one located above the empirical curve corresponds to AGN galaxies, while the one below corresponds to star-forming galaxies. Transition objects (equivalent to composite galaxies in the \citetalias{Kewley} diagrams) are located in the other small region between the two empirical curves. It can be observed from this figure that only 1 AGN (SXDF021834.7-050432) was detected among our 6 AGN candidates selected from the MIR colors and emission line diagrams. This raises the question whether the MEx diagram is in fact effective for detecting AGNs in low-mass hots. For this reason, we used the Greene $\&$ Ho \citeyearpar{Greene2}, and Barth, Greene and Ho \citeyearpar{Barth} data, and specifically we chose those galaxies with total masses lower than 10$^{10}M_{\odot}$ in order to check the relative number of AGN detected using the MEx diagram. Figure \ref{maex} also shows that the MEx diagram missed a high percentage (more than 70$\%$) of the AGNs in low-mass host galaxies. 

\subsection{\textit{MIR Star-forming$-$AGN Diagram}}\label{mir}

The \textit{Spitzer} data was used to select galaxies with strong emission in the MIR regime relative to stellar emission. From these data three types of emission could be distinguished: (1) galaxies dominated by the AGN, which had a power law like emission in the 3$-$10$\mu$m band that originated from very hot dust heated by the intense AGN radiation field, (2) star-forming galaxies rich in gas and dust and largely dominated by Poly Aromatic Hydrocarbon (PAH) features\footnote{PAHs are aromatic molecules ubiquitous in the interstellar medium (ISM) of our own galaxy and nearby galaxies with ongoing or recent star formation.}, and (3) starbursts, which had a very steeply rising continuum at 12$-$16$\mu$m (\citeauthor{Laurent,Weedman,Smith}). However, there were galaxies with a power law SEDs in the MIR that also showed a slight increase in the NIR, which might have been associated to the redshifted 1.6$\mu$m peak from the stellar continuum, suggesting that both AGN and star formation activity may coexist in these galaxies (\citeauthor{Alonso,Donley,Lacy}).

There were 19 galaxies in our sample with available photometry in all IRAC/\textit{Spitzer} bands. For these galaxies we constructed color$-$color diagrams, with figure \ref{iracds} showing the Lacy et al. \citeyearpar{Lacy} log($S_{8.0}/S_{4.5}$) versus log($S_{5.8}/S_{3.6}$) diagram (left panel) and the Stern et al. \citeyearpar{Stern} (3.6$-$4.5$\mu$m) versus (5.8$-$8.0$\mu$m) diagram (right panel). It is interesting to note that while the Stern et al \citeyearpar{Stern} diagram revealed 6 AGN candidates, the Lacy et al. \citeyearpar{Lacy} diagram only showed 1 AGN candidate. It should also be noted, that 3 of these 6 candidates were classified as AGN through the \citetalias{Kewley} (squares) and \citetalias{Lamareille} (triangles) diagrams, namely SXDF021834.7-050432, SXDF021721.6-050245 and SXDF021821.7-044659. These results are consistent with low redshift Seyfert 2 galaxies with SEDs being dominated by the host rather than the AGN light, resulting in then appearing a bluer color (3.6$-$4.5$\mu$m) than high luminosity AGNs (\citeauthor{Brusa,Eckart,Donley}), with SEDs dominated by the AGN. In addition, there were also 3 (50$\%$) AGN candidates that were not classified as AGN according to the \citetalias{Kewley} and \citetalias{Lamareille} diagrams, namely SXDF021758.7-05003, SXDF021836.9-045046 and SXDF021838.3-050410. These were interpreted by Lacy et al. \citeyearpar{Lacy2} as obscured AGN as they found a similar result (46$\%$) from a sample of 35 galaxies. Moreover, Goulding $\&$ Alexander \citeyearpar{Goulding} showed that in a volume-limited sample of the most bolometric luminous galaxies, 50$\%$ of the IR AGN candidates were not identified as AGN using optical spectroscopy. Finally\footnote{Another possibility is that some MIR AGN candidates may be star-forming contaminants.}, it is very well known that star formation activity could dilute the AGN optical spectral signatures, with $z$ $\sim$ 0.5, 1" corresponding to $\sim$5 kpc, which is 2.5 times the average size for the circumnuclear star-forming regions (\citeauthor{Pastoriza,Greene}). Similar examples of this effect can be found in Colina et al. \citeyearpar{Colina}, Seth et al. \citeyearpar{Seth} and Wright et al. \citeyearpar{Wright}.

\subsection{\textit{X-Ray Counterparts?}}\label{Xray}

The SXDF has been observed for the XMM survey through its different missions (for instance, \citeauthor{Watson}). Barcons et al. \citeyearpar{Barcons} used the XMM Medium sensitivity Survey (XMS) to describe the population responsible for these intermediate fluxes ($\sim$10$^{-14}$ erg cm$^{-2}$ s$^{-1}$) in various X-ray energy bands. In the SXDF they found 30 AGNs at redshift 0.04 $<$ $z$ $<$ 2.209, of which 14 were at $z$ $<$ 1.0, with seven of these being broad line AGNs. A search was made in public X-ray catalogues for counterparts of our AGN candidates. Ueda et al. \citeyearpar{Ueda} published X-ray data, observed with XMM, with sensitivity limits of 6$\times$10$^{-16}$, 8$\times$10$^{-16}$, 3$\times$10$^{-15}$ and 5$\times$10$^{-15}$ erg cm$^{-2}$ s$^{-1}$ in the 0.5$-$2, 0.5$-$4.5, 2$-$10 and 4.5$-$10 keV bands. However, no counterparts were found in this catalogue or in the Barcons sample, even for our massive AGN SXDF021834.7-050432 that satisfied the AGN criteria for the \citetalias{Kewley}, \citetalias{Lamareille} and Lacy et al. \citeyearpar{Lacy} diagrams. This result would suggest that the AGN candidates in our sample belonged to the low luminosity class ($L_{X}<$10$^{43}$ erg s$^{-1}$) , and also that both the X-ray and optical surveys may have suffered from biases against the most obscured phases of BH growth. Similar examples can also be found in Poletta et al. \citeyearpar{Polletta} and Caccianiga et al. \citeyearpar{Caccianiga}.

An estimation was made of the fluxes that our sources could had in the 2$-$10 keV band, based on the relation found between color $(R-K)_{Vega}$ vs log($L_{X}$) at 2$-$10 keV for type 2 AGN (\citeauthor{Brusa}). The most massive AGN in the sample had $R-K$=3.0, which corresponded to an $L_{X}\sim$ 6$\times$10$^{42}$ erg s$^{-1}$ (see fig. 7, right panel from \citeauthor{Brusa}). At the redshift of the galaxy, this luminosity implied $S_{X}\sim$ 3$\times$10$^{-15}$ erg cm$^{-2}$ s$^{-1}$, corresponding to the detection limit flux of the XMM survey. The remaining AGN candidates had lower $R-K$ values and similar redshifts as the most massive AGN in our sample, which implied even lower X-ray fluxes.

\section{PHYSICAL PROPERTIES OF THE SAMPLE}\label{prop}

\subsection{\textit{Reddening}}\label{dust}

Since we expected some of our galaxies to be significantly affected by dust (\citeauthor{Calzetti2,Hopkins,Papovich}), the internal stellar reddening was estimated for each galaxy by adopting the dust extinction law from Calzetti et al. \citeyearpar{Calzetti2}, which is a simple but reasonable approximation to better known extinction laws such as those of Fitzpatrick (\citeyear{Fitzpatrick}, SMC type), Fitzpatrick (\citeyear{Fitzpatrick2}, Milky Way type) and Bouchet et al. (\citeyear{Bouchet}, LMC type). A value of Rv=4.05, instead of the typical value of Rv=3.1 (Milky Way) was used as this was the most suitable value for our sample galaxies, given that moving to higher redshifts would have implied galaxies with higher luminosity, greater star formation, and higher dust amounts than those observed in local galaxies (i.e. actively star-forming galaxies, \citeauthor{Calzetti2}). Using this extinction law with a set of $E(B-V)$ values to run the $\chi^{2}$ SED fitting program, it was observed that the sample galaxies had stellar $E(B-V)$ values in the range [0.0; 0.3].

For galaxies that showed the necessary emission lines, the nebular $E(B-V)$ was also measured using the Balmer decrement or the [OII]$\lambda$3727 luminosity recipe given by Kewley, Geller and Jansen \citeyearpar{Kewley2}. The ratio H$\alpha$/H$\beta$=2.85 was adopted for the star-forming galaxies, while for AGNs we chose H$\alpha$/H$\beta$=3.1 (\citeauthor{Gaskell,Veilleux,Osterbrock}). Figure \ref{red} shows a comparison between the stellar reddening (obtained from SED fitting) versus nebular extinction (obtained from the Balmer decrement, left panel; and from the [OII] luminosity, right panel), where in general terms, the nebular reddening can be observed to be higher than the stellar reddening. This can be explained by considering that ionizing stars are close to dusty molecular clouds, while non-ionizing stars responsible for the UV$-$optical continuum reside in regions with smaller dust amounts (\citeauthor{Calzetti}). However, we also had a few exceptions that revealed the opposite behaviour, with 2 galaxies showing low S/N spectra together with the H$\alpha$ emission lines redshifted to 9000-10000$\AA{}$, a region where the sky line subtraction was subject to large residuals.

\subsection{\textit{Ages}}\label{Edad}

Galaxies are composed of stellar populations of different ages and metallicities related to their SFH. The $\chi^{2}$ statistics were used to find the age of the dominant stellar population, for two types of SFHs: passive evolution (a single burst, i.e. SSP) and exponentially declining histories (SFR($t$)$\sim$ $e^{-t/\tau}$, for 0.25 Gyr$<\tau<$15 Gyr). In general terms, we observed that the exponentially declining model showed much better $\chi^{2}$ values than the SSP models. However, for a few sample galaxies both models provided similar $\chi^{2}$ values and ages, but it should be noted that the age parameter was no limited by the age of the Universe. Figure \ref{ages} shows the age distribution of the sample galaxies (top panel), which peaked at 10$^{9}$ yr. The bottom panel shows age versus spectroscopic redshift, where it can be observed that younger galaxies had higher redshifts.

In the left panel of figure \ref{colorage}, the age values were compared with the rest frame color index $(u-B)$, a good indicator of galaxy age (\citeauthor{Rudnick,Kriek}). The right panel shows how the dispersion of the data disappeared almost completely when a dust extinction correction was applied. However, there were two distinct branches, which can possible be explained as an effect of a degeneracy in the color$-\tau$ (SFR mean lifetime) parameter space of the SED model. It is also interesting to note that the fits obtained before and after dust extinction correction were very similar (slopes $\sim$1.75).
 
\subsection{\textit{Stellar Masses}}\label{mass}

Total magnitudes were used in all available bands of the sample galaxies to calculate their total stellar masses. For this purpose, we used the same spectral library generated with the GALAXEV code (see section \ref{SED}), and applied a least $\chi^{2}$ fitting method to obtain the best template that matched the calculated total magnitudes. Figure \ref{histm} shows the mass distribution of our galaxies (top panel). This plot peaked at 10$^{9.25}M_{\odot}$ and it ranged from 10$^{7.5}M_{\odot}$ to 10$^{11}M_{\odot}$. Most of the galaxies (72$\%$) had masses in the range 10$^{8.5}-$10$^{10}M_{\odot}$, and therefore the sample was mainly composed of low mass galaxies. This could imply a problem when comparing AGN in other samples that included massive (M$_{\ast}\gtrsim$ 10$^{10}M_{\odot}$) galaxies. The bottom panel shows stellar masses as a function of redshift, where it can be observed that galaxies with M$_{\ast}>$ 10$^{9}M_{\odot}$ were located throughout the redshift range probed by our sample, whereas the massive ones (M$_{\ast}\gtrsim$ 10$^{10}M_{\odot}$) had redshift values of 0.35 $<$ $z$ $<$ 0.75. Furthermore, these 4 galaxies with M$_{\ast}\gtrsim$ 10$^{10}M_{\odot}$ were the only objects of the sample with a FIR detection at 24 $\mu$m. Note that AGNs (black squares) was detected at $z$ $>$ 0.4. 

Figure \ref{color} shows the total stellar mass versus rest frame color $(u-B)_{AB}$ corrected by reddening, where solid lines delimit the so called ``blue cloud'' (lower), ``green valley'' (intermediate) and ``red sequence'' (upper) regions (\citeauthor{Baldry,Weiner,Mendez}). It can be observed that there were no galaxies located in the red sequence region and that all but one galaxy was located in the blue cloud region. It is interesting to note that the only object located in the green valley region was SXDF021834.7-050432, which was the most massive AGN (black squares) of the sample. 
This result clearly indicates that most of the sample galaxies had strong star formation and therefore this might also explain the low number of AGNs detected ($\sim8\%$) by the optical diagnostic diagrams. In other words, the starburst was conspicuous (\citeauthor{Pastoriza,Greene}).

\subsection{\textit{Black Hole Masses}}\label{BH}

It is widely accepted that all galaxies with a massive bulge component contain a central massive BH. The good correlations between the mass of the central BH and the physical properties of the surrounding stellar bulge have provided evidence that BHs play a key role in the evolution of galaxies. Bennert et al. \citeyearpar{Bennert} also found the following correlation between the BH mass ($M_{\text{BH}}$) and the total stellar mass of the host ($M_{\text{host},\ast}$):

\begin{equation}
\log\left( \frac{M_{\text{BH}}}{M_{\odot}} \right) = 1.12 \log\left( \frac{M_{\text{host},\ast}}{M_{\odot}} \right) +1.15 \log (1+z) - 19.88.
\end{equation}

Moreover, the X-ray luminosity ($L_{X}$) produced by the BH of our AGN candidates in the 2$-$10 keV range can be derived using the equation of Kiuchi et al. \citeyearpar{Kiuchi}:

\begin{equation}
\log \left( \frac{L_{2-10 \text{keV}}}{\text{erg s$^{-1}$}} \right) = \log \left( \frac{M_{\text{BH}}}{M_{\odot}} \right) - \log \left( \frac{BC}{30} \right) + \log \left( \frac{\lambda}{0.1} \right) + 35.6.
\end{equation}

This is particularly interesting since this is an independent way to calculate X-ray luminosities for our galaxies and these values can then be compared to those obtained in section \ref{Xray}. Using a mean redshift of 0.5, for a total stellar mass range of 10$^{8.2}-$10$^{9.7}M_{\odot}$ (calibrated to Chabrier IMF), a BH mass range of 10$^{5.5}-$10$^{7.2}M_{\odot}$ was obtained. Similar values were found in low-mass Seyfert 2 by Barth, Greene and Ho \citeyearpar{Barth}. 

We have assumed conservative values for the bolometric correction (BC=30) and for the Eddington ratio ($\lambda$=0.1), (\citeauthor{Kiuchi,Ballo,Goulding2}). Figure \ref{Lx} shows the $L_{X}$ obtained from the BH mass $L_{X, 2-10 \text{keV}, M_{\text{BH}}}$ against the $L_{X}$ predicted by the $(R-K)_{Vega}$ color $L_{X, 2-10 \text{keV}, R-K_{Vega}}$ (\citeauthor{Brusa}, section \ref{Xray}), with the solid line defining a 1:1 relation. The arrows show how $L_{X, 2-10 \text{keV}, M_{\text{BH}}}$ would vary for other values of $\lambda$ and BC. In general terms, there is a fair agreement between both X-ray luminosity estimates. The differences found might be attributed to the $\lambda$ and BC values adopted, which are not necessarily the same for all galaxies.

\subsection{\textit{Star Formation Rates}}\label{SFR}

The emission line fluxes from H$\alpha$ 6563 $\AA{}$, and [OII]$\lambda$3727 lines were used to estimate the SFRs of our galaxy sample. To carry this out, the new estimators corrected by the dust attenuation of Kennicutt et al. \citeyearpar{Kennicutt2} were utilized. In addition, the SFR were determined in an independent way using the UV continuum obtained from the fitted templates and the estimators of Kennicutt (\citeauthor{Kennicutt,Kennicutt2}). Line fluxes were reddening corrected using the Balmer decrement or the Kewley, Geller and Jansen \citeyearpar{Kewley2} recipe depending on the emission lines available in the spectrum. The UV fluxes were corrected using the $E(B-V)$ values estimated from SED fitting (see section \ref{dust}), with the left panel in figure \ref{sfr} showing SFR$_{\text{H}\alpha,\text{[OII]}}$ versus SFR$_{\text{UV}}$. In this plot, there are 4 duplicate galaxies (connected points) for which it was possible to compute the SFR values from both H$\alpha$ and [OII]. As can be observed, the SFR derived from both methods correlated fairly well (the dashed line shows a linear fit). However, the plot shows that the dust-corrected UV estimator may in fact overestimate the SFR with respect to that calculated via emission lines. Indeed, as the UV fluxes were estimated via SED extrapolation from UV$-$optical$-$NIR wavelengths to FUV/NUV regions, the UV fluxes may also have been overestimated. A much better correlation was observed for the SFR$_{\text{H}\alpha,\text{[OII]}}$ versus SFR$_{\text{SFH}}$, with the SFR from an exponentially declining star formation history (right panel) indicating that the SFR obtained by both methods were remarkably similar.

\subsubsection{SFR Cosmic Evolution}

\defcitealias{Silverman}{S09}

Our galaxies spanned the redshift range 0.04 $<$ $z$ $<$ 1.0, which therefore allowed us to study the SFR cosmic evolution. For this purpose, the SFR surface density $\Sigma_{\text{SFR}}$ was estimated in the inner 2.5 kpc region of the sample galaxies. This radius represented the mean projected size for a 1" aperture at the mean distance of the sample. In the same way, the stellar mass per unit area and time as $\Sigma_{M_{\ast}/\tau}$ was defined, by dividing the stellar mass per unit area by the age of the dominant stellar population calculated in the SED fitting. This value represented the average SFR with which the galaxies gathered their stellar mass, under the hypothesis of a constant SFR. Figure \ref{sfrevol} (left panel) shows the $\Sigma_{\text{SFR}_{\text{[OII],H}\alpha}}$ (blue squares) obtained from [OII]$\lambda$3729 or H$\alpha$ lines, with $\Sigma_{M_{\ast}/\tau}$ (red squares) showing our galaxies as a function of redshift. The solid blue line is the linear fit for $\Sigma_{\text{SFR}_{\text{[OII],H}\alpha}}$, while the long-dashed red line shows the fit for $\Sigma_{M_{\ast}/\tau}$. The figure also shows the $\Sigma_{\text{SFR}_{\text{[OII]}}}$ values for zCOSMOS and SDSS star-forming galaxies (\citeauthor{Silverman}) with redshifts in the range 0 $<$ $z$ $<$ 1.02 (triangles). These values were calculated using Figure 10 from Silverman et al. (\citeyear{Silverman}, hereafter S09), and then the \citetalias{Silverman} SFR values were divided by the projected area corresponding to the slit aperture at the galaxy redshift. The short-dashed line shows the fit for this sample. As can be observed, the slope of the linear fit for $\Sigma_{M_{\ast}/\tau}$ (1.06$\pm$0.38) was very similar to that obtained for the \citetalias{Silverman} $\Sigma_{\text{SFR}_{\text{[OII]}}}$ values ($\sim$1.13), and was lower than the one obtained for our $\Sigma_{\text{SFR}_{\text{[OII],H}\alpha}}$ values. This result encouraged us to investigate possible bias effects. We determined whether this effect was consistent with a possible bias due to the low mass galaxies present in our sample, as galaxies with higher stellar mass values tended to have higher SFRs. In order to avoid this possible effect, the SFR values for those galaxies with M$_{\ast}<$ 1$\times$10$^{9}M_{\odot}$ were scaled using the relation found by \citeauthor{Elbaz} (SFR $\propto$ $M_{\ast}^{0.9}$), and this value was used considering our mass detection limit shown in fig. \ref{histm}. In figure \ref{sfrevol} (right panel), it can be seen that the slope of the new fit (1.18$\pm$0.29) was now consistent with \citetalias{Silverman}, which implied a similar increasing rate of $\Sigma_{\text{SFR}}$ as a function of redshift.

\subsubsection{SFR and Specific SFR Relations}

\defcitealias{Elbaz}{E07}
\defcitealias{Brinchmann}{B04}

The relation between SFR and total stellar mass in the galaxy sample was now investigated. Finlator et al. \citeyearpar{Finlator} studied the physical properties of LBGs at z=4 in cosmological hydrodynamics simulations, and found a strong correlation between SFR and stellar mass, with a slope of $\sim$1.14 in their simulated galaxy sample. Although Weinberg, Hernquist and Katz \citeyearpar{Weinberg} had already reported this correlation in their simulated galaxies in LCDM models at z=3, no attempt was made to calculate its slope. Daddi et al. \citeyearpar{Daddi2} and Elbaz et al. \citeyearpar{Elbaz} also observed this relation using data from the Great Observatories Origins Deep Survey (GOODS) at 1.4 $<$ $z$ $<$ 2.5 and 0.8 $<$ $z$ $<$ 1.2, respectively. Figure \ref{massfr} shows log(SFR$_{\text{[OII],H}\alpha}$) versus log($M_{\ast}$) for the galaxies separated into two groups: intermediate redshift galaxies (0.5 $<$ $z$ $<$ 1.0, gray triangles) and low redshift galaxies (0 $<$ $z$ $<$ 0.5, black squares). The obtained linear fits are shown for both subsamples as a long dashed line (intermediate redshift galaxies) and a solid line (low redshift galaxies). For comparison, the fit obtained for the SDSS galaxies with $z$ $<$0.2 (\citeauthor{Brinchmann}) is also given (dashed line). Similarly, the dotted line shows the trend observed in the GOODS sample by Elbaz et al. (\citeyear{Elbaz}, hereafter E07), while the dot-dashed line shows the fit calculated by Daddi et al. \citeyearpar{Daddi2}. It is remarkable that all samples had similar slopes and also a zero point shift that correlated with the redshift range of the galaxy sample. Our calculated slopes were 0.76$\pm$0.30 for the intermediate redshift subsample, comparable to the value reported by \citetalias{Elbaz} (0.9); and 0.77$\pm$0.1 for the low redshift subsample, quite similar to that obtained by Brinchmann et al. (\citeyear{Brinchmann}, hereafter B04; 0.77). Similar values were also found by Santini et al. \citeyearpar{Santini} using GOODS-MUSIC galaxies at 0.3 $<$ $z$ $<$ 1.0, who found slopes of 0.73 and 0.70 for their subsamples at 0.6 $<$ $z$ $<$ 1.0 and 0.3 $<$ $z$ $<$ 0.6, respectively. Even though our sample was relatively small, it gives additional support to the Daddi et al. \citeyearpar{Daddi2} and \citetalias{Elbaz} results, i.e. for a given stellar mass the SFR is on average larger at higher redshifts.

The specific SFR (SSFR) is a measure of the rate at which new stars are added relative to the total stellar mass of a galaxy. Brinchmann $\&$ Ellis \citeyearpar{Brinchmann2}, and Bauer et al. \citeyearpar{Bauer} studied starforming galaxies at 0 $\lesssim$ $z$ $\lesssim$ 1.5, and found an anticorrelation between stellar mass and SSFR. Figure \ref{ssfr} shows this relation for the sample galaxies, separated as shown previously. We have also included in the plot the fits obtained by \citetalias{Brinchmann}, Rodighiero et al. \citeyearpar{Rodighiero} for GOODS-N galaxies at 0.5 $<$ $z$ $<$ 1.0 and Karim et al. \citeyearpar{Karim} for COSMOS galaxies at 1.6 $<$ $z$ $<$ 2.0. It should be noted that the SSFR values were computed using aperture SFR values. Our hypothesis is that SFR takes place in the few central kiloparsecs and beyond this region the star formation is negligible. In fact, for the mean distance of our sample of galaxies the aperture is $\sim$ 6 kpc. It can be observed that our subsamples followed a very similar trend to the \citetalias{Brinchmann} and Rodighiero et al. \citeyearpar{Rodighiero} samples. We calculated a slope of $-$0.24$\pm$0.30 for the intermediate redshift subsample, that was comparable to that obtained by Rodighiero et al. (\citeyear{Rodighiero}; $-$0.28). In addition, our calculated slope of $-$0.23$\pm$0.10 for the low redshift subsample was quite similar to the one obtained by \citetalias{Brinchmann} ($-$0.23). A similar slope was also found by Rodighiero et al. (\citeyear{Rodighiero}; $-$0.24) in their subsample at 0 $<$ $z$ $<$ 0.5. Our data therefore gives additional support to the idea that, for galaxies with a given mass, the specific SFR is on average larger for higher redshifts.

\subsubsection{SSFR$-$Color Relation}

The correlation between morphology and broadband colors was studied by Driver et al. \citeyearpar{Driver} and Pannella et al. \citeyearpar{Pannella}, with the latter authors finding that at higher redshift both early type and late type galaxies have on average higher SSFR values than their counterparts at low redshift. Figure \ref{cssfr} shows the SSFR$_{\text{[OII],H}\alpha}$ versus dust attenuation corrected restframe color $(u-B)_{AB}$ diagram for our galaxy sample. As in the previous figures, we discriminated intermediate redshift galaxies and low redshift ones. A linear fit to these subsamples gave slopes of $-$1.32$\pm$0.69 and $-$1.46$\pm$0.66, respectively. For comparison, we also included the Twite et al. \citeyearpar{Twite} sample at 0.55 $<$ $z$ $<$ 1.23 and the Bauer et al. \citeyearpar{Bauer2} sample at 1.5 $<$ $z$ $<$ 2. It can be observed that our results are in good agreement. For the Twite et al. \citeyearpar{Twite} and Bauer et al. \citeyearpar{Bauer2} samples, slopes were calculated of $-$1.34$\pm$0.44 and $-$1.18$\pm$0.29, respectively, indicating that younger galaxies (bluer) have higher SSFR values than older (redder) galaxies. Note that in section \ref{Edad} a clear correlation was obtained between age and color, with younger galaxies having a bluer color than older galaxies. Moreover, it can be also observed that there was a trend with redshift in the sense that for a given $(u-B)$ color, the high redshift galaxies revealed on average higher SSFR than low redshift galaxies. One possible physical interpretation for this result is that in the past galaxies were more efficient at forming stars, when compared to more recent galaxies of the same color (age) and stellar mass.  This could imply a larger reservoir of cold gas and higher star formation rates. The latter could explain the double branch in the color-age plot, with a double sequence in $\tau$ values.  Alternatively, the IMF could have been closer to top heavy, which would also explain this difference.

\section{SUMMARY}\label{Sum}

We investigated the physical properties, star formation and AGN activity in a sample of 37 Balmer break galaxies with emission lines at redshift 0.045 $<$ $z$ $<$ 0.93. Using SED fits, it was calculated that most of the galaxies had masses in the range 10$^{8.5}-$10$^{10}M_{\odot}$. This sample forms part of a spectroscopic survey pilot project in the Subaru XMM deep field (SXDF) out to z $\sim$ 2.5, with the purpose of investigating the end of star formation in massive galaxies.

According to the classical BPT (\citeauthor{Kewley}) and Lamareille \citeyearpar{Lamareille} diagnostic diagrams, 31 (84$\%$) star-forming galaxies, 2 (5$\%$) composite galaxies and 3 (8$\%$) AGNs (2 Seyfert 2 types and 1 LINER) were found. In addition, the MEx diagram (\citeauthor{Juneau}) detected one out of the three AGN, found by alternative emission line diagnostics (a Seyfert 2).

Nineteen of our sample galaxies had available photometry in all the IRAC/\textit{Spitzer} bands. According to the MIR diagnostic diagrams (\citeauthor{Stern,Lacy}) we found that 6 galaxies (32$\%$) could host an AGN, with three of these objects being the aforementioned Seyfert 2 and LINER, while the remaining 3 galaxies were classified in the optical diagrams as composite or starburst. This suggests that they could have hosted an obscured AGN or that the nuclear star-forming activity was masking the AGN optical signatures.

For each of the sample galaxies, the SFR were calculated using 4 different parameters: (1) [OII]$\lambda$3727$\AA{}$ luminosity, (2) H$\alpha$ luminosity, (3) UV luminosity, and (4) SED fitting with an exponentially declining SFH. All these methods were found to give very similar results, with a range of 0.01$-$100 $M_{\odot}$/yr. 

The cosmic evolution of the SFR surface density $\Sigma_{\text{SFR}}$ was compared with the cosmic evolution of the stellar mass surface density per unit time ($\Sigma_{M_{\ast}/\tau}$), and it was found that both these quantities increased for higher redshift galaxies. Despite our sample being small, these results are in good agreement with the trends found in other surveys, such as SDSS and zCOSMOS (\citeauthor{Silverman}). A strong correlation was observed when SFR versus stellar mass was compared, with more massive galaxies revealing higher SFR values. The same trend was observed when comparing different samples at different redshifts, with higher redshift galaxies showing on average higher SFR values. This result has already been reported for SDSS galaxies (\citeauthor{Brinchmann}) and GOODS galaxies (\citeauthor{Elbaz,Daddi2}).

When the SSFR versus stellar mass relation was investigated, it was found that less massive galaxies had higher SSFR values than more massive galaxies. Moreover, for a particular mass, high redshift galaxies revealed on average larger SSFR values. This trend was also reported by Brinchmann et al. \citeyearpar{Brinchmann}, Rodighiero et al. \citeyearpar{Rodighiero} and Karim et al. \citeyearpar{Karim}, among others. Finally, we compared SSFR versus $(u-B)$ color and found that bluer galaxies, corresponding to younger ages from SED fitting results, had larger SSFR values. This result was also reported by Twite et al. \citeyearpar{Twite} and Bauer et al. \citeyearpar{Bauer2}, but we also showed that for a particular $(u-B)$ color, the high redshift galaxies had on average higher SSFR values.

Even though our sample was small and therefore possible exposed to strong selection effects, the results obtained were in good agreement with other authors using much larger samples. Nevertheless, a larger sample is necessary to be able to produce more statistically robust conclusions.

\setcounter{secnumdepth}{0}
\section{Acknowledgments}

We thank to the anonymous referee that gave a very detailed and constructive report that helped to greatly improve the manuscript. We also thank Y. Ohyama for his help in to overcoming some bugs found in the COSMOS program routines, and Paul Hobson, native english speaker, for revision of the manuscript.

This research was partially supported by Consejo Nacional de Investigaciones Científicas E Técnicas (CONICET, Argentina), the Japan Society for the Promotion of Science (JSPS) and Comisión Nacional de Investigaciones Científicas Y Tecnológicas (CONICYT, Chile) through FONDECYT $\#$1110328 and BASAL PFB-06 ``Centro de Astrofísica y Tecnologías Afines". We are specially grateful to CONICET for the fellowship awarded to the first author. 

This study is based on data mostly collected by the Subaru telescope operated by the National Astronomical Observatory of Japan (NAOJ); the United Kingdom Infrared Telescope (UKIRT) operated by the Joint Astronomy Centre (JAC) on behalf of the Science and Technology Facilities Council of the U. K.; the \textit{Spitzer Space Telescope} operated by the Jet Propulsion Laboratory, California Institute of Technology (Caltech) under contract with NASA; the Blanco telescope of Cerro Tololo Inter-American Observatory (CTIO) operated by the Association of Universities Research in Astronomy (AURA), under cooperative agreement with the National Science Foundation as part of the National Optical Astronomy Observatories (NOAO); the Magellan telescope operated by the Carnegie Institution of Washington, Harvard University, Massachusetts Institute of Technology (MIT), University of Michigan and University of Arizona; and the Gemini telescope operated by AURA, Inc., under a cooperative agreement with the NSF on behalf of the Gemini partnership: the National Science Foundation (United States), the Science and Technology Facilities Council (United Kingdom), the National Research Council (Canada), CONICYT (Chile), the Australian Research Council (Australia), Ministerio da Ciência e Tecnologia (Brazil) and Ministerio de Ciencia, Tecnología e Innovación Productiva (Argentina).

{\it Facilities:} \facility{Magellan:Baade (IMACS)}, \facility{Gemini:South (GMOS)}, \facility{Blanco (MOSAICII)}, \facility{Subaru (SUPRIME-CAM)}, \facility{UKIRT (WFCAM)}, \facility{Spitzer (IRAC)}. 


\addtocounter{figure}{-1}
\begin{figure}
\begin{center}
\includegraphics[width=0.32\textwidth]{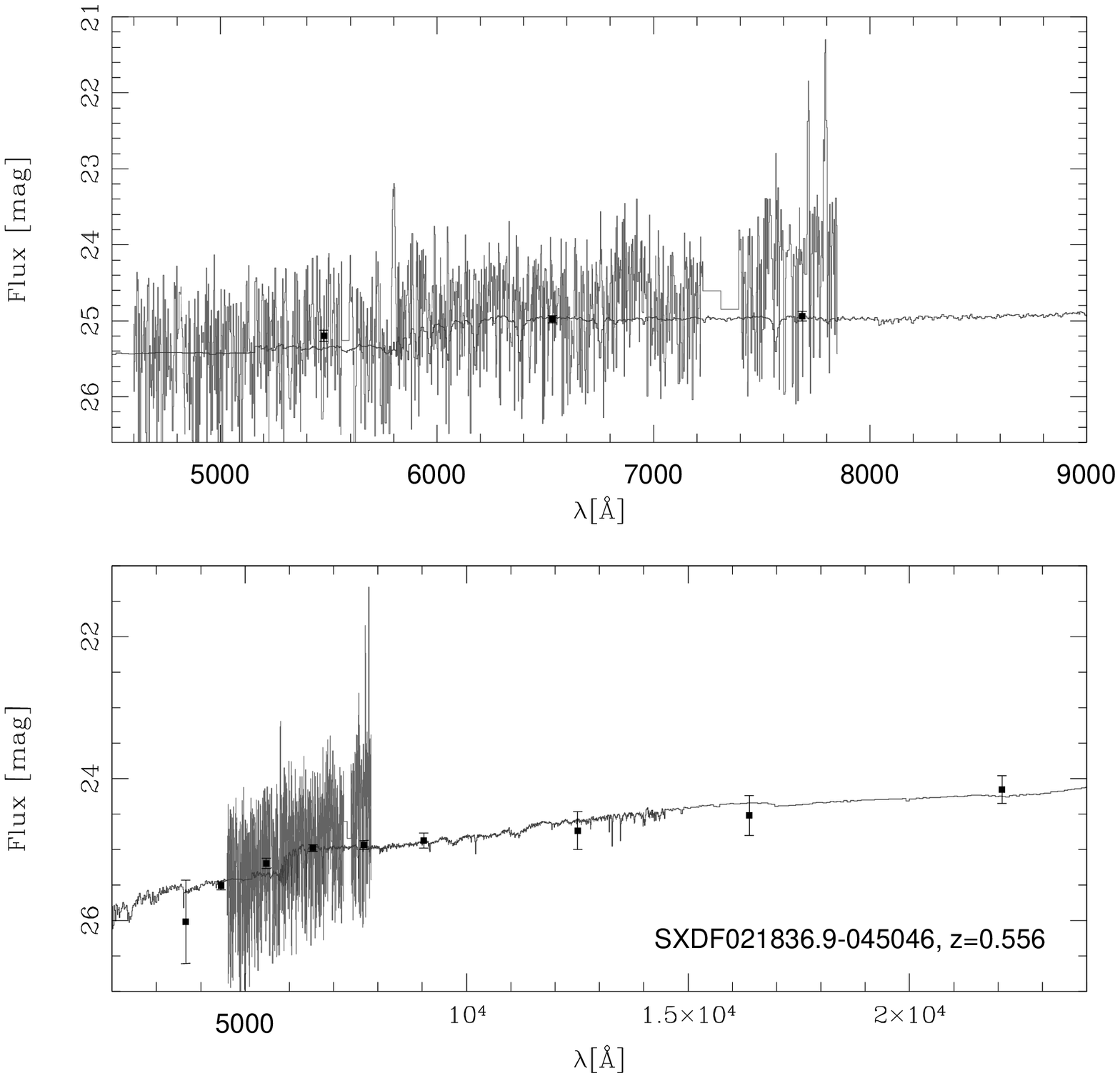}
\includegraphics[width=0.32\textwidth]{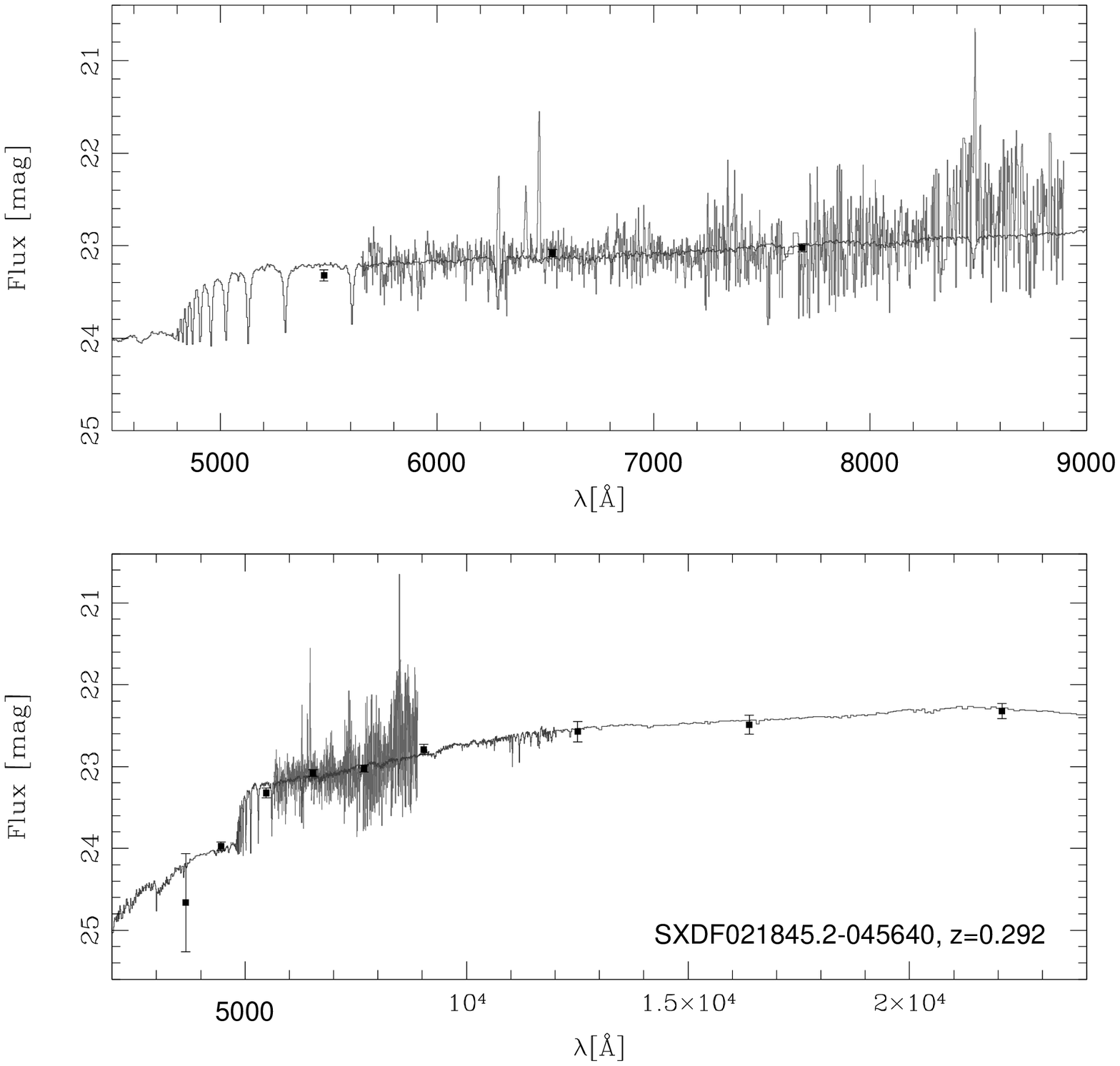}
\includegraphics[width=0.32\textwidth]{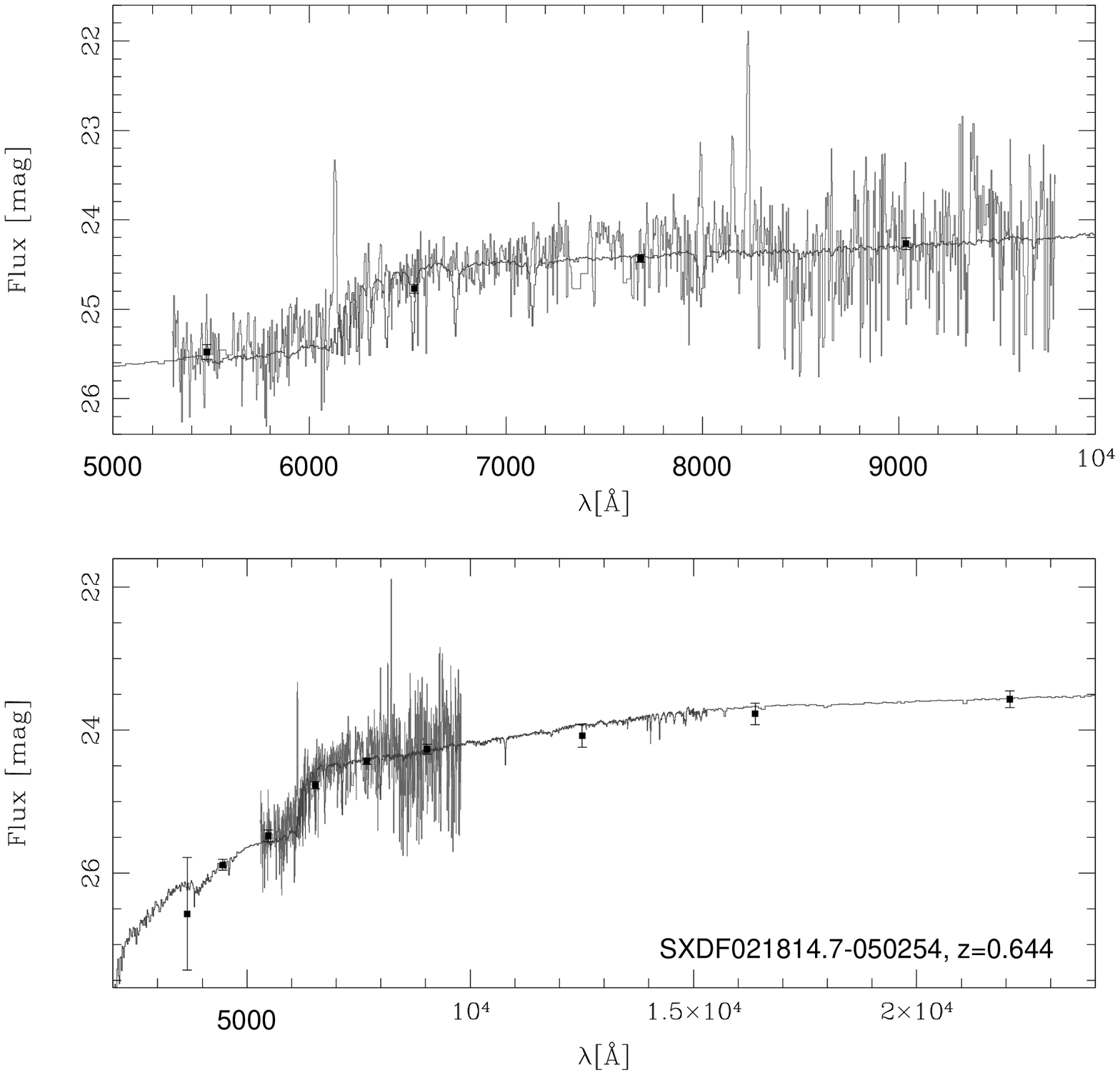}
\includegraphics[width=0.32\textwidth]{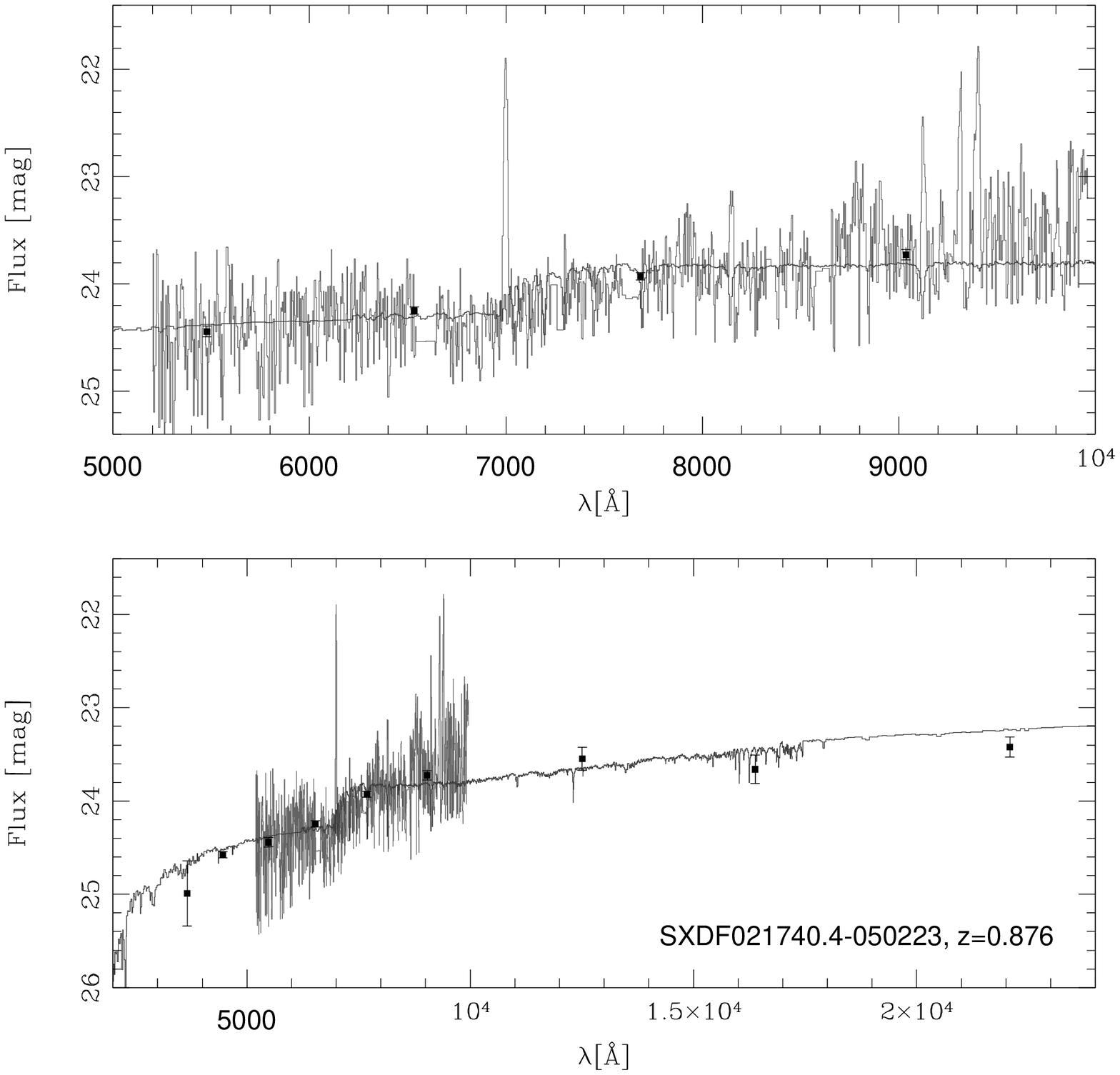}
\includegraphics[width=0.32\textwidth]{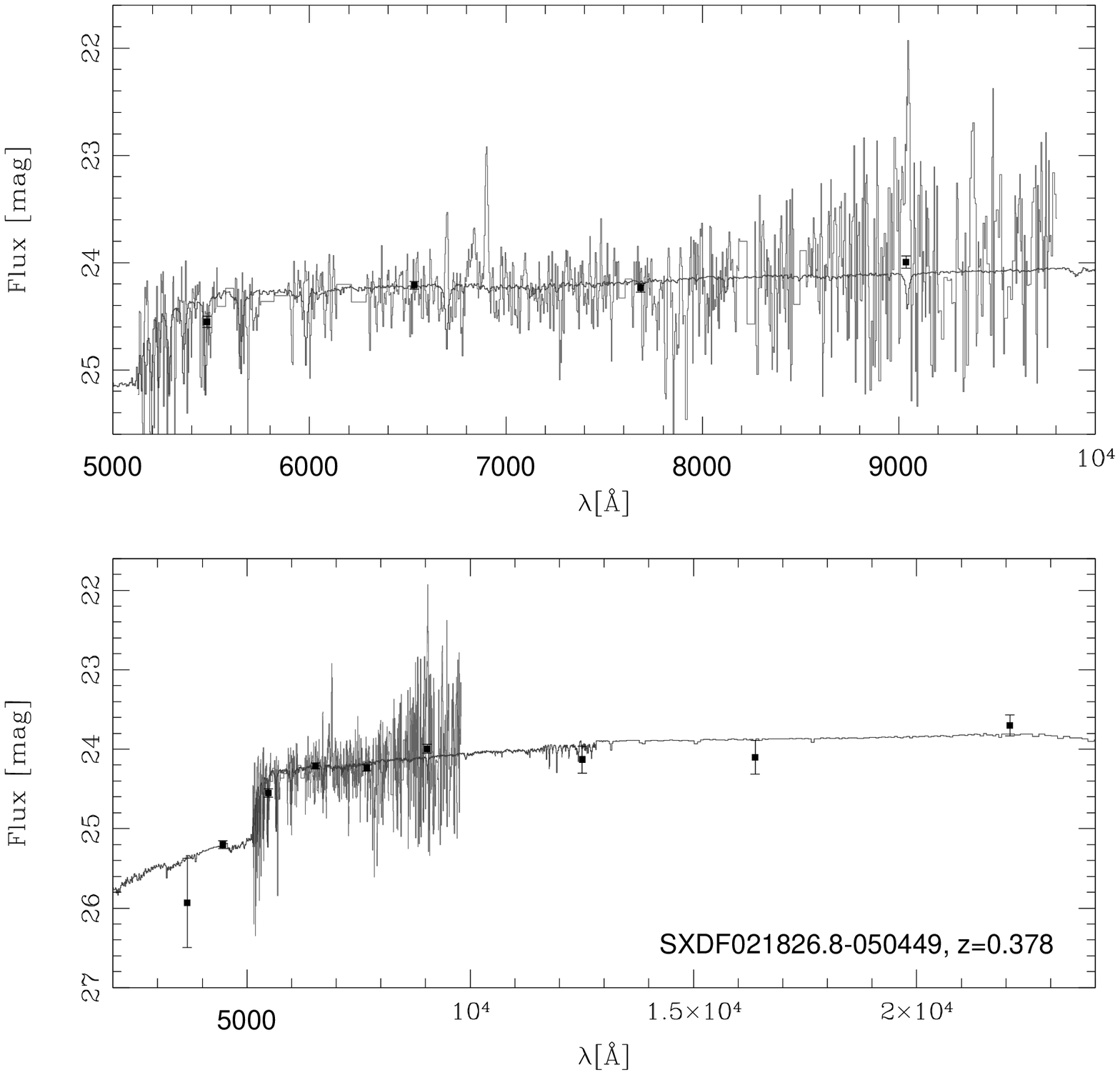}
\includegraphics[width=0.32\textwidth]{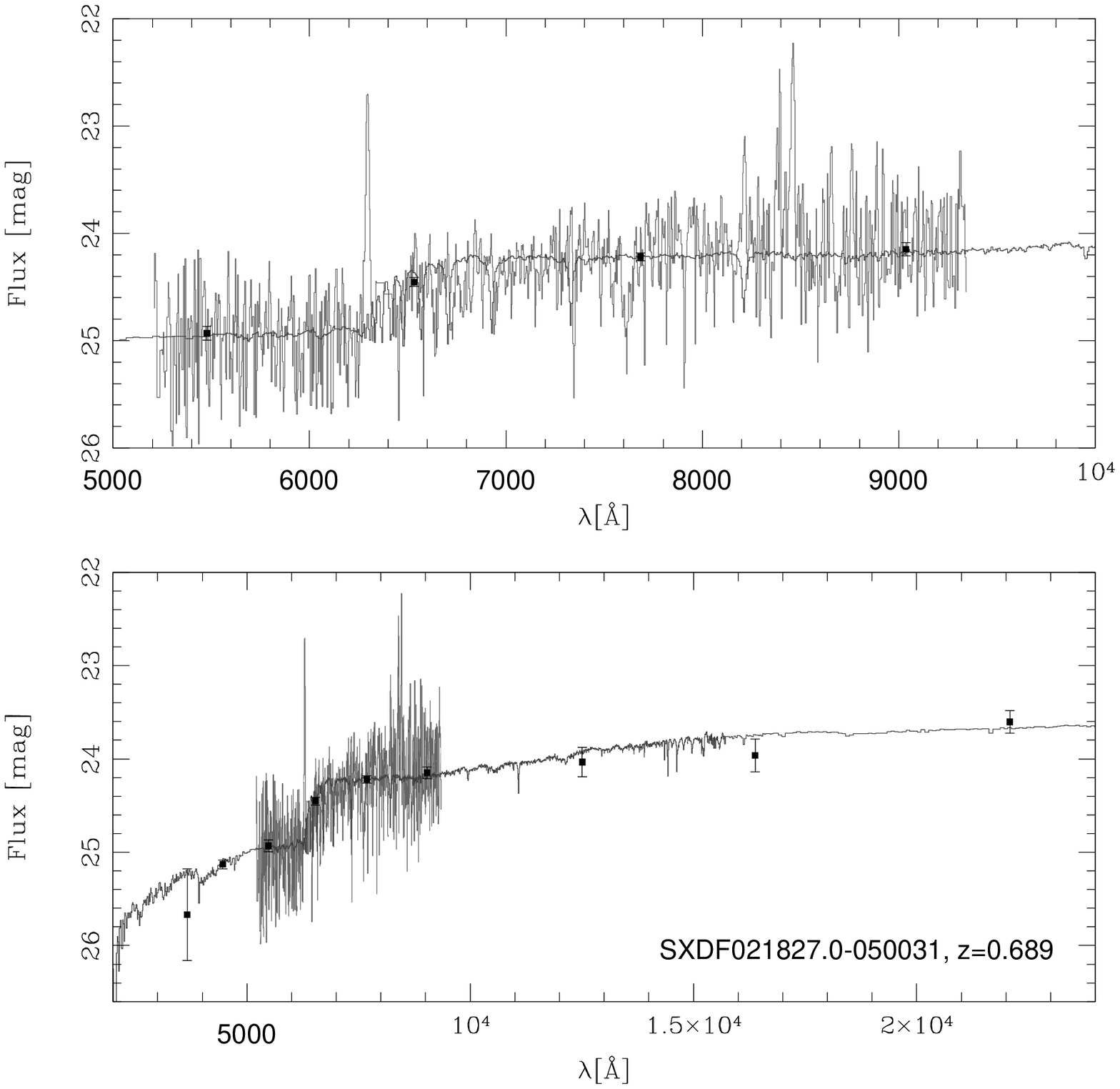}
\includegraphics[width=0.32\textwidth]{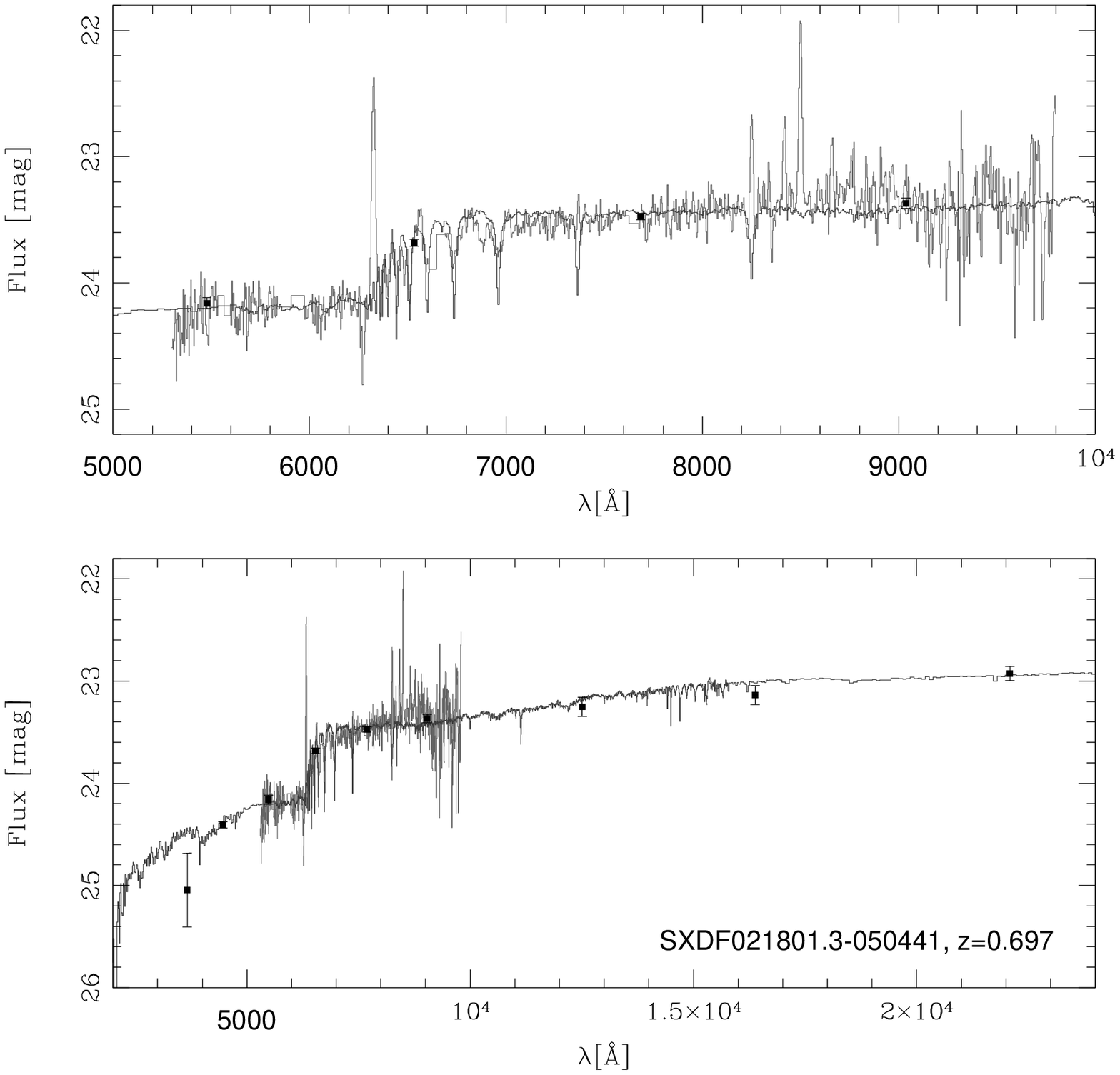}
\includegraphics[width=0.32\textwidth]{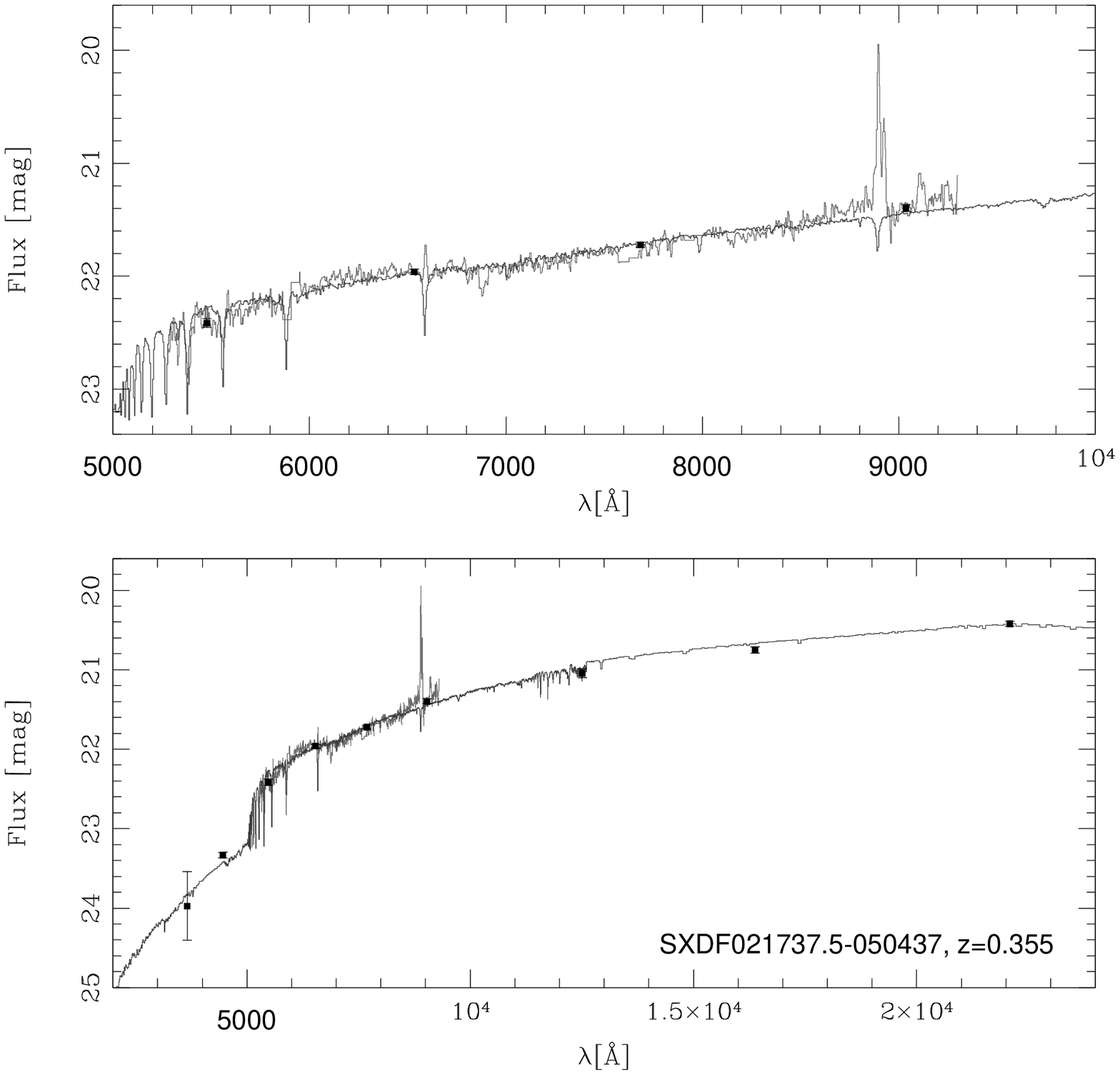}
\includegraphics[width=0.32\textwidth]{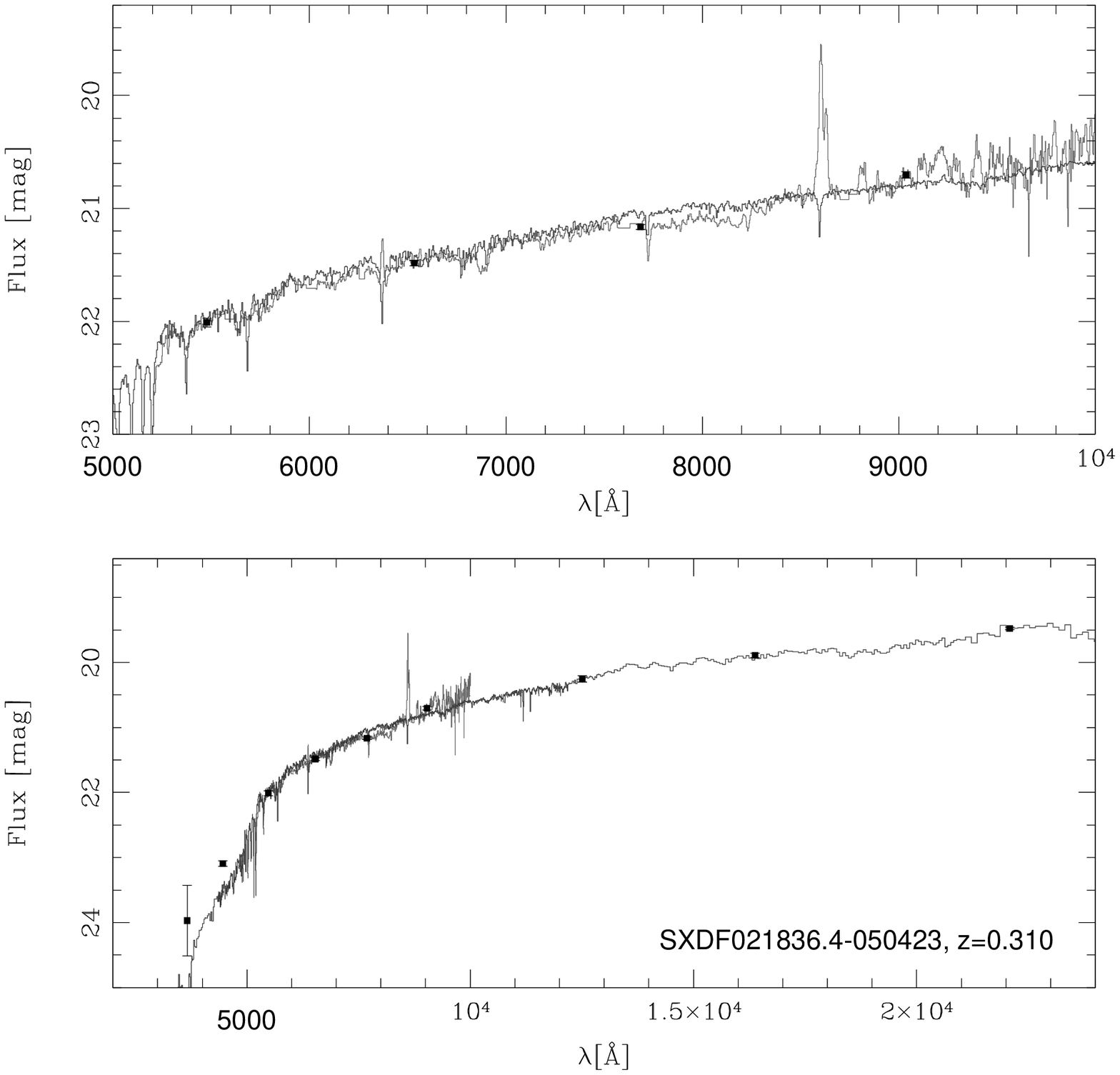}
\includegraphics[width=0.32\textwidth]{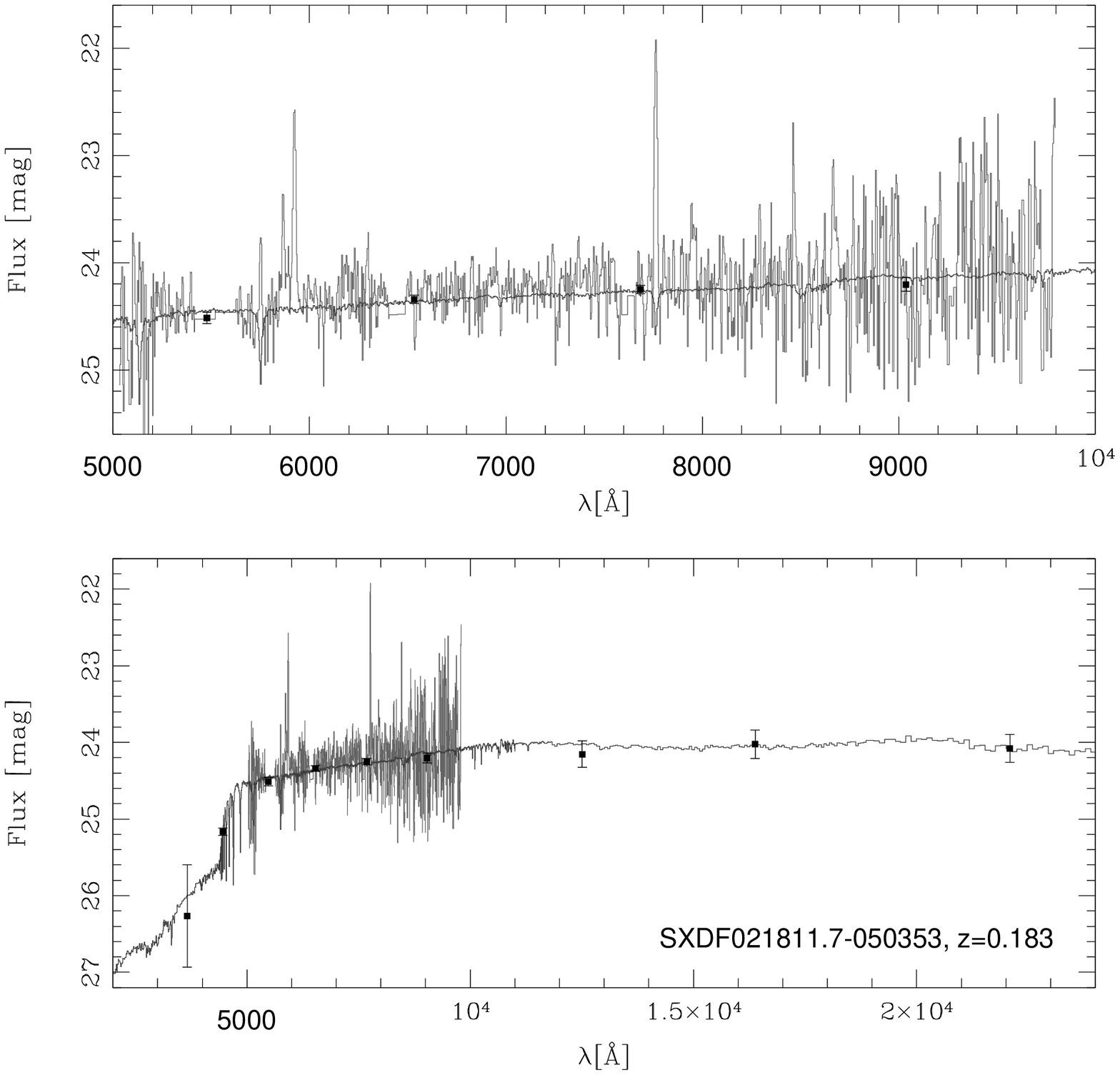}
\includegraphics[width=0.32\textwidth]{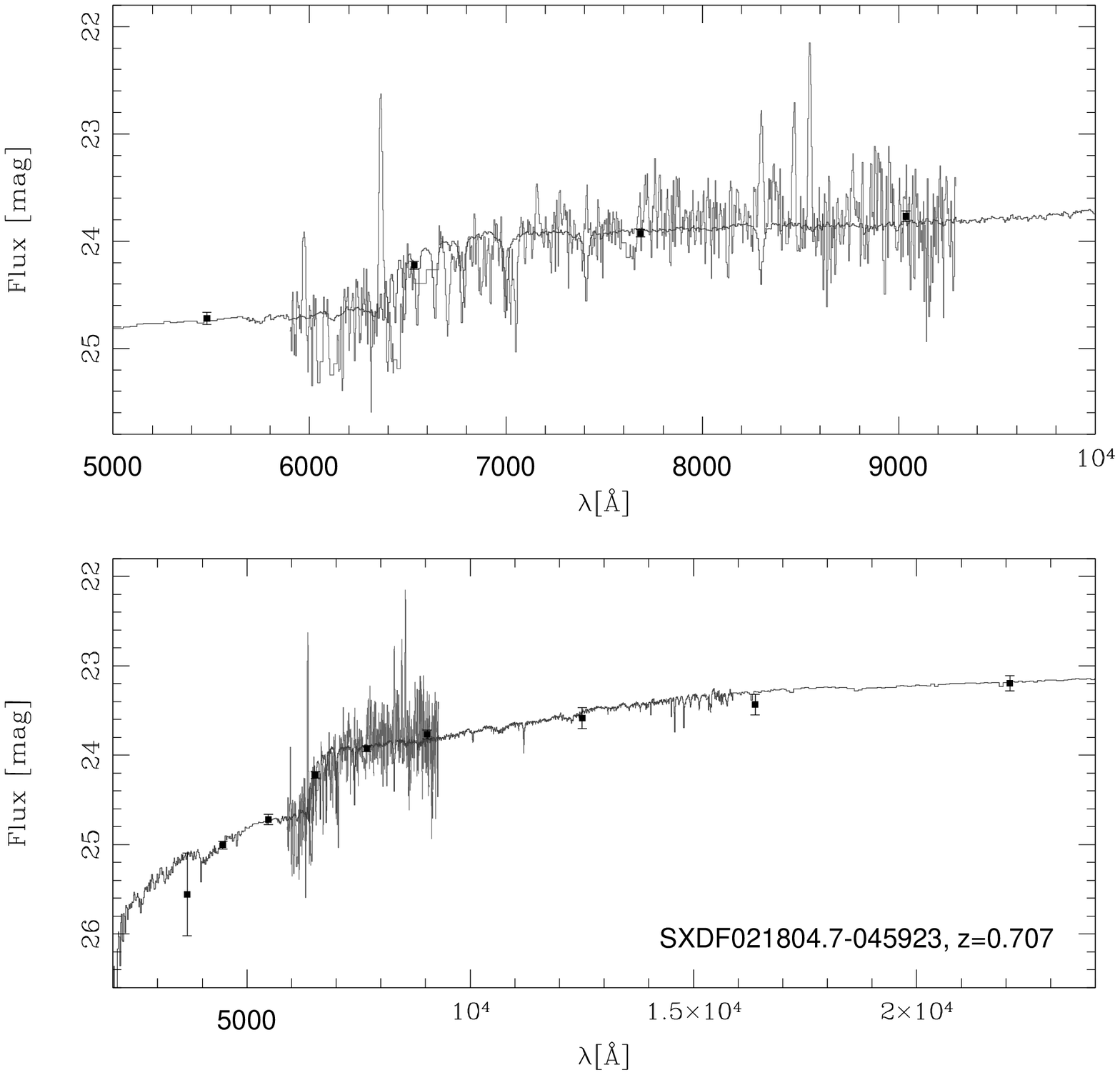}
\includegraphics[width=0.32\textwidth]{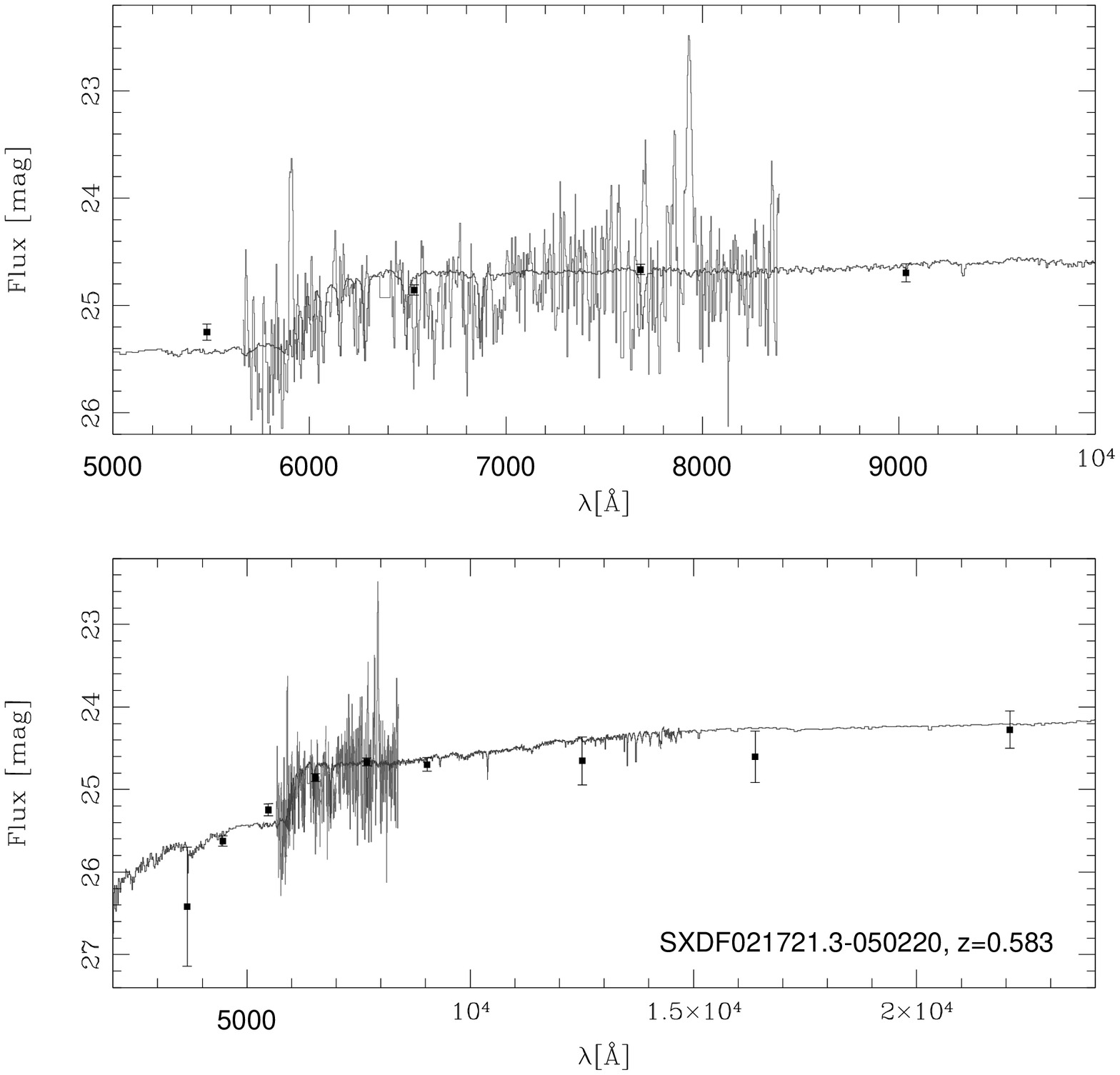}
\caption{\footnotesize{\textit{$-$Continued.}}}
\end{center}
\end{figure}

\addtocounter{figure}{-1}
\begin{figure}
\begin{center}
\includegraphics[width=0.32\textwidth]{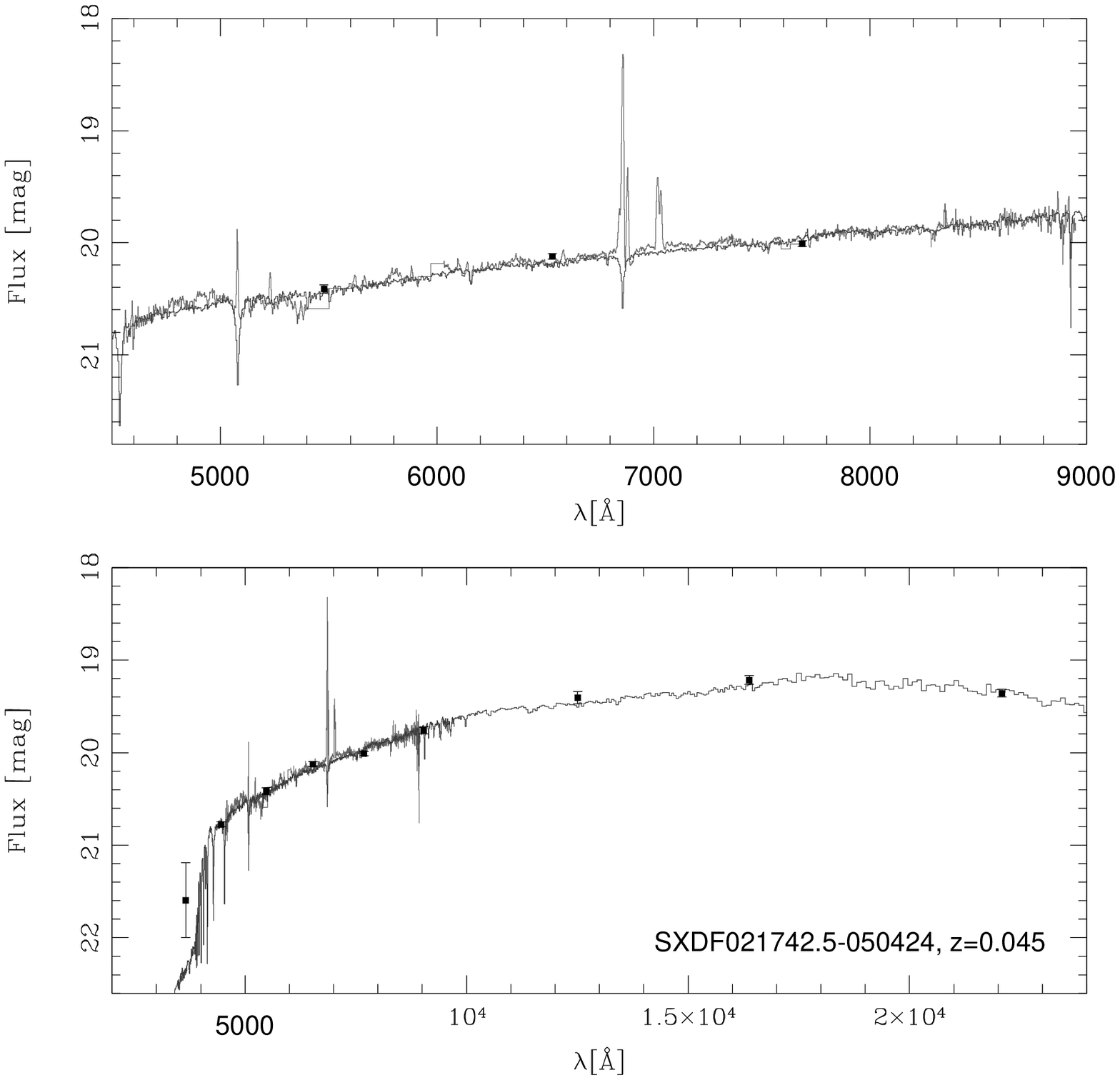}
\includegraphics[width=0.32\textwidth]{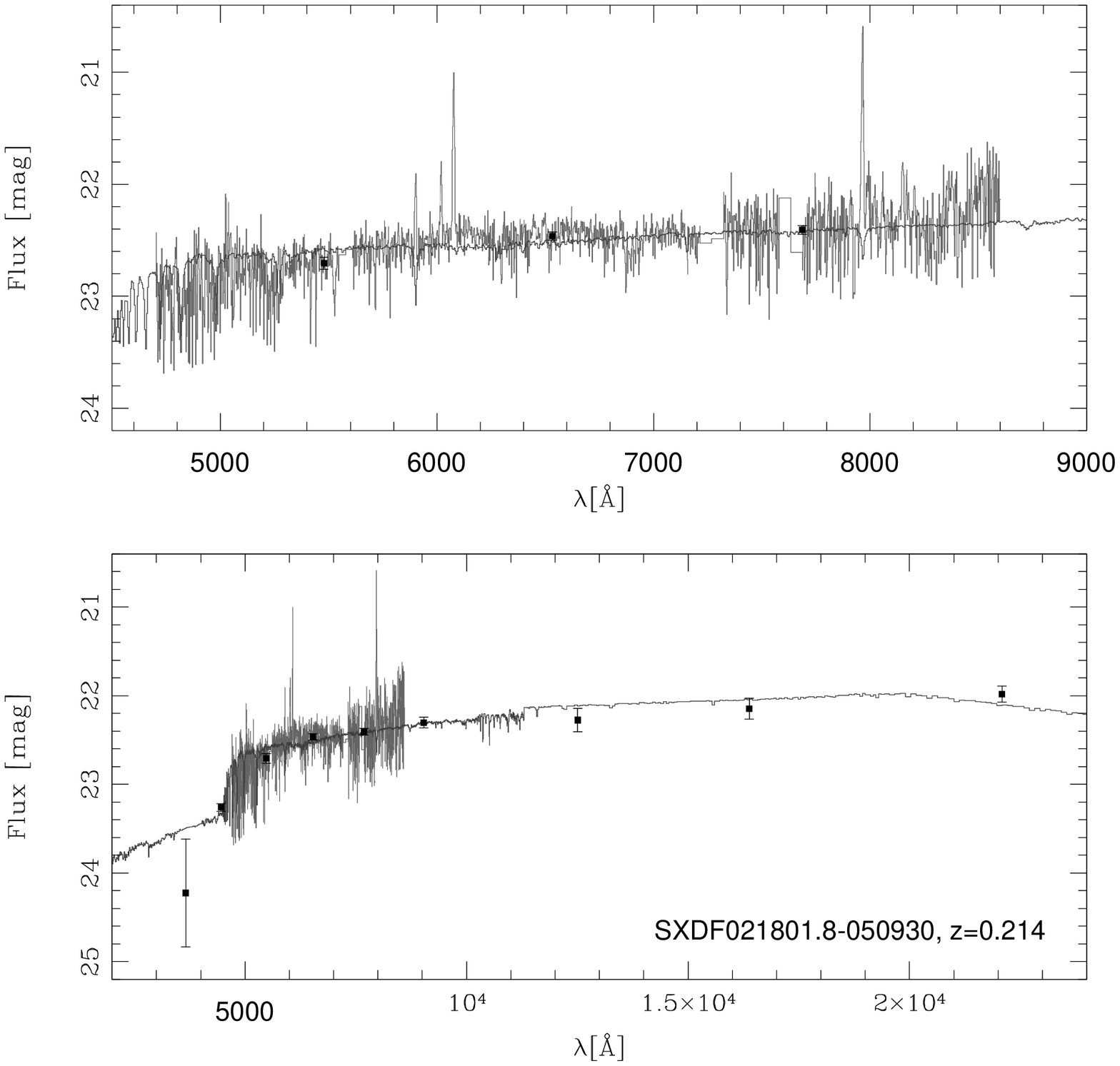}
\includegraphics[width=0.32\textwidth]{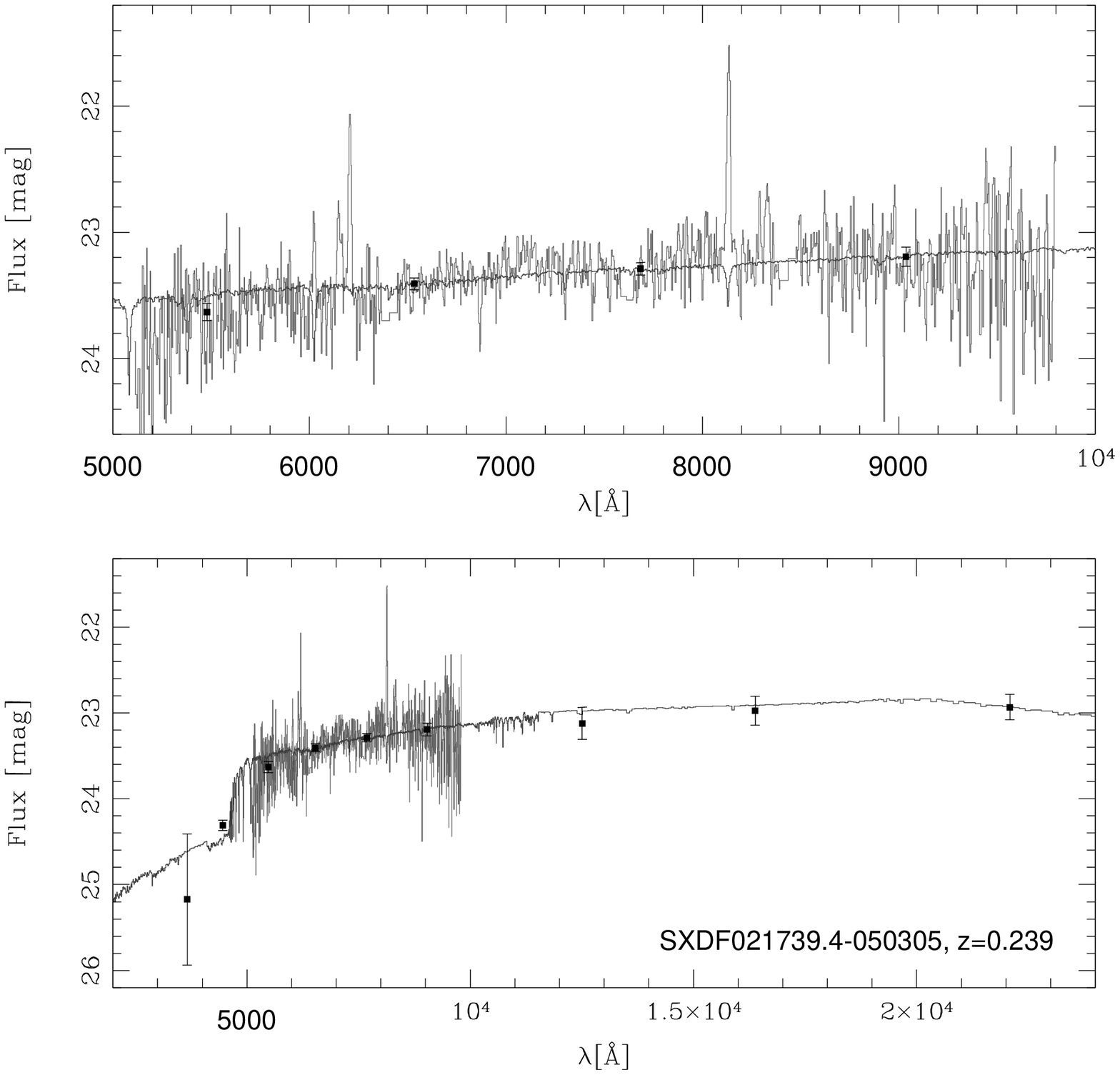}
\includegraphics[width=0.32\textwidth]{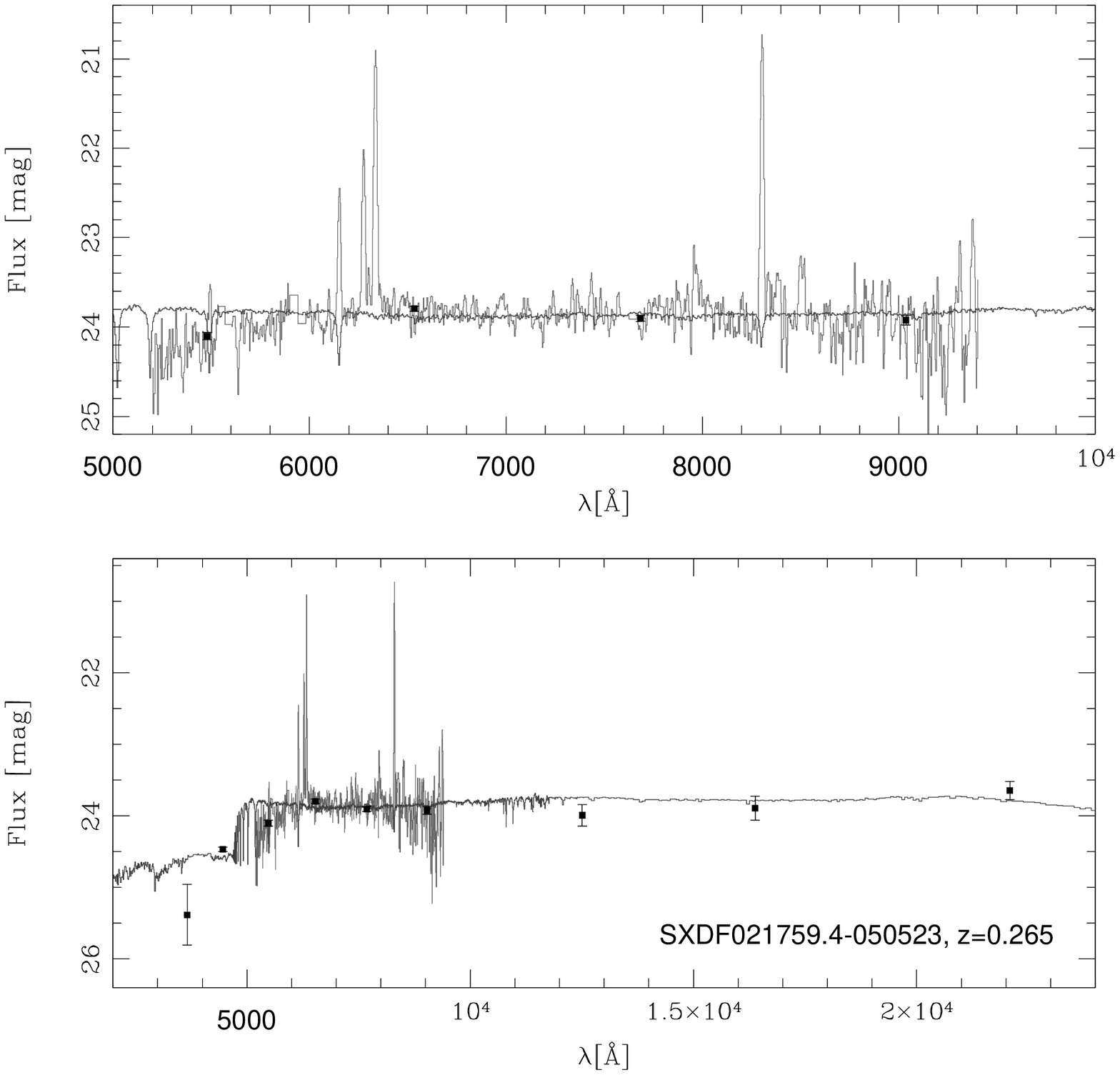}
\includegraphics[width=0.32\textwidth]{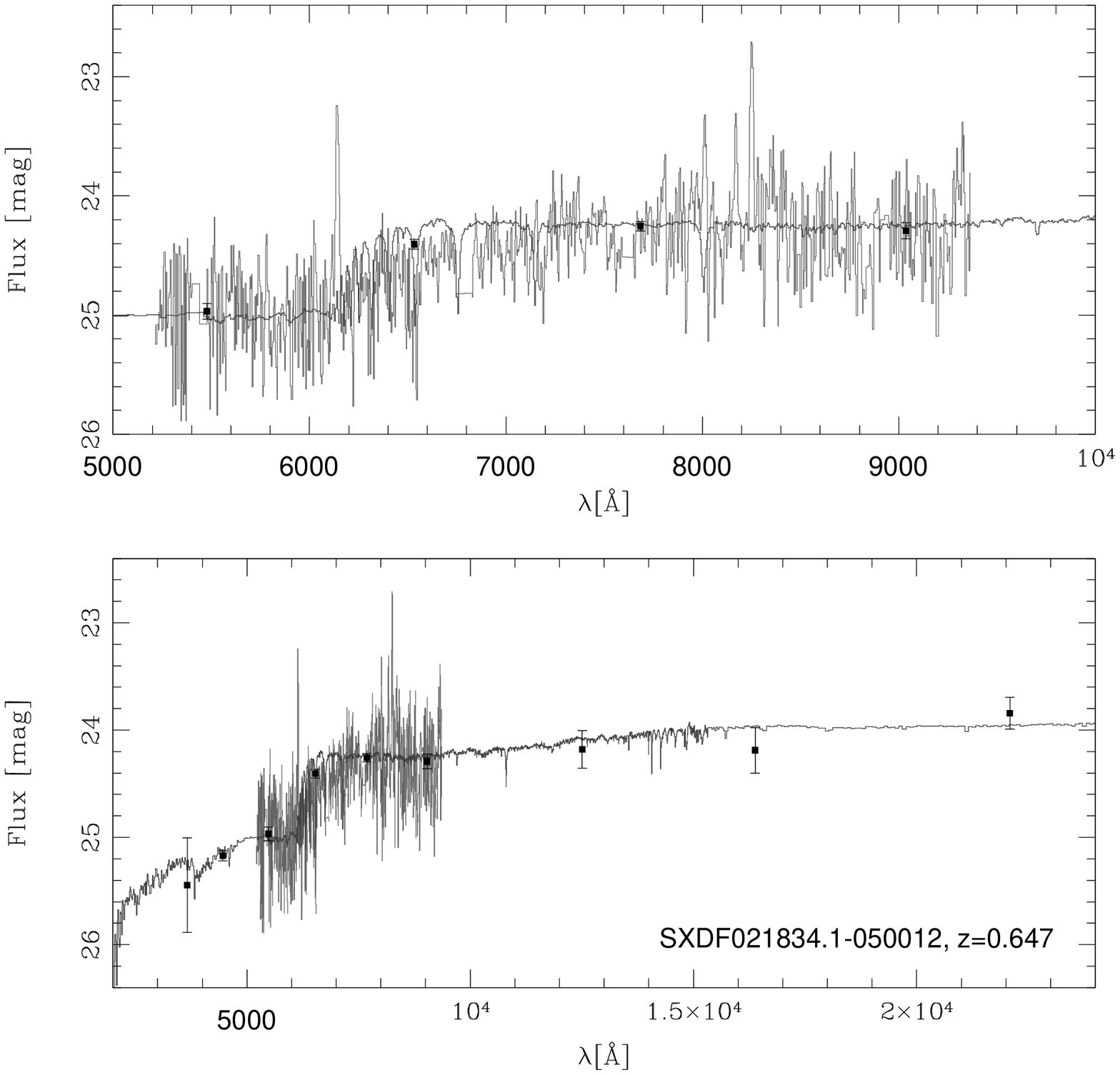}
\includegraphics[width=0.32\textwidth]{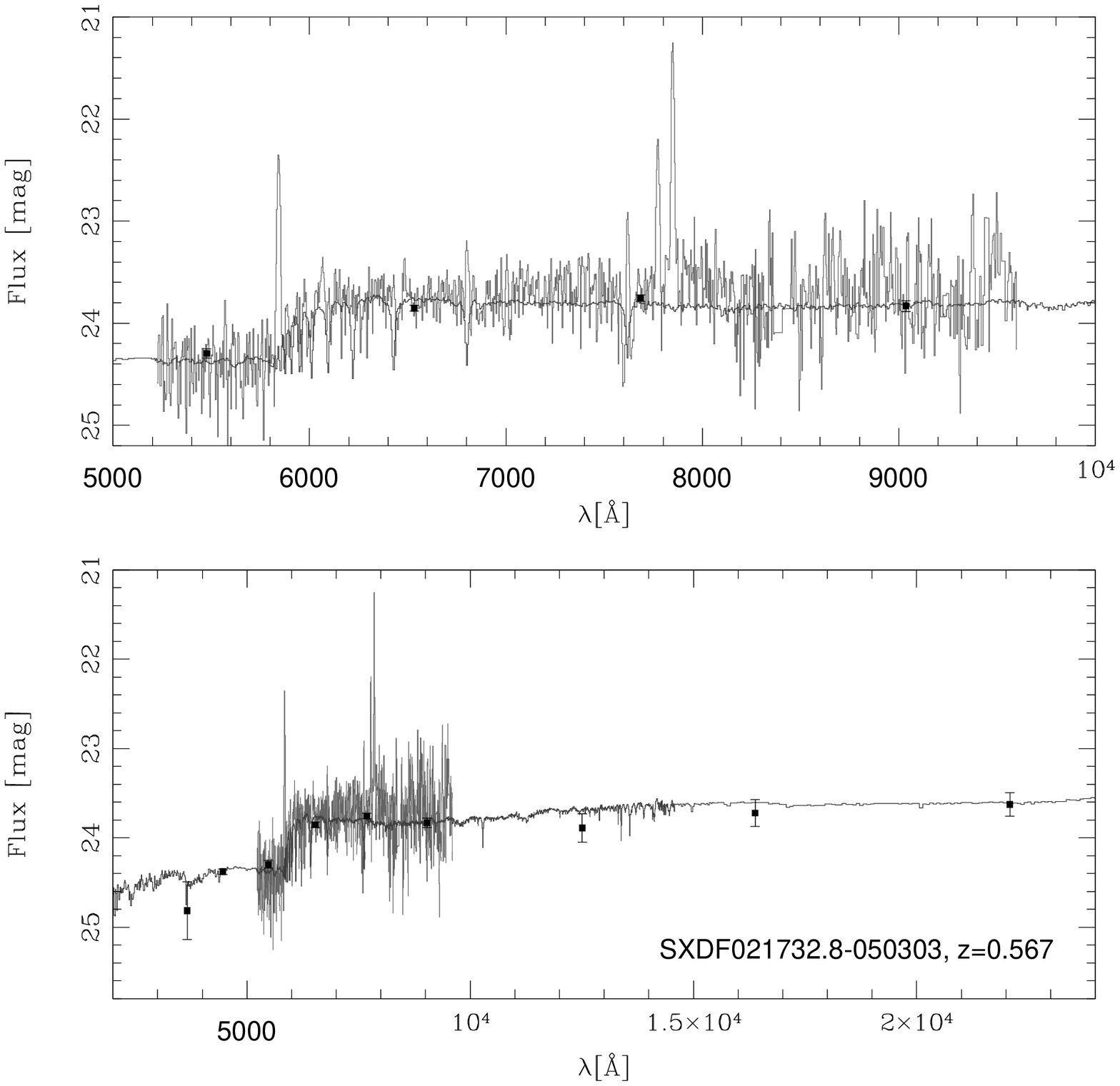}
\includegraphics[width=0.32\textwidth]{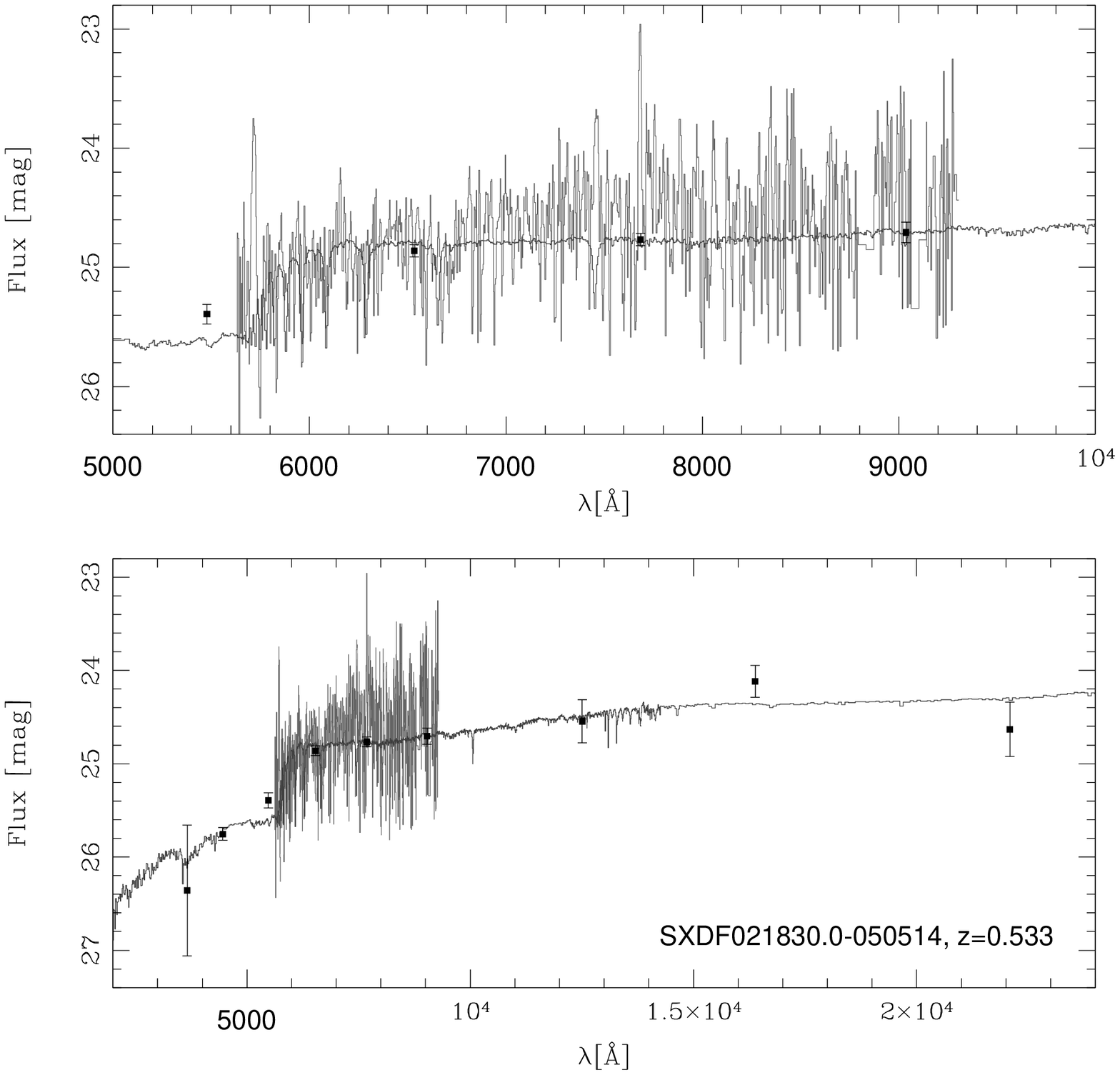}
\includegraphics[width=0.32\textwidth]{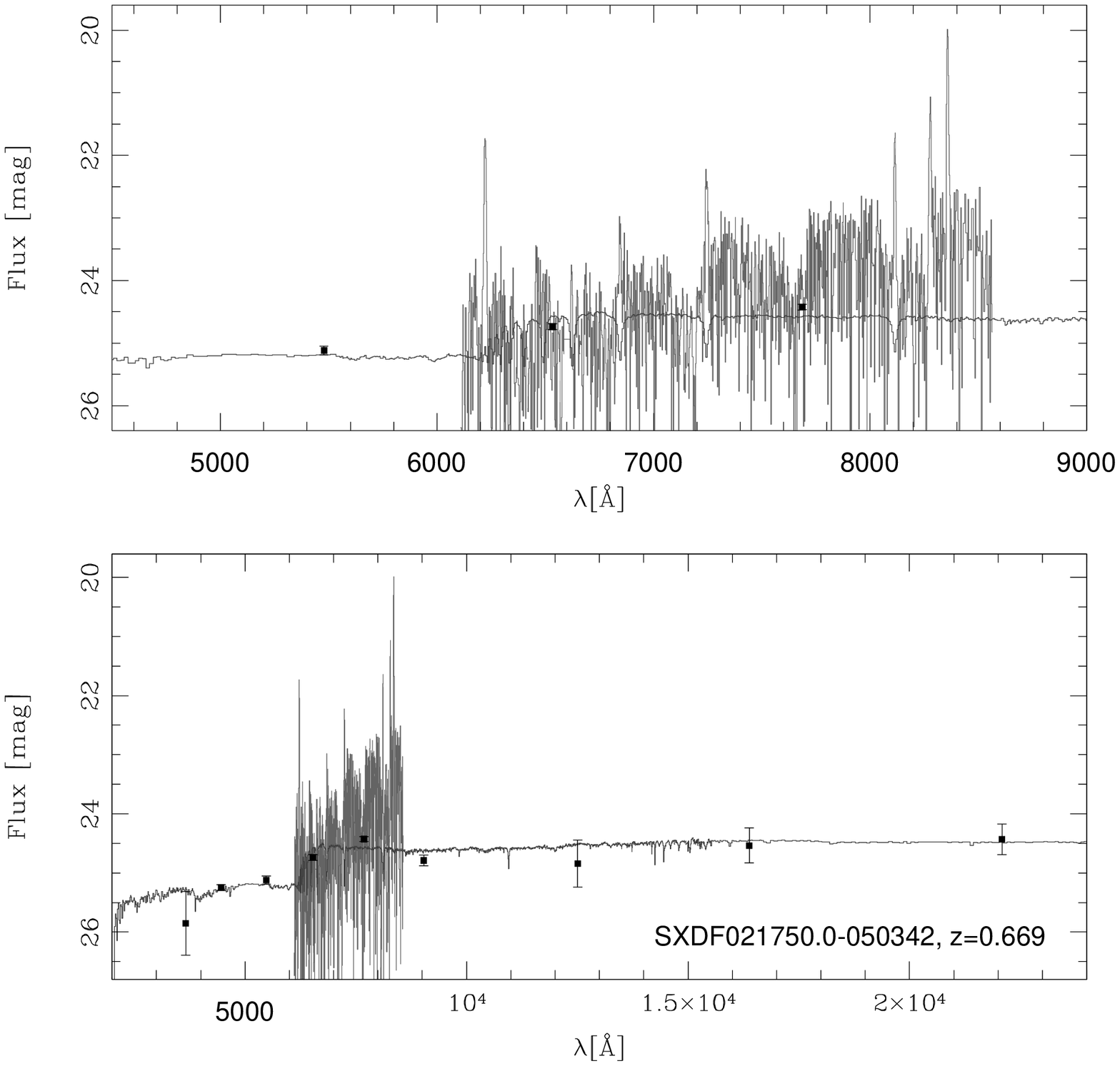}
\includegraphics[width=0.32\textwidth]{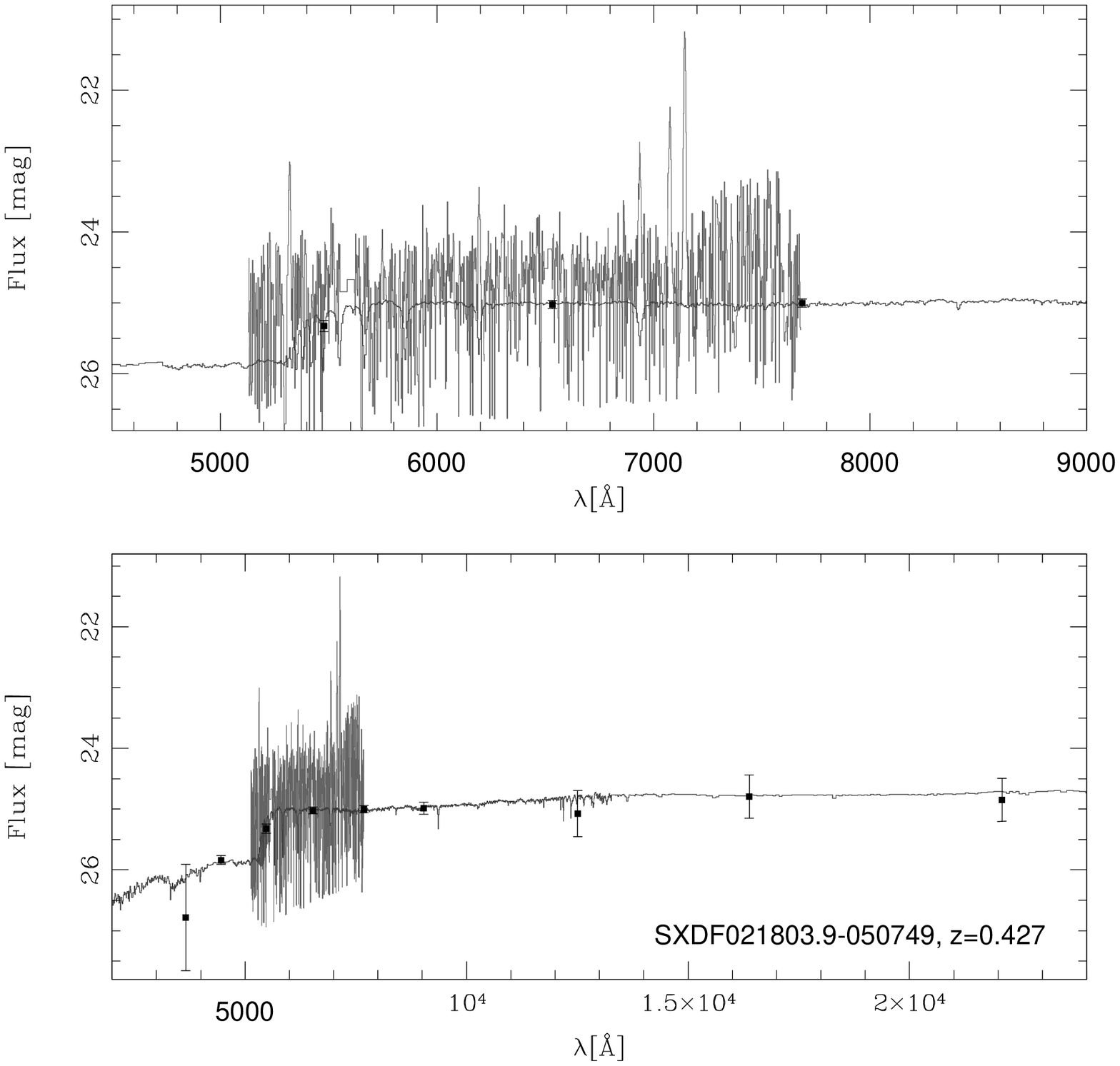}
\includegraphics[width=0.32\textwidth]{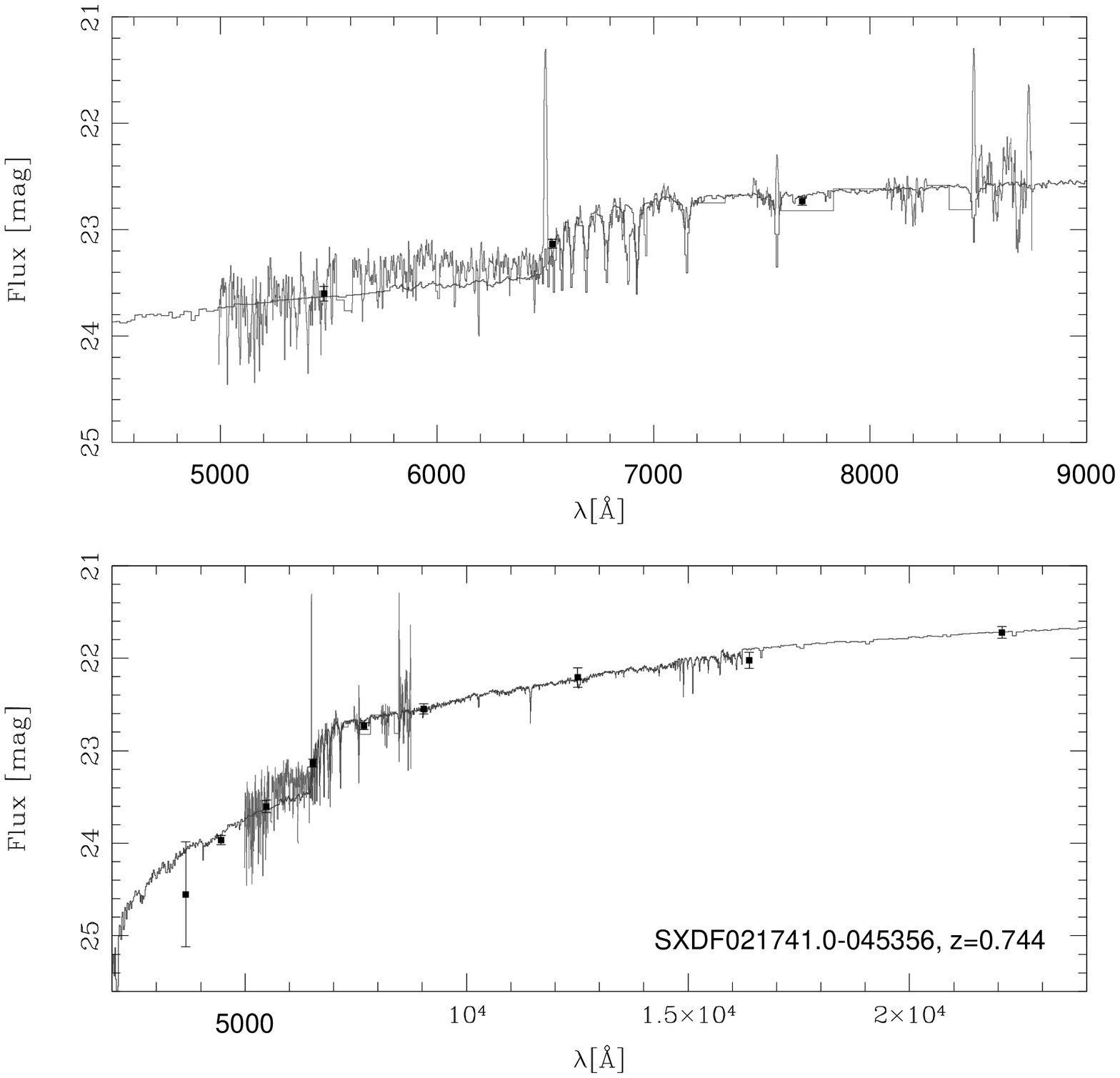}
\includegraphics[width=0.32\textwidth]{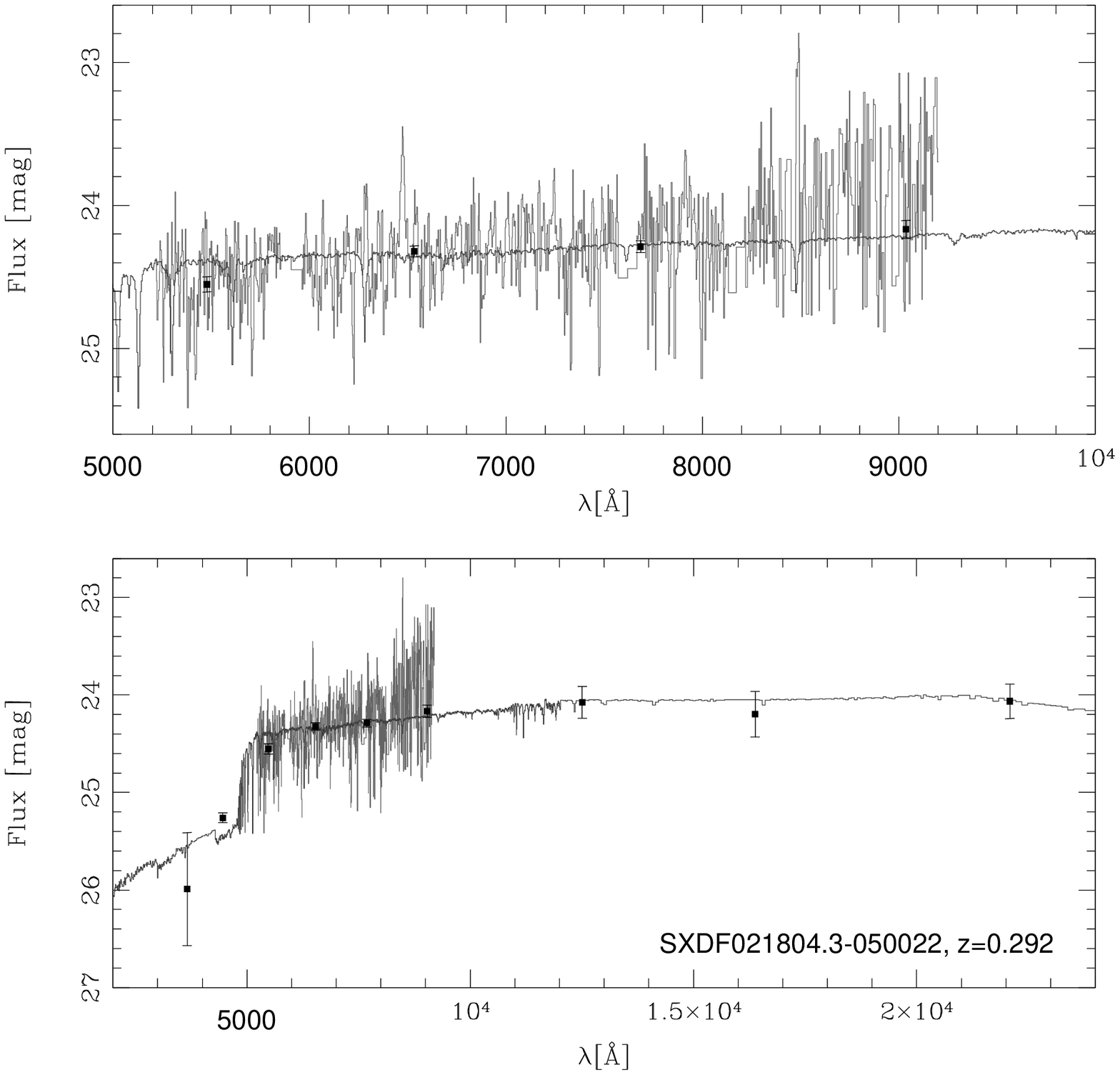}
\includegraphics[width=0.32\textwidth]{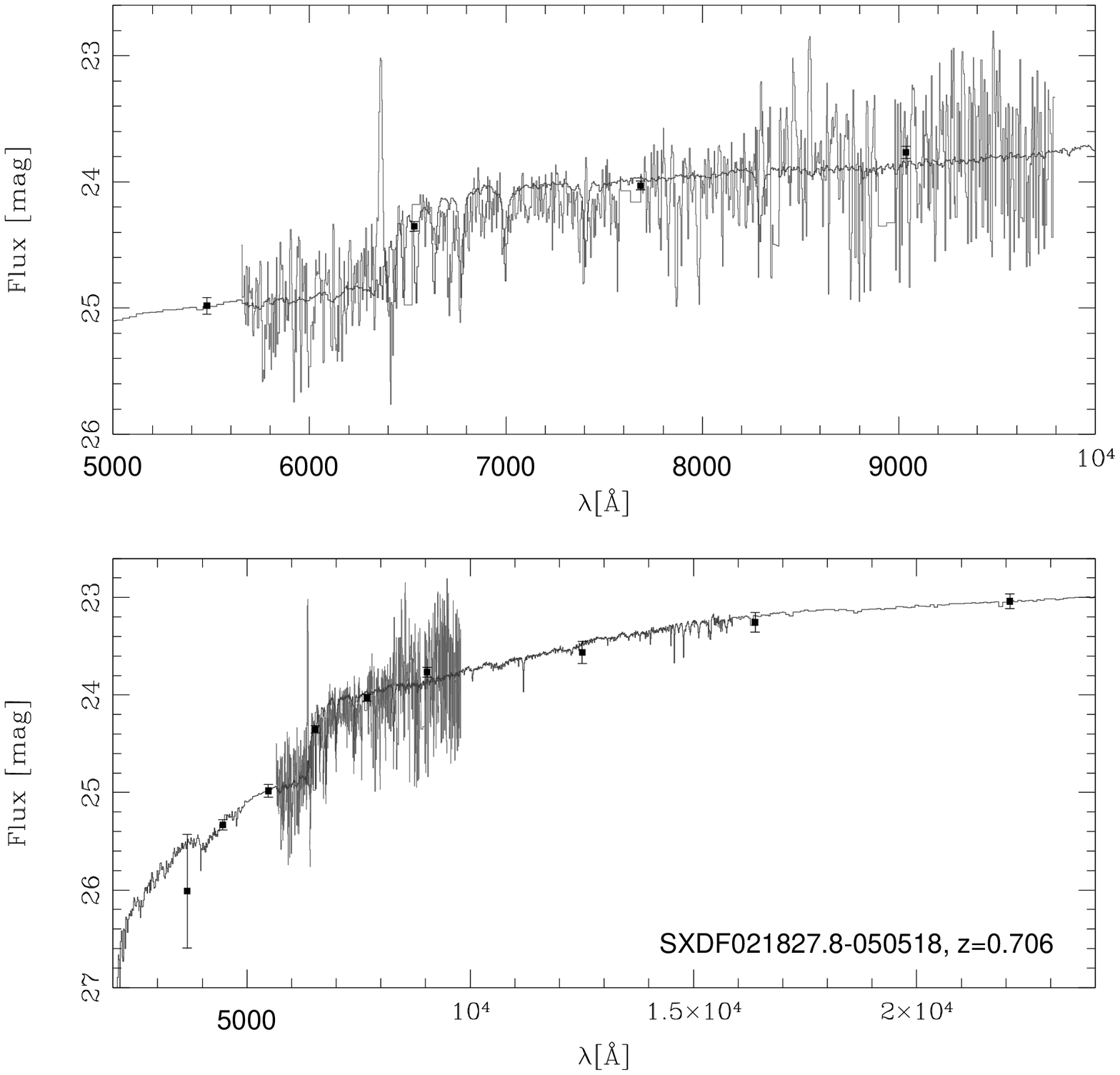}
\caption{\footnotesize{\textit{$-$Continued.}}}
\end{center}
\end{figure}

\addtocounter{figure}{-1}
\begin{figure}
\begin{center}
\includegraphics[width=0.32\textwidth]{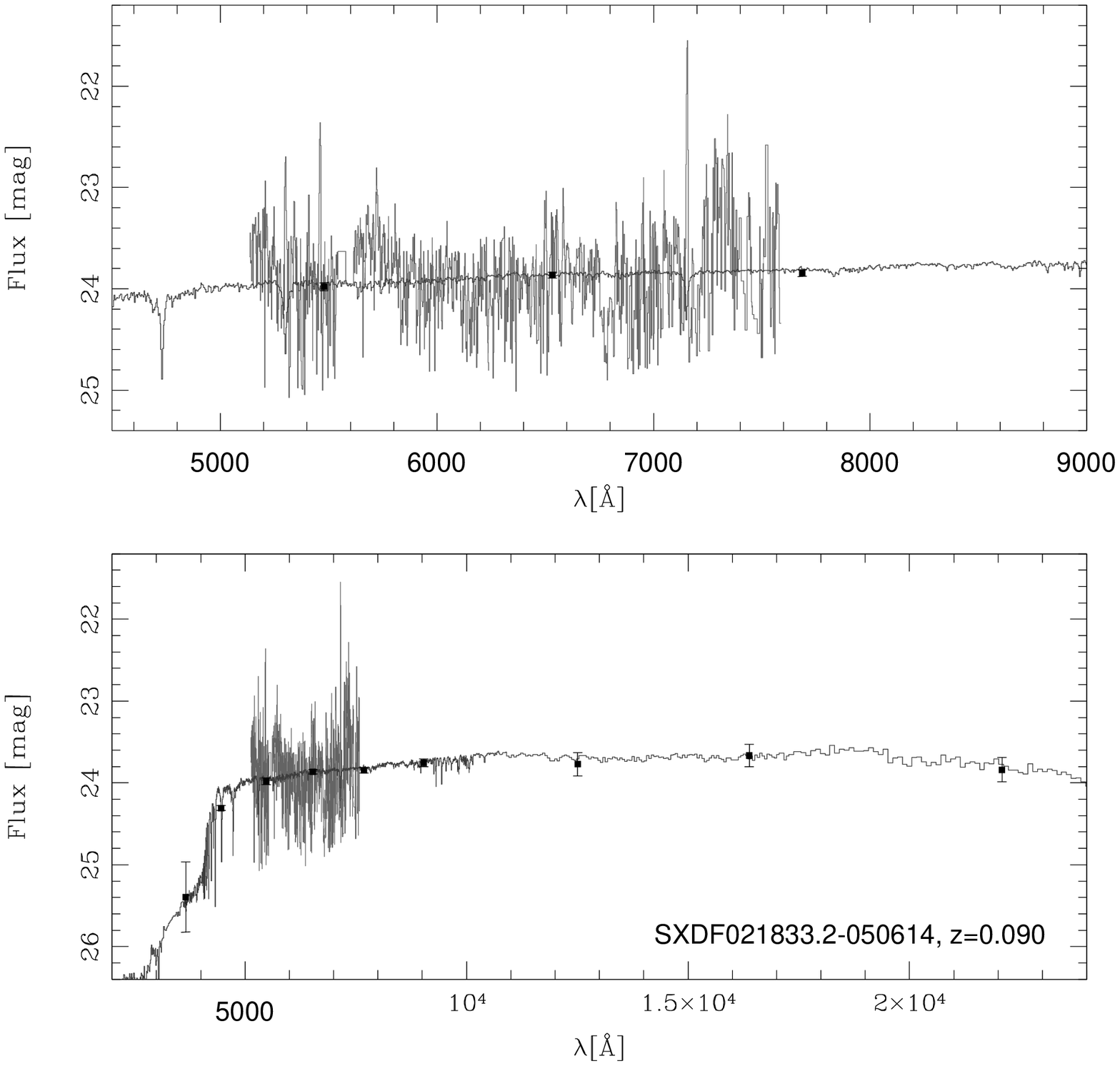}
\includegraphics[width=0.32\textwidth]{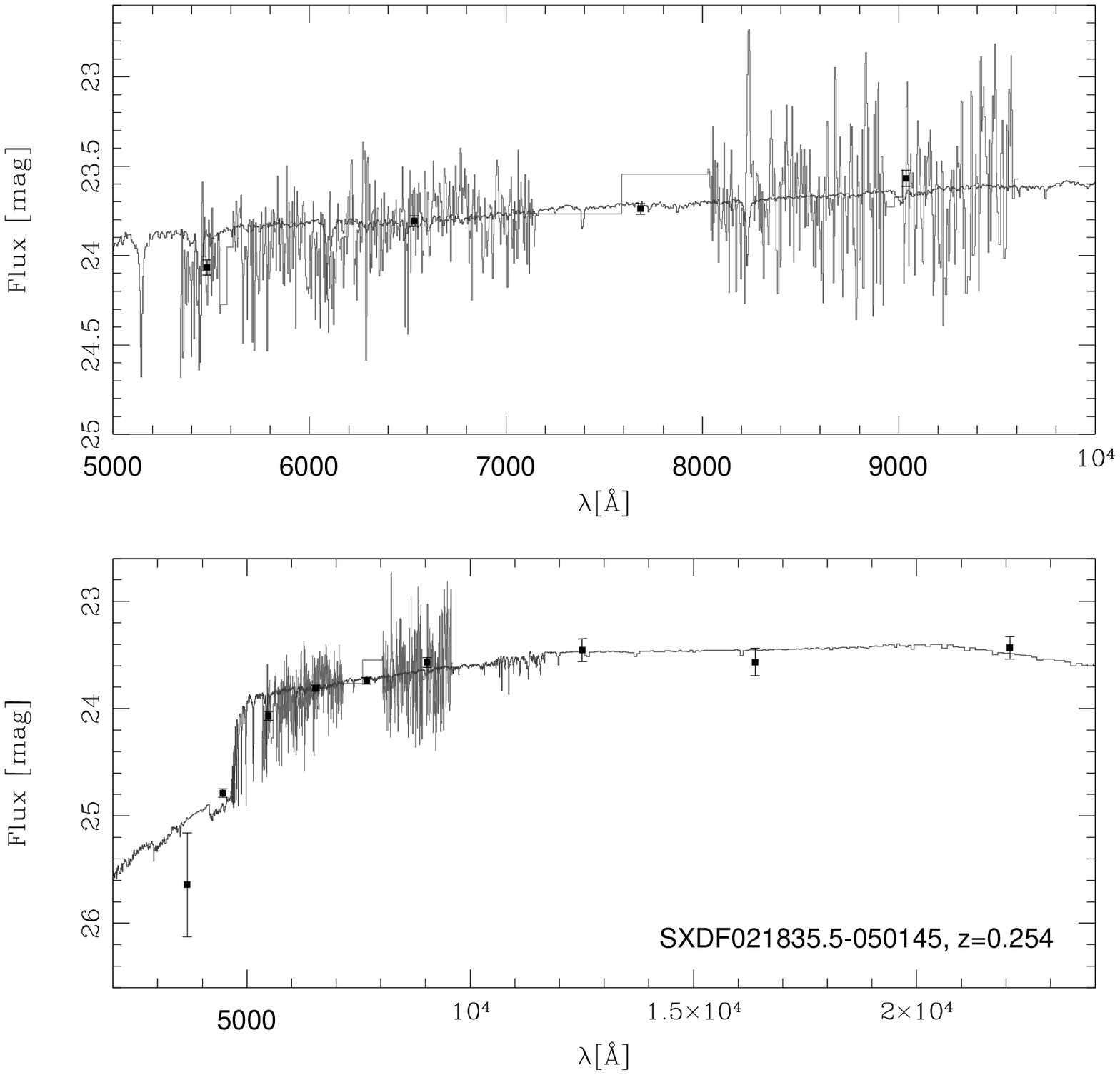}
\includegraphics[width=0.32\textwidth]{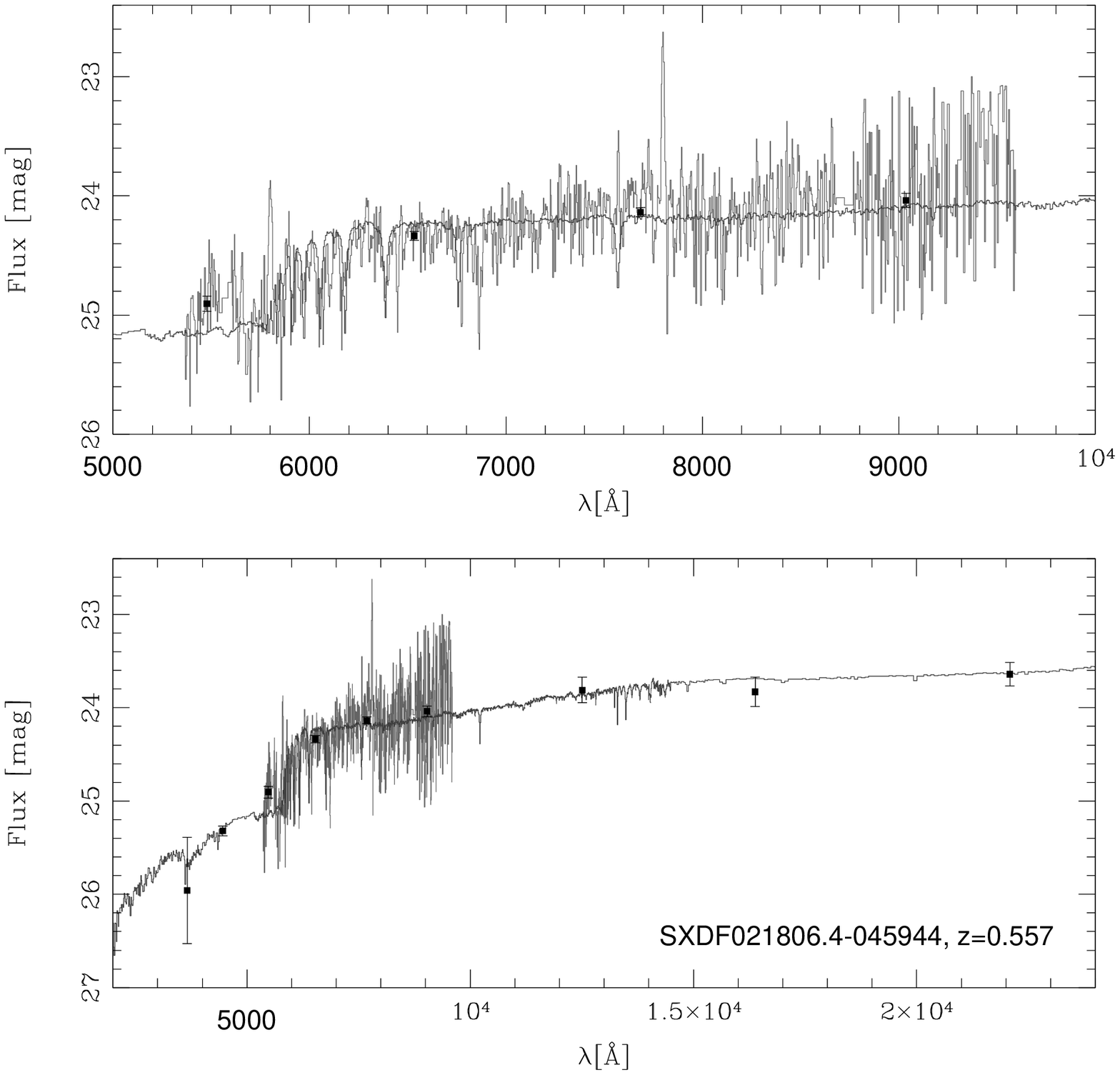}
\includegraphics[width=0.32\textwidth]{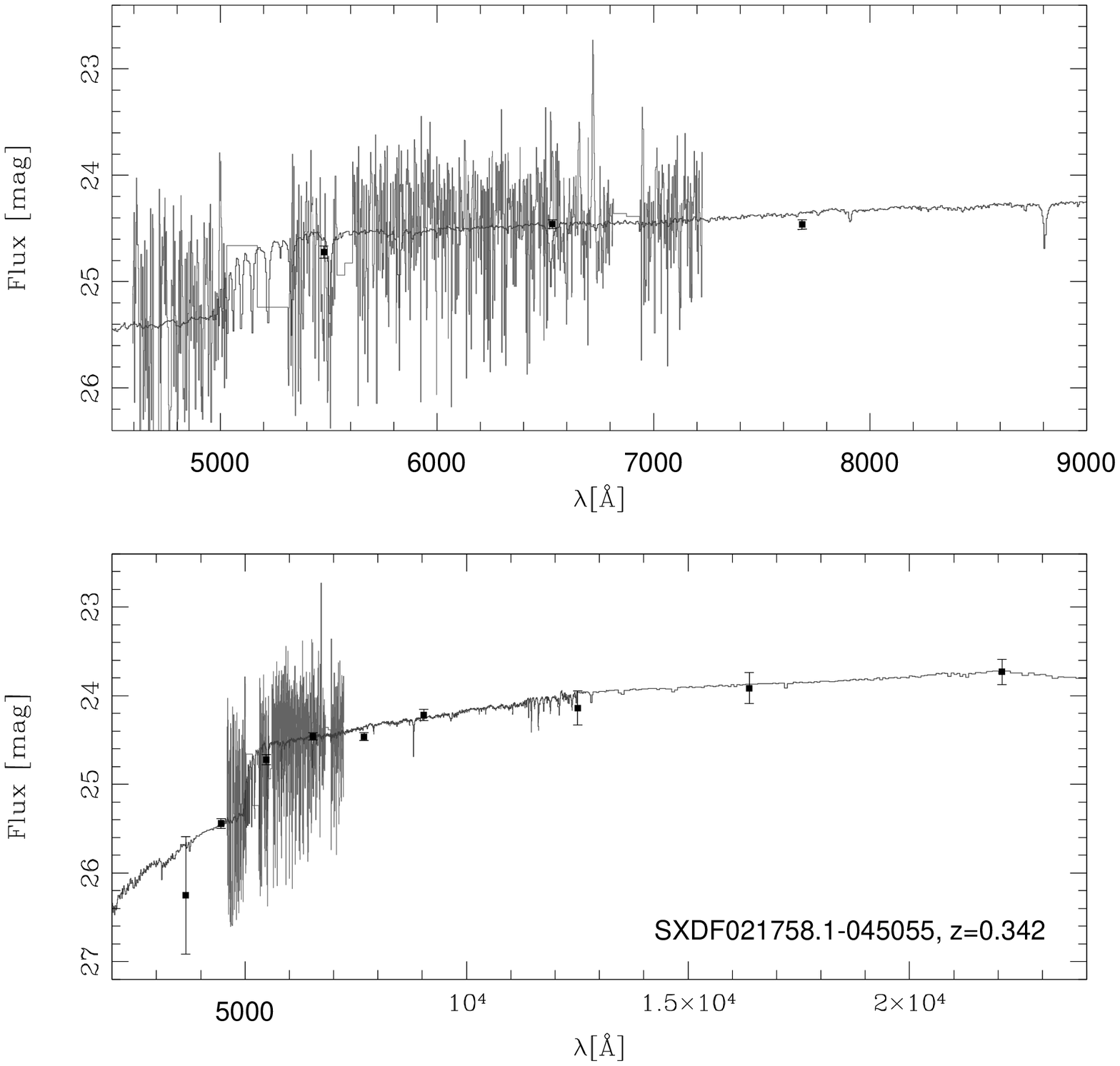}
\caption{\footnotesize{\textit{$-$Continued.}}}
\end{center}
\end{figure}
\clearpage

\begin{figure}
\begin{center}
\includegraphics[width=0.5\textwidth]{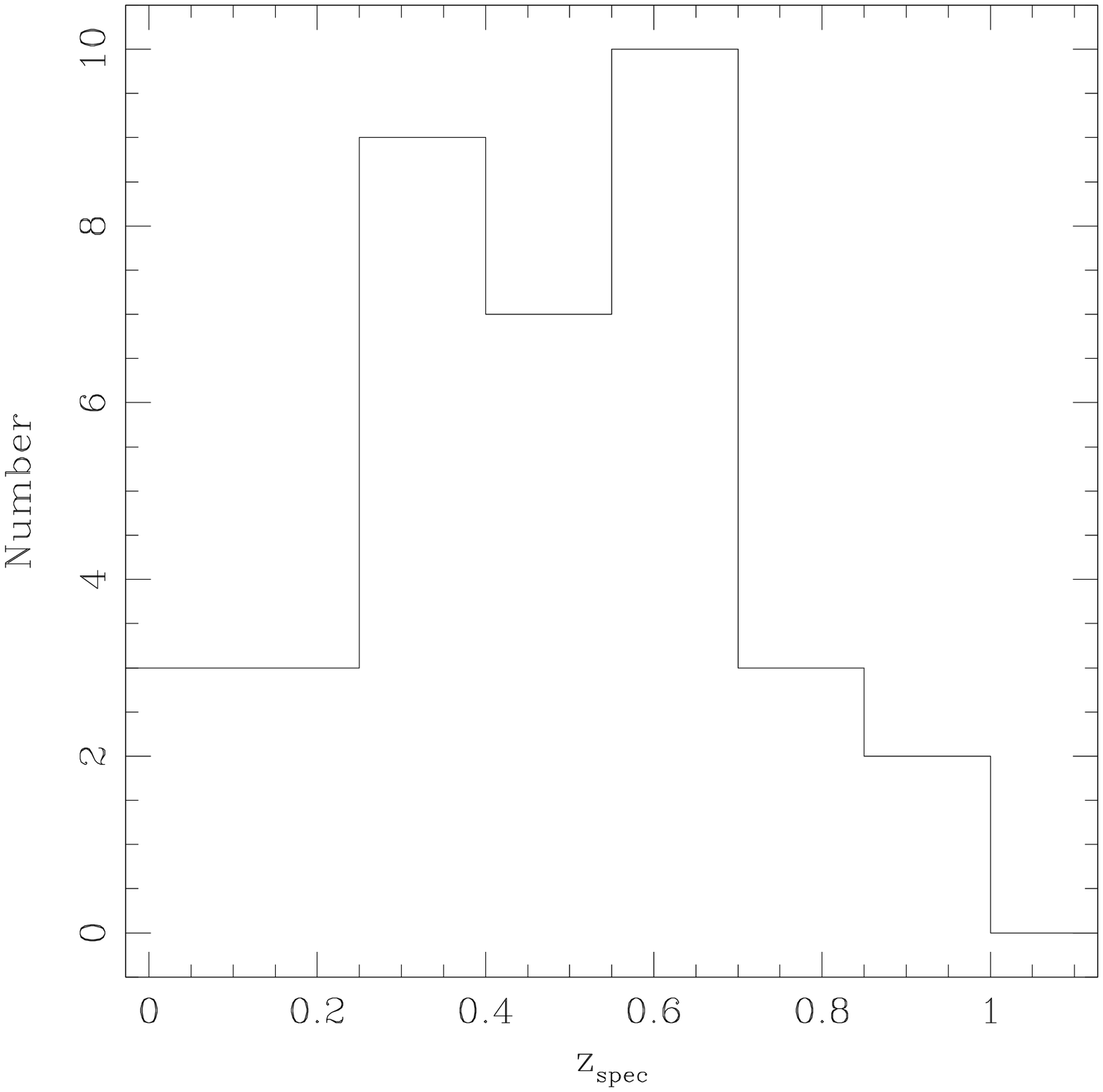}
\caption{\footnotesize{Redshift distribution of galaxies studied in this paper.}}\label{redshift}
\end{center}
\end{figure}

\begin{figure}
\begin{center}
\includegraphics[width=0.49\textwidth]{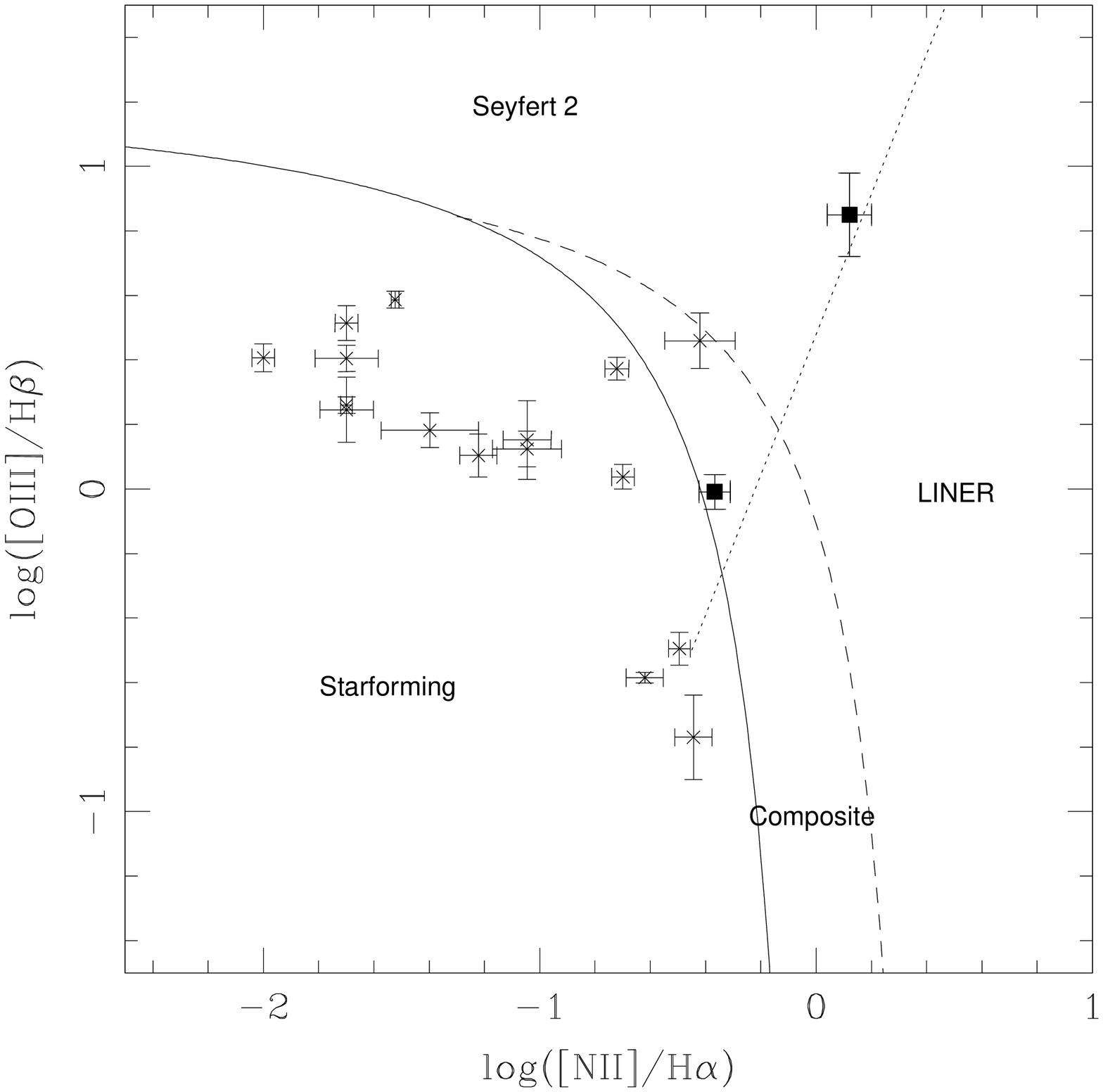}
\includegraphics[width=0.49\textwidth]{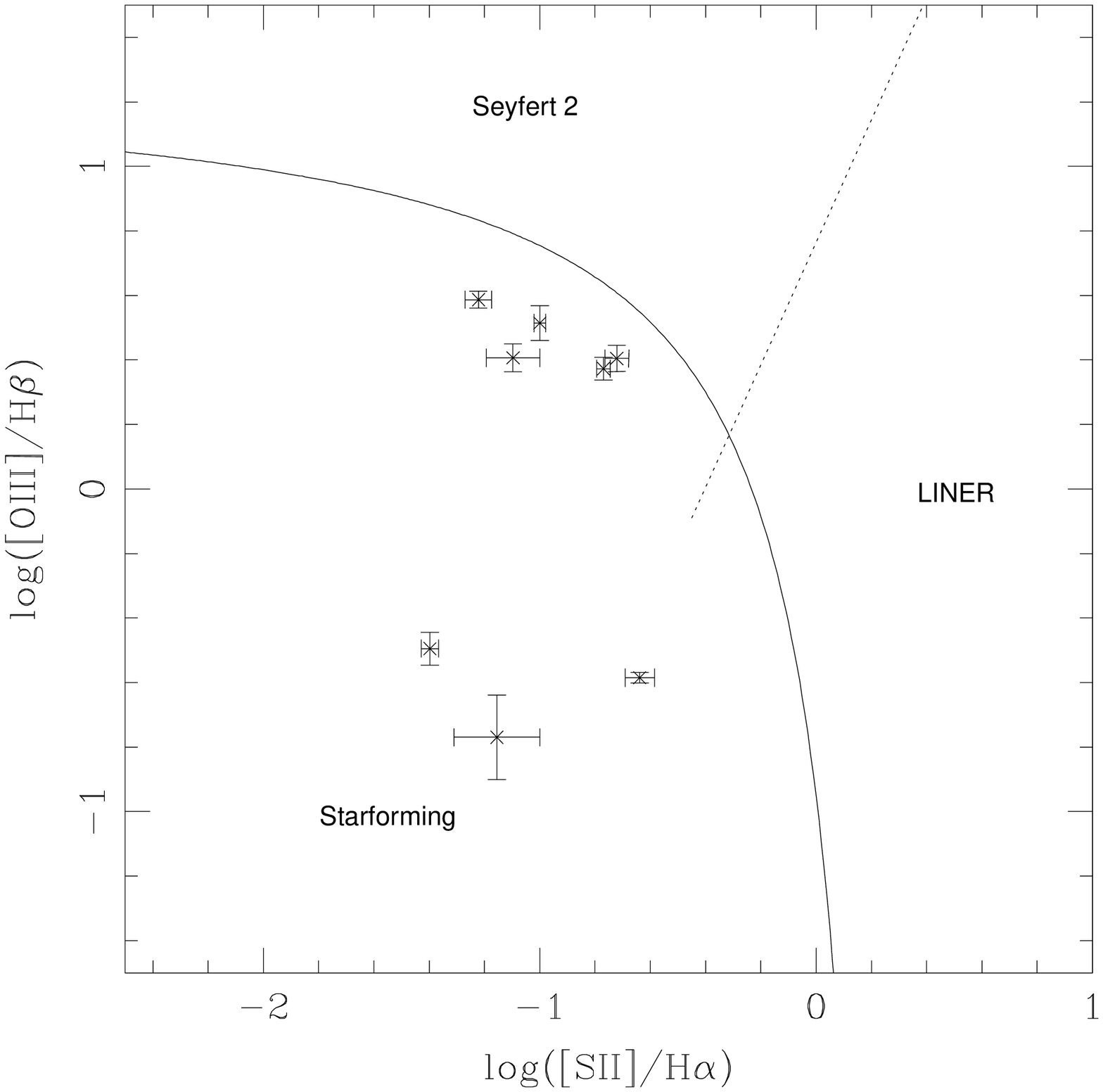}
\caption{\footnotesize{Left panel: [OIII]/H$\beta$ vs [NII]/H$\alpha$ diagnostic diagram. The conservative boundary line for AGNs was taken from Kewley et al. \citeyearpar{Kewley3}. Seyfert 2$-$LINER and star-forming$-$composite (AGN+star-forming) limits were taken from Kauffmann et al. \citeyearpar{Kauffmann}. Black square symbols are galaxies that could host an obscured AGN if their MIR colors are explored (see section \ref{mir} for more details).$-$ Right panel: [OIII]/H$\beta$ vs [SII]/H$\alpha$ diagnostic diagram. Boundary lines were taken from Kewley et al. \citeyearpar{Kewley3,Kewley}.}}\label{bpts}
\end{center}
\end{figure}
\clearpage

\begin{figure}
\begin{center}
\includegraphics[width=0.5\textwidth]{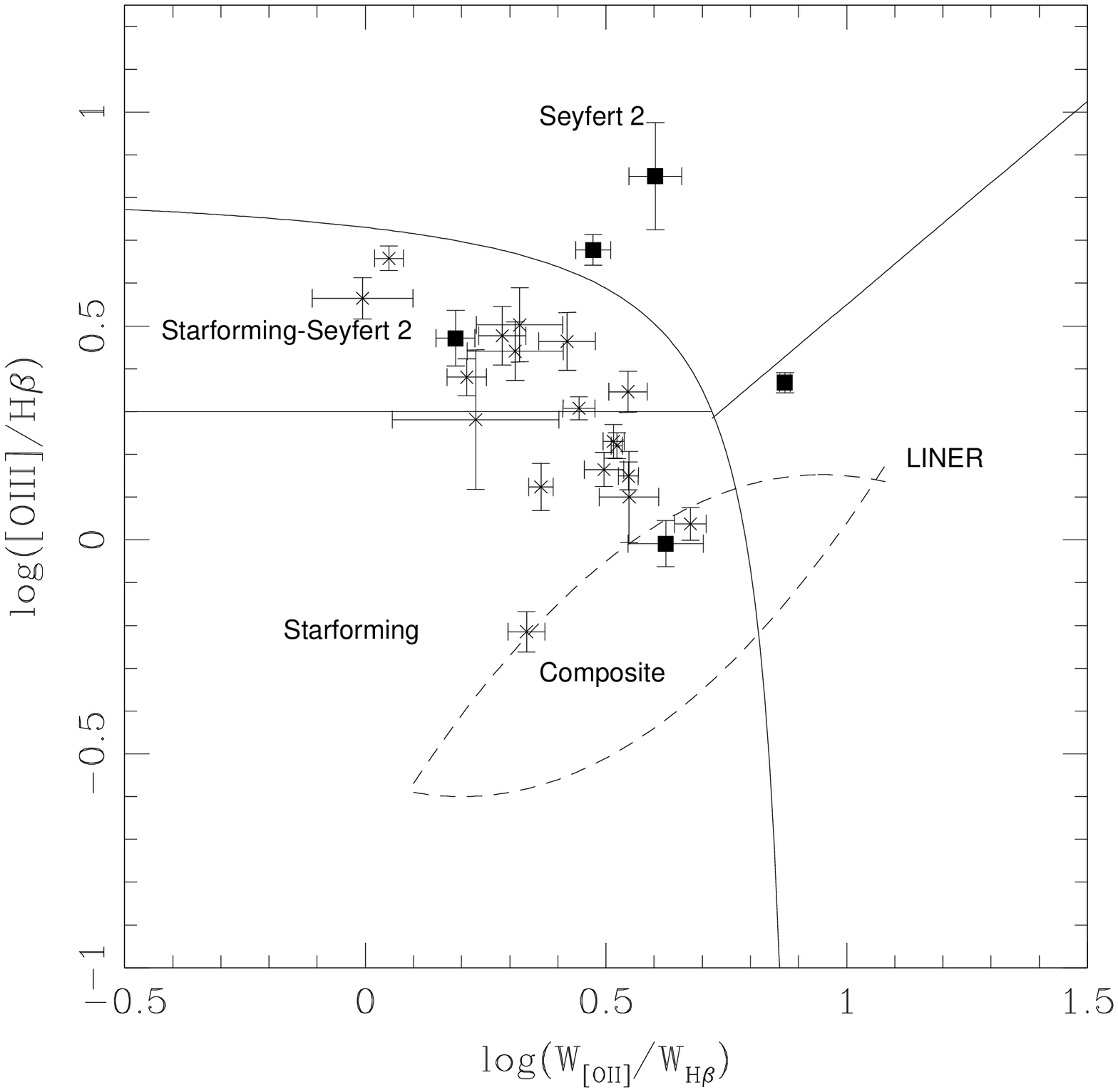}
\caption{\footnotesize{Lamareille [OIII]/H$\beta$ vs W$_{\text{[OII]}}$/W$_{\text{H}\beta}$ diagnostic diagram. AGN$-$star-forming boundaries are presented as Lamareille \citeyearpar{Lamareille}. Black square symbols are galaxies that could host an AGN according to their MIR colors.}}\label{hzbpt}
\end{center}
\end{figure}

\begin{figure}
\begin{center}
\includegraphics[width=0.5\textwidth]{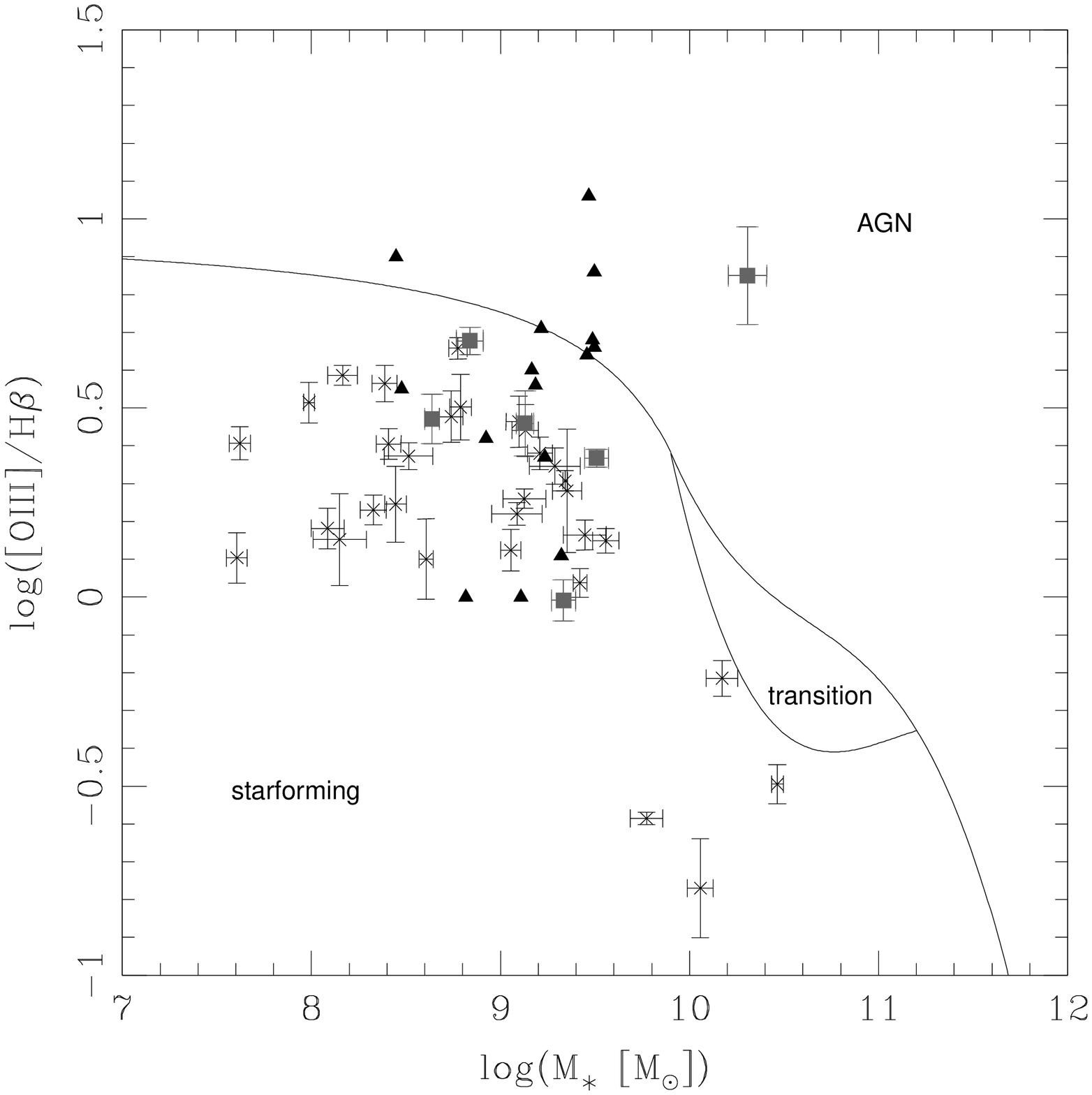}
\caption{\footnotesize{Mass-Excitation star-forming$-$AGN diagnostic diagram. The empirical curves were taken from Juneau et al. \citeyearpar{Juneau}, which separate between star-forming and AGNs. The region located between the two empirical curves on the MEx diagram contains composites galaxies. Gray square symbols are our AGN candidates, either selected in the optical or MIR diagnostic diagrams. Black triangles are low mass Seyfert 2 galaxies of Barth, Greene and Ho \citeyearpar{Barth} and narrow line Seyfert 1 galaxies of Greene $\&$ Ho \citeyearpar{Greene}.}}\label{maex}
\end{center}
\end{figure}
\clearpage

\begin{figure}
\begin{center}
\includegraphics[width=0.49\textwidth]{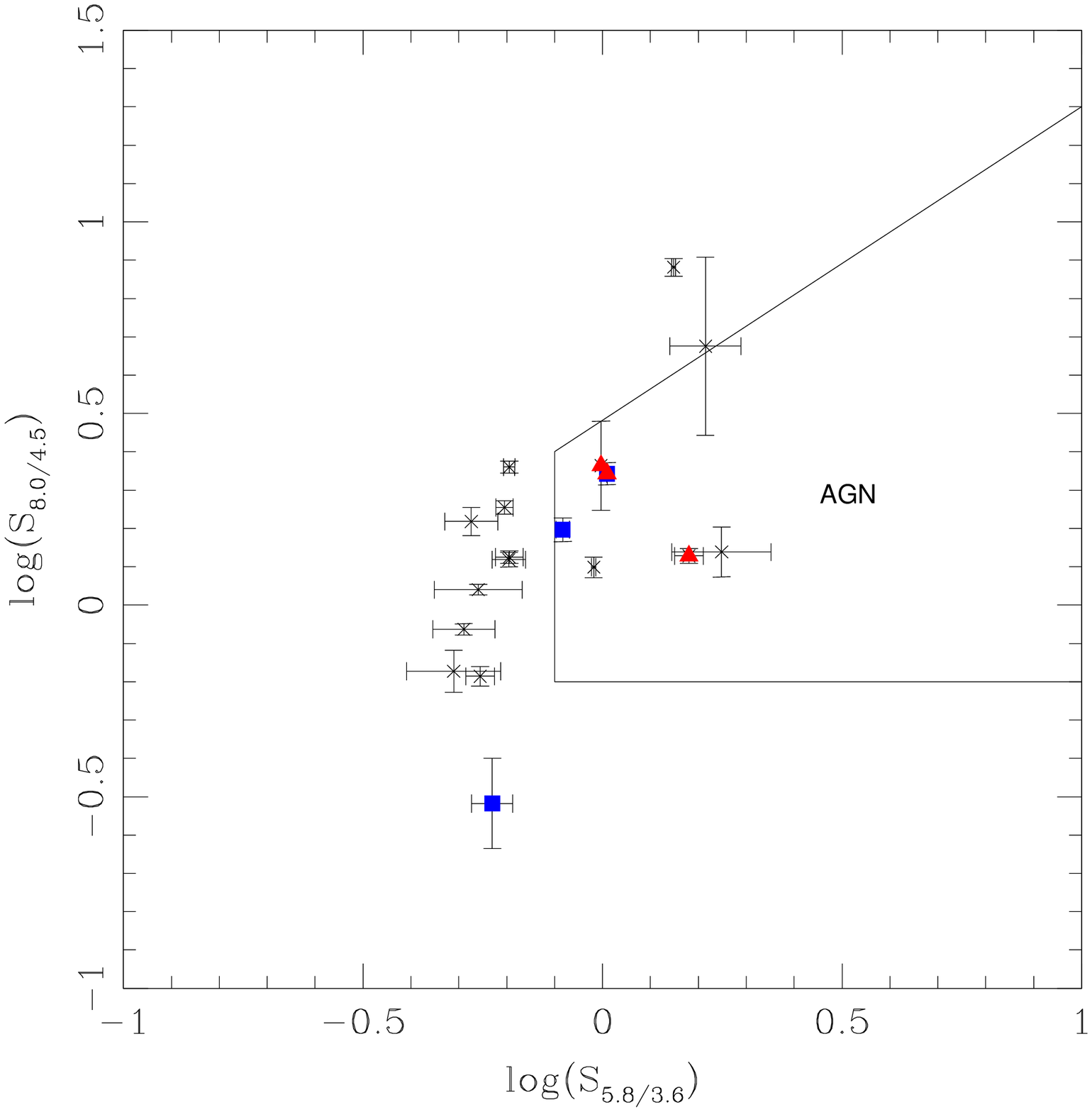}
\includegraphics[width=0.49\textwidth]{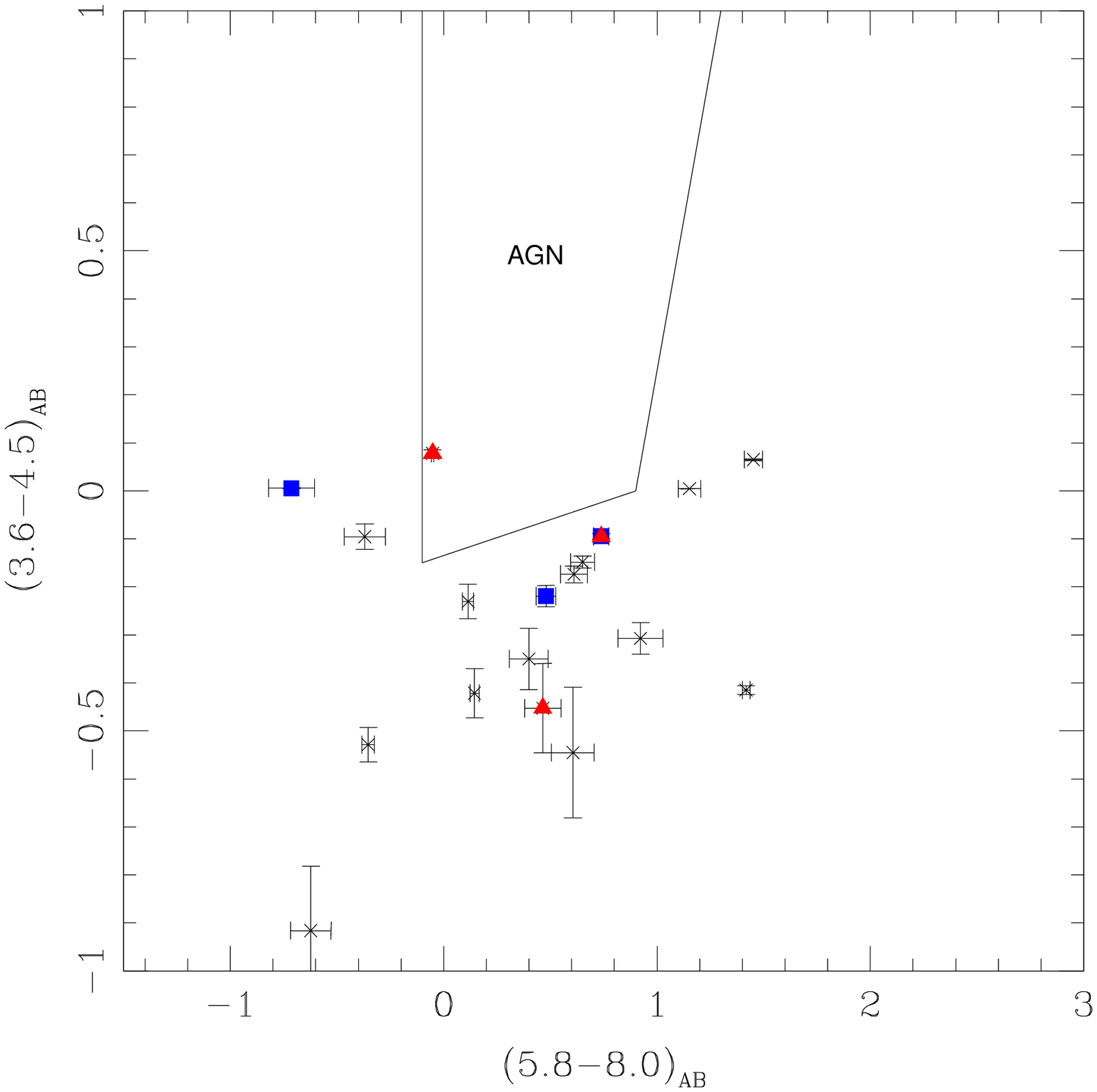}
\caption{\footnotesize{Left panel: log($S_{8.0}/S_{4.5}$) vs log($S_{5.8}/S_{3.6}$) diagram.$-$ Right panel: (3.6$-$4.5$\mu$m) vs (5.8$-$8.0$\mu$m) diagram. Black solid lines show the limits adopted by Stern et al. \citeyearpar{Stern} and Lacy et al. \citeyearpar{Lacy} respectively. Blue squares are AGN candidates selected by the Kewley's diagrams, while red triangles are AGN candidates selected from Lamareille's diagram (see section \ref{opt}).}}\label{iracds}
\end{center}
\end{figure}

\begin{figure}
\begin{center}
\includegraphics[width=0.49\textwidth]{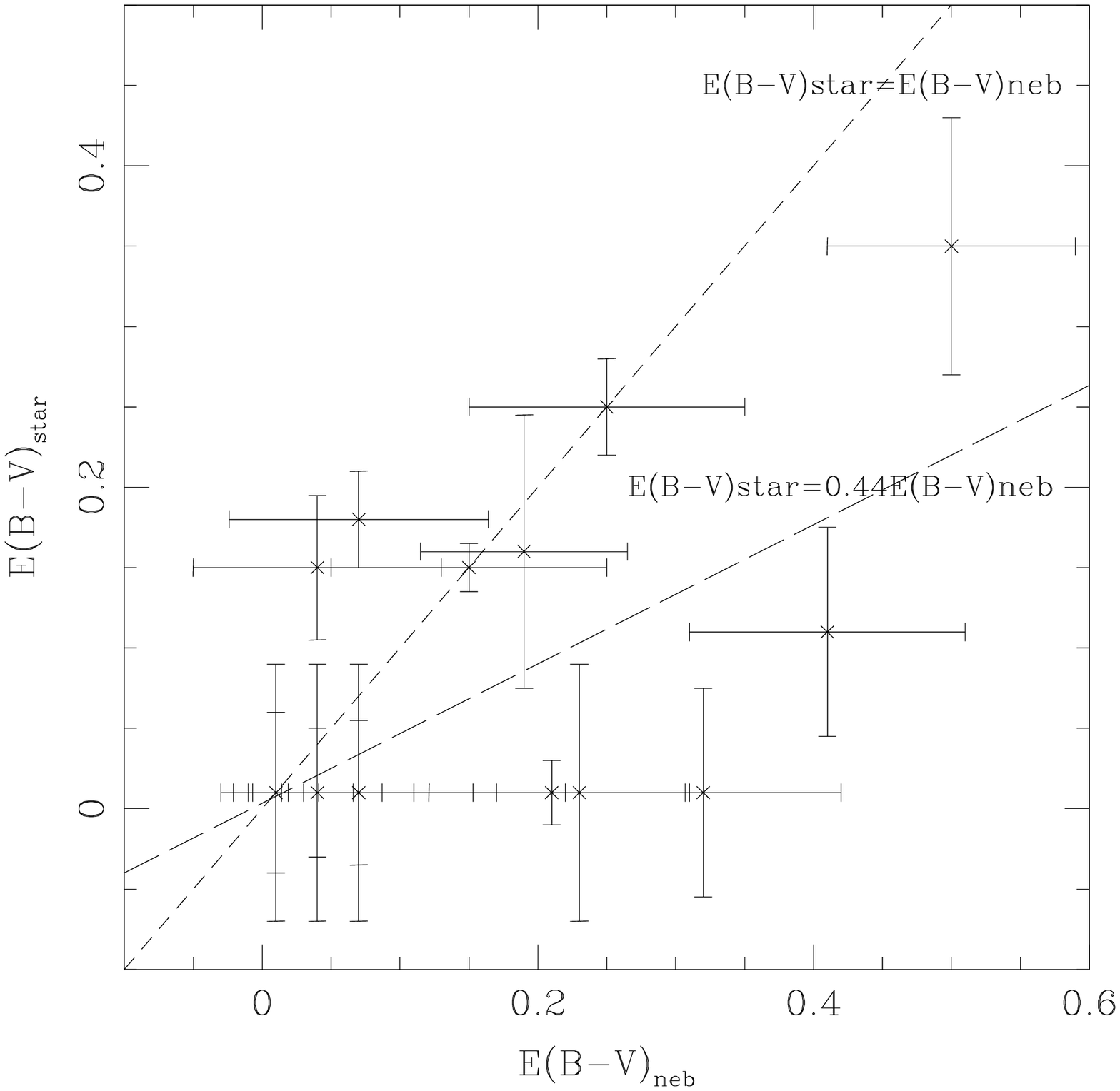}
\includegraphics[width=0.49\textwidth]{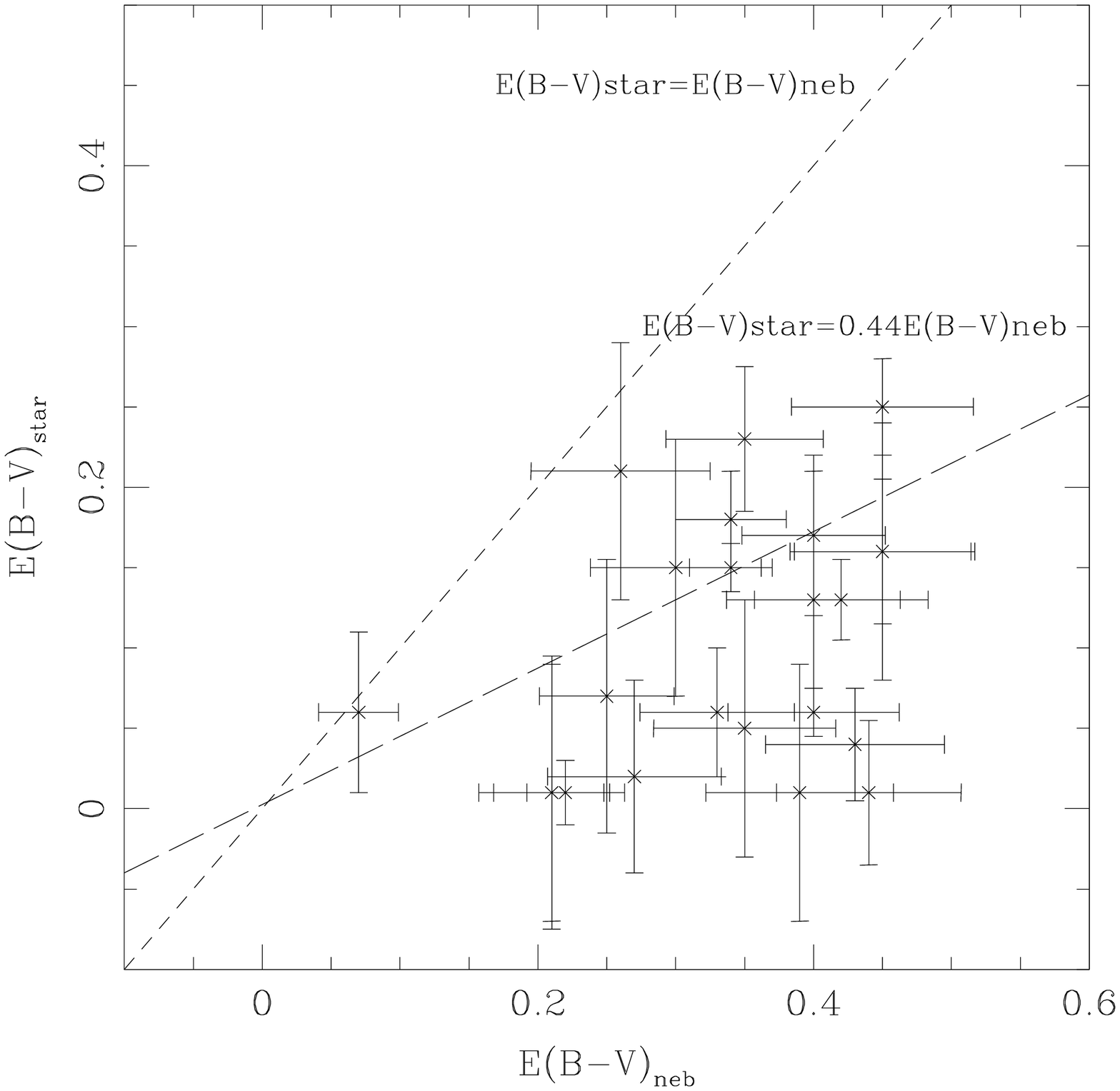}
\caption{\footnotesize{Stellar reddening vs nebular reddening. The left panel shows nebular extinction obtained from the Balmer decrement,
while the right panel shows the nebular extinction calculated through the [OII]$\lambda$3727 luminosity (\citeauthor{Kewley2}). Stellar extinction was obtained from SED fitting to aperture magnitudes. The short-dashed line shows the 1:1 ratio, while the long-dashed line shows the ratio found by Calzetti et al. \citeyearpar{Calzetti2} for local starburst galaxies, with the nebular reddening being more significant than the internal stellar reddening.}}\label{red}
\end{center}
\end{figure}
\clearpage

\begin{figure}
\begin{center}
\includegraphics[width=0.5\textwidth]{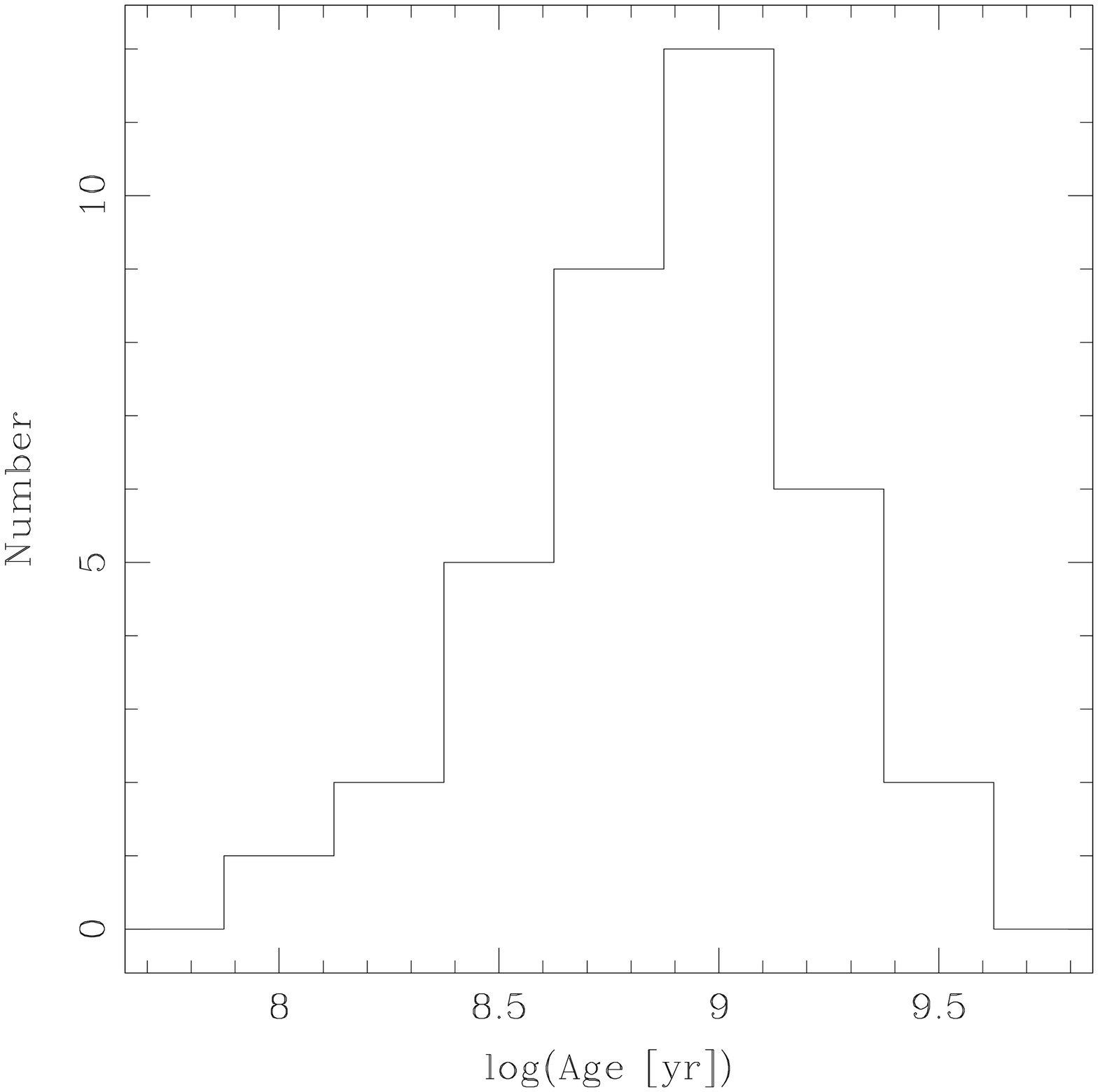}
\includegraphics[width=0.5\textwidth]{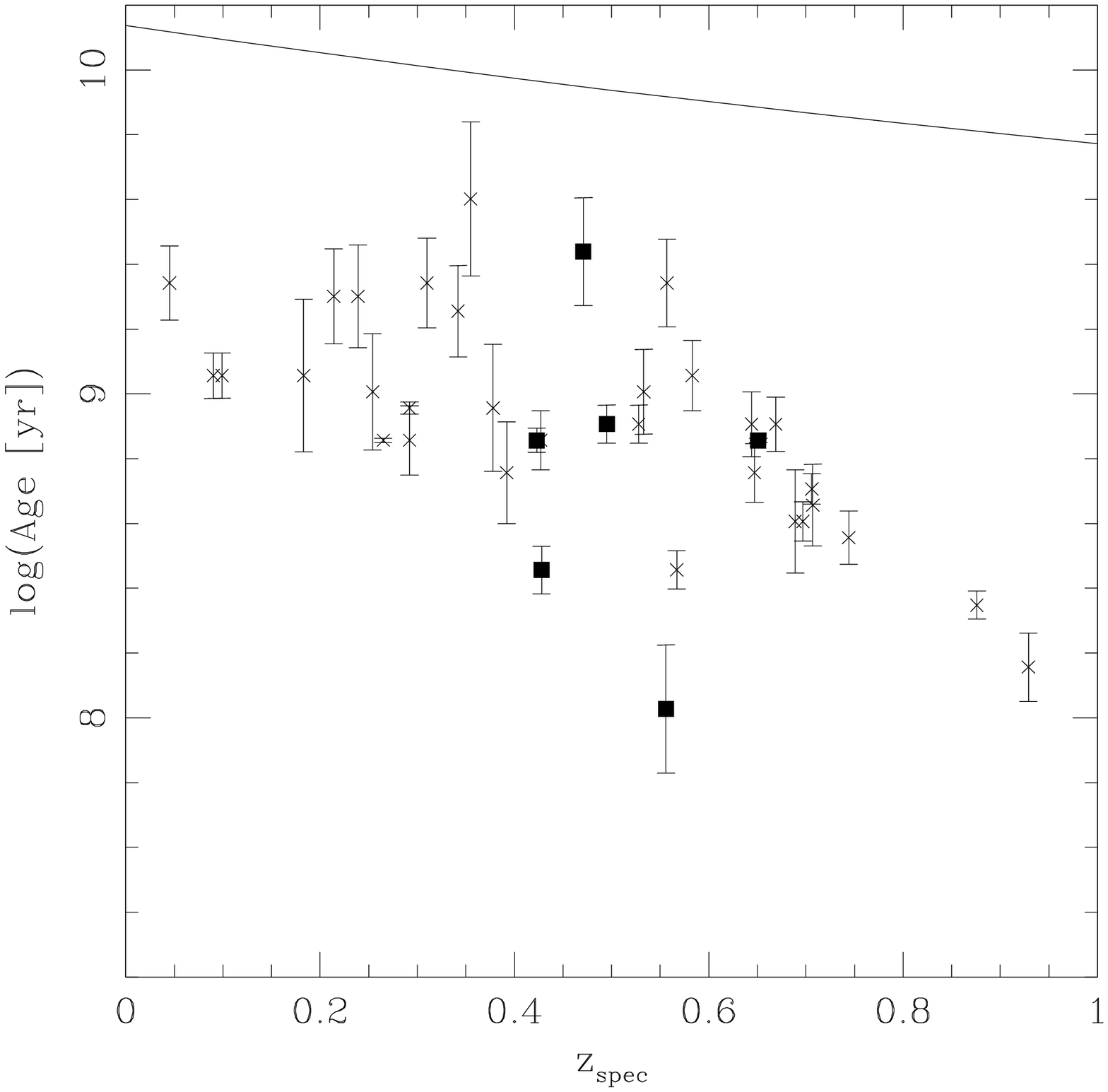}
\caption{\footnotesize{Top panel: age distribution for the sample galaxies.$-$ Bottom panel: age versus redshift. Black square symbols are our AGN candidates. The solid line shows the age of the universe as a function of the redshift.}}\label{ages}
\end{center}
\end{figure}
\clearpage

\begin{figure}
\begin{center}
\includegraphics[width=0.49\textwidth]{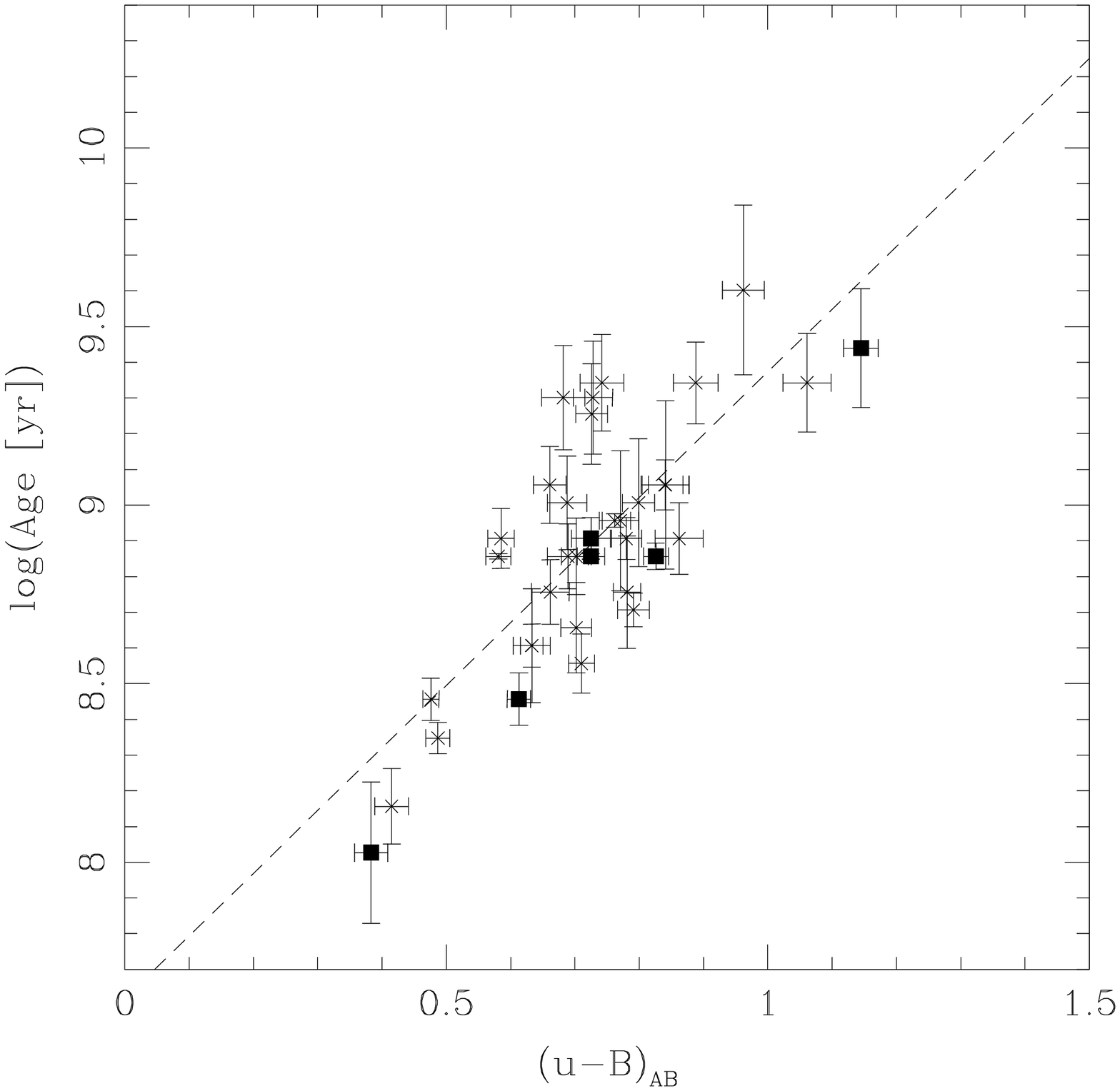}
\includegraphics[width=0.49\textwidth]{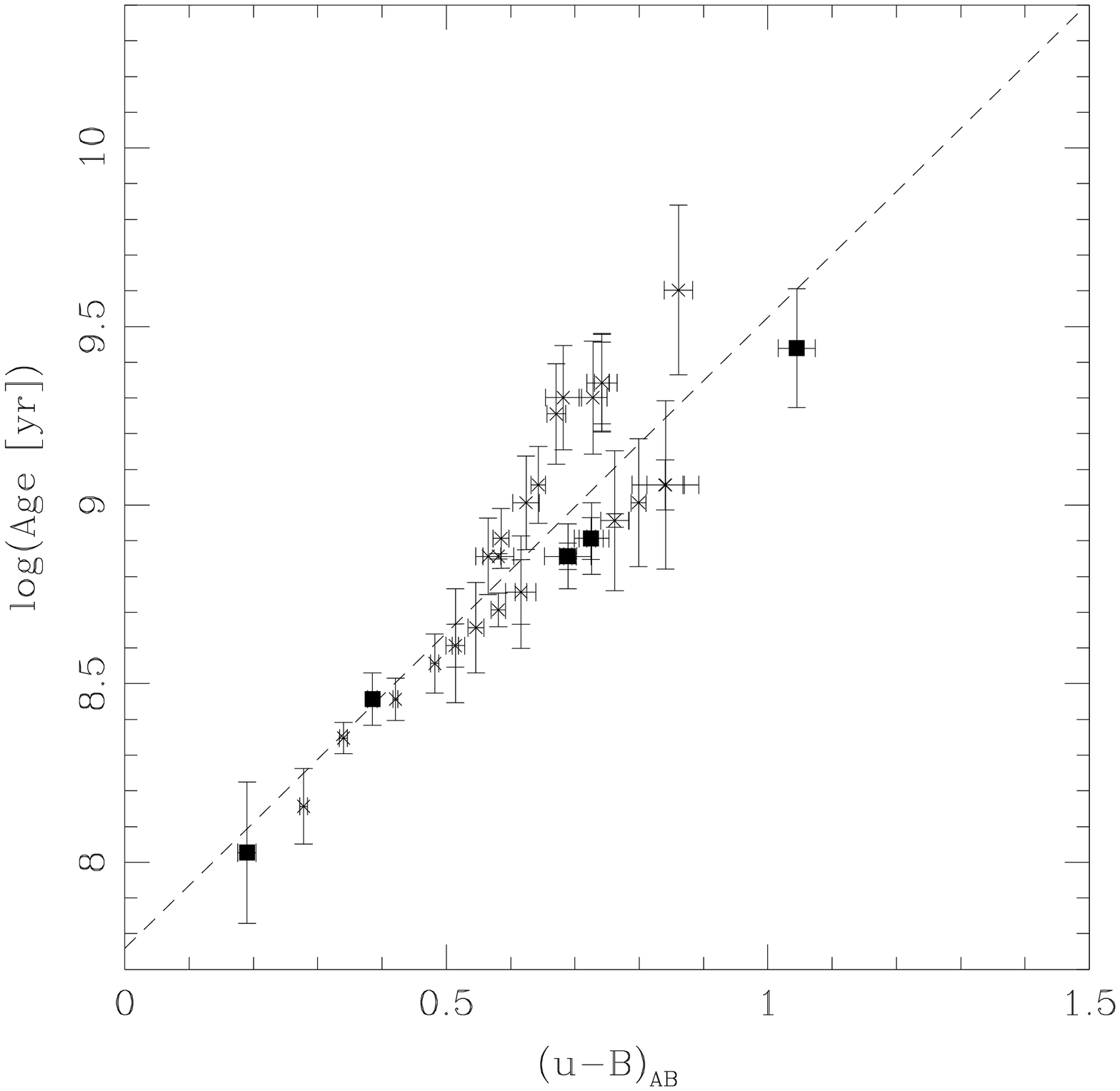}
\caption{\footnotesize{Age vs rest frame color index $(u-B)_{AB}$, assuming an exponentially declining star formation history. The left panel shows the color index $(u-B)_{AB}$ without reddening correction, while the right panel shows the comparison after the reddening correction. The dashed line shows the best linear fit between both quantities. Black square symbols are our AGN candidates.}}\label{colorage}
\end{center}
\end{figure}
\clearpage

\begin{figure}
\begin{center}
\includegraphics[width=0.5\textwidth]{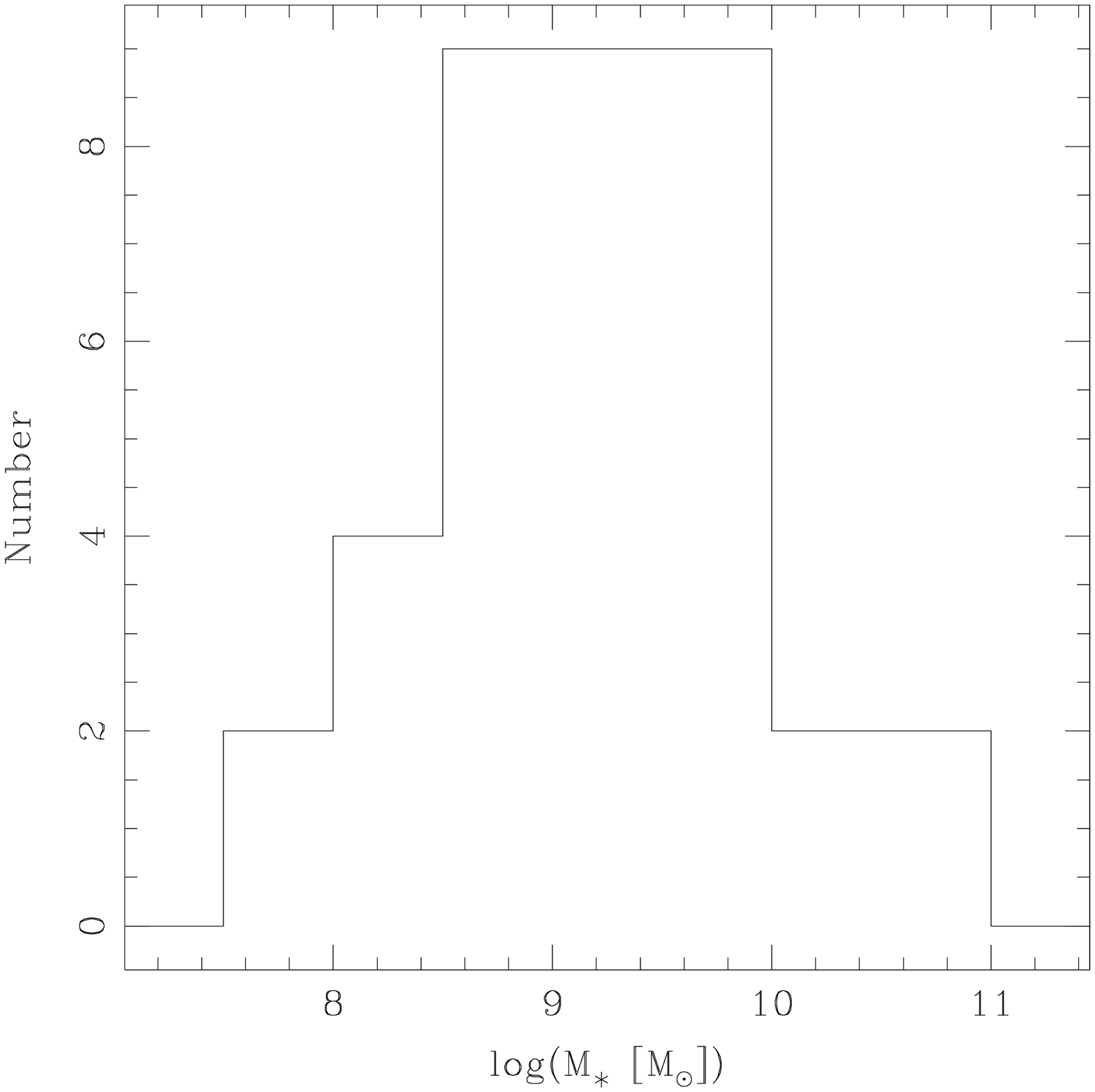}
\includegraphics[width=0.5\textwidth]{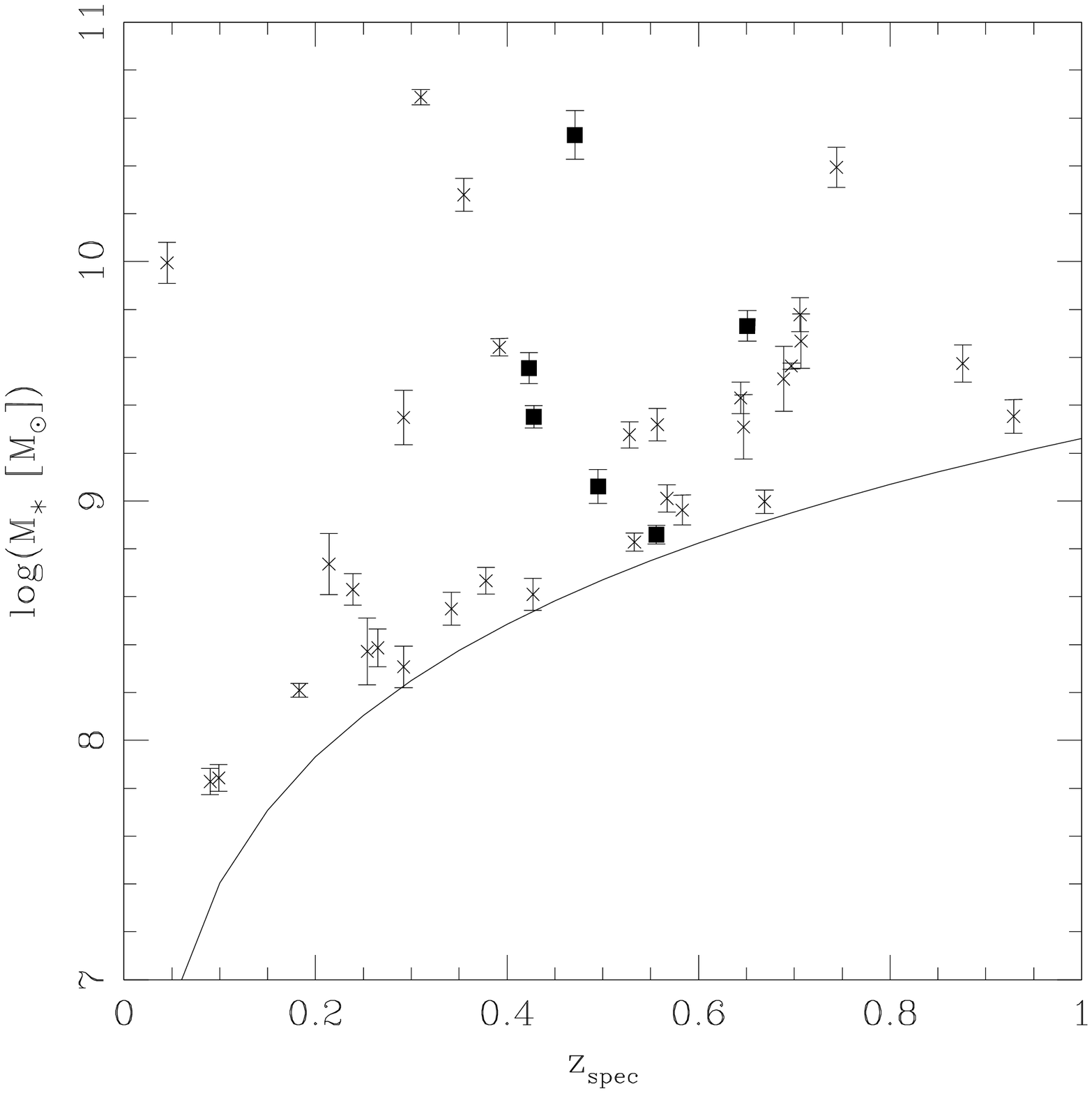}
\caption{\footnotesize{Top panel: total stellar mass distribution of galaxies.$-$ Bottom panel: stellar masses of our galaxies as a function of their redshifts. Black square symbols are our AGN candidates. The solid line represents our mass limit calculated using the M/L ratio of a young starburst (M/L$_{V}$=0.02).}}\label{histm}
\end{center}
\end{figure}
\clearpage

\begin{figure}
\begin{center}
\includegraphics[width=0.5\textwidth]{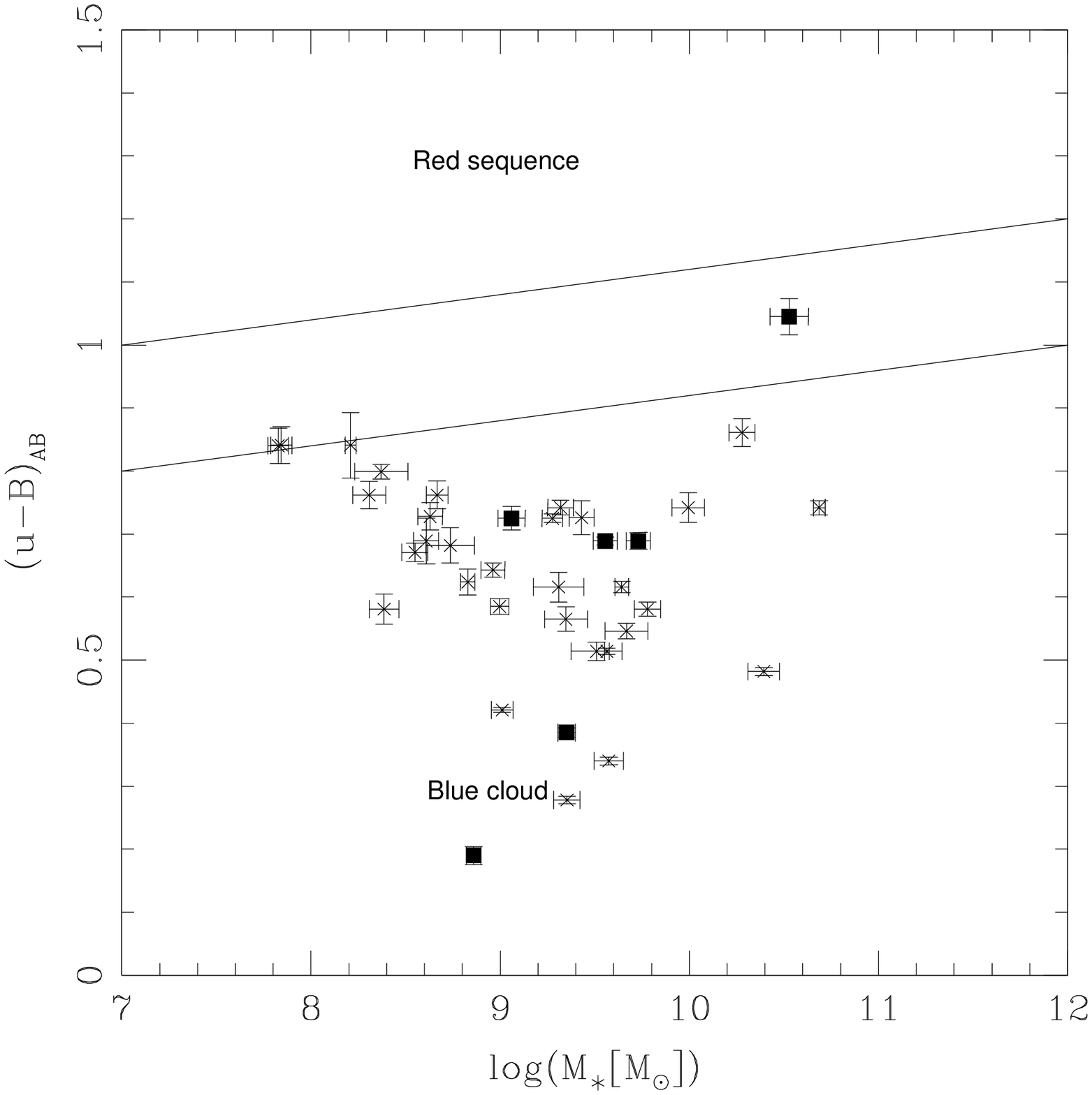}
\caption{\footnotesize{Rest frame dust attenuation corrected color $(u-B)_{AB}$ vs total stellar mass. Black square symbols represent our AGN candidates selected either using optical or MIR diagnostic diagrams. The solid lines indicate the regions where passive (upper), post-starburst (middle) and star-forming (lower) galaxies would be located.}}\label{color}
\end{center}
\end{figure}

\begin{figure}
\begin{center}
\includegraphics[width=0.5\textwidth]{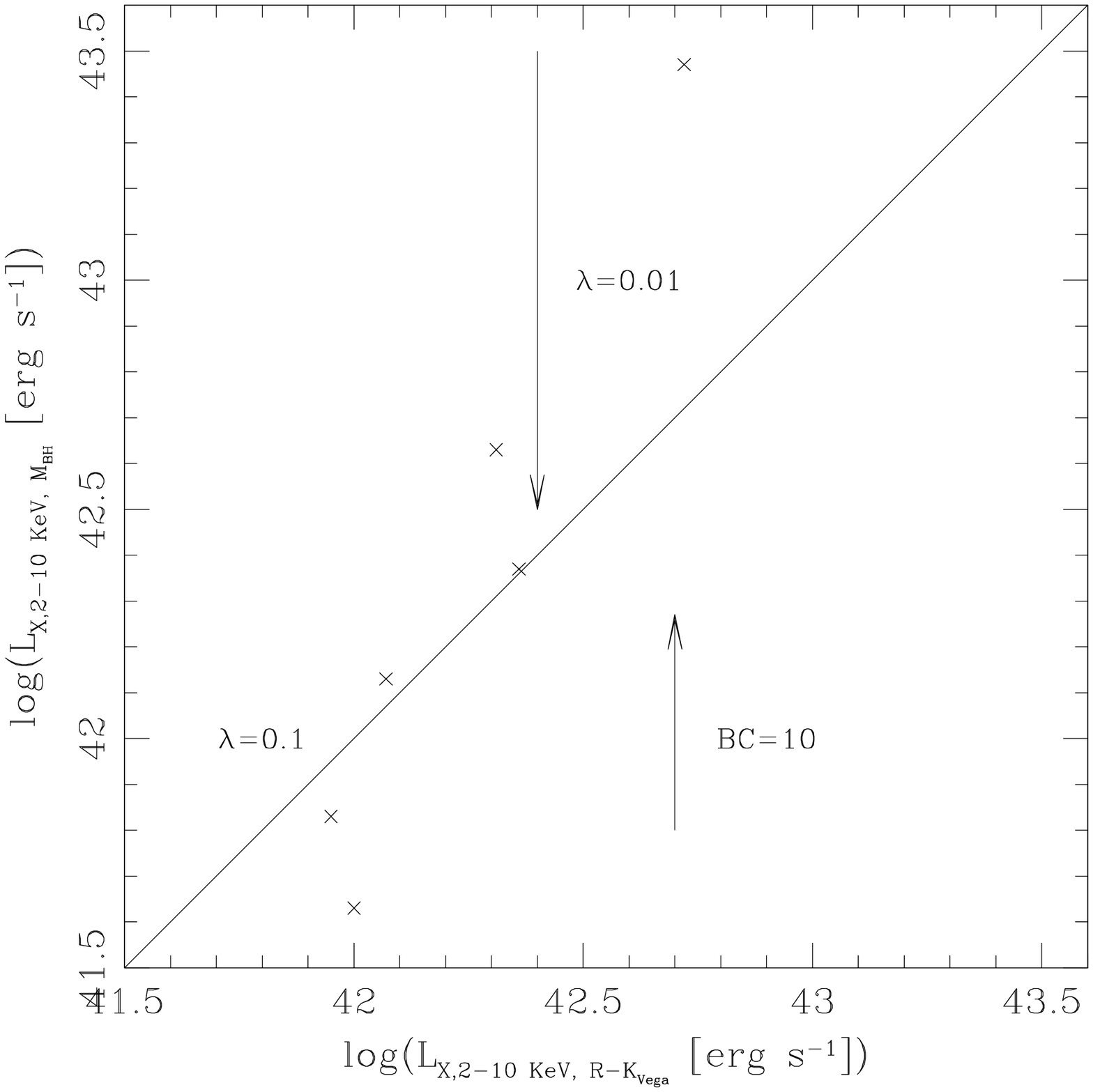}
\caption{\footnotesize{X-ray luminosities derived from $M_{\text{BH}}$, $L_{X,2-10 \text{keV}, M_{\text{BH}}}$ against X-ray luminosities derived from their $(R-K)_{\text{Vega}}$ colors, $L_{X,2-10 \text{keV}, R-K_{\text{Vega}}}$ for the six AGN candidates. The solid line shows a 1:1 relation between both quantities, with the $\lambda$ and BC values adopted (0.1 for the Eddington ratio, and 30 for the bolometric correction) in the L$_{X,M_{\text{BH}}}$ estimation. Arrows show how $L_{X,M_{\text{BH}}}$ would vary if other $\lambda$ and BC values were adopted.}}\label{Lx}
\end{center}
\end{figure}
\clearpage

\begin{figure}
\begin{center}
\includegraphics[width=0.42\textwidth]{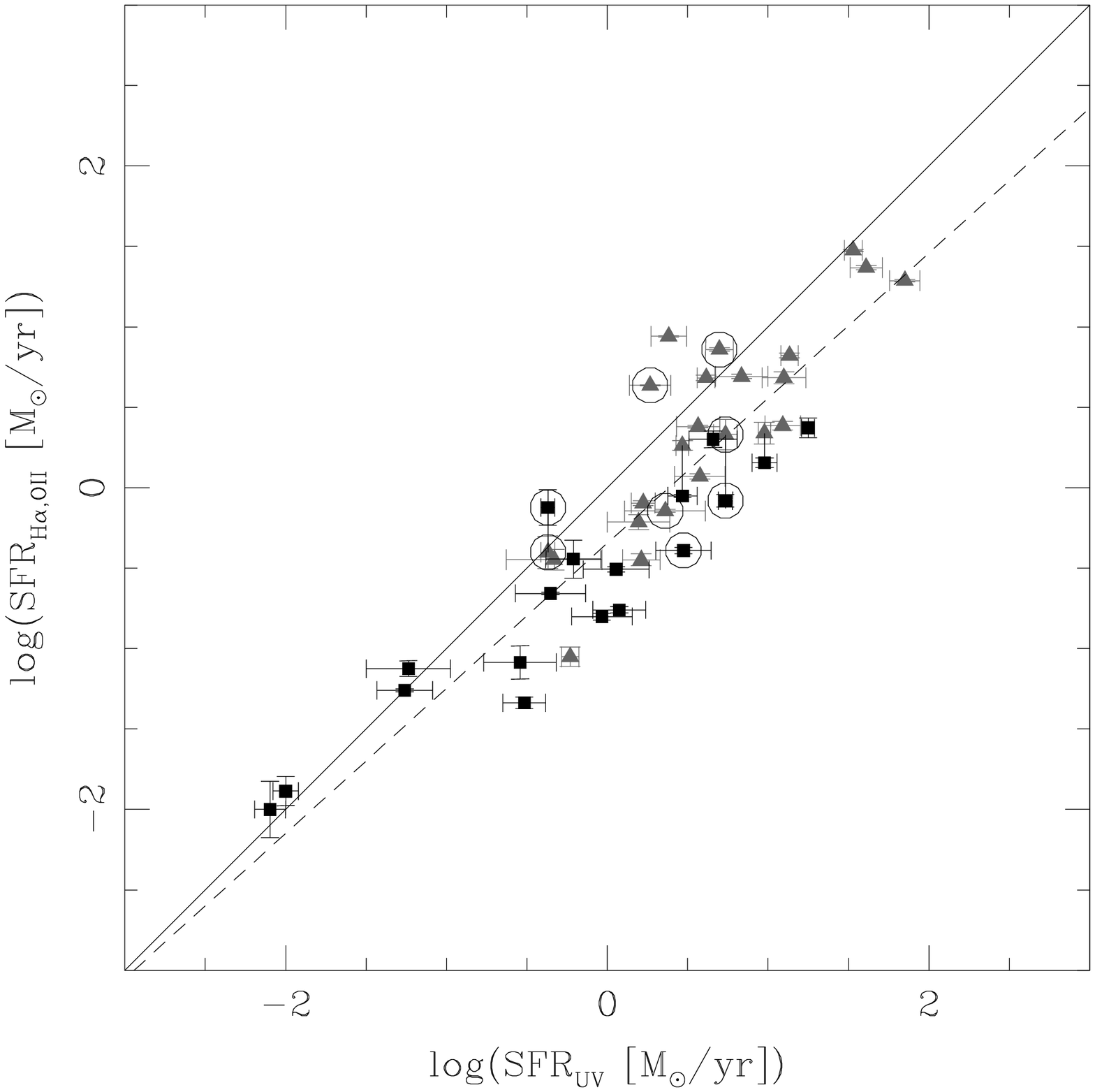}
\includegraphics[width=0.42\textwidth]{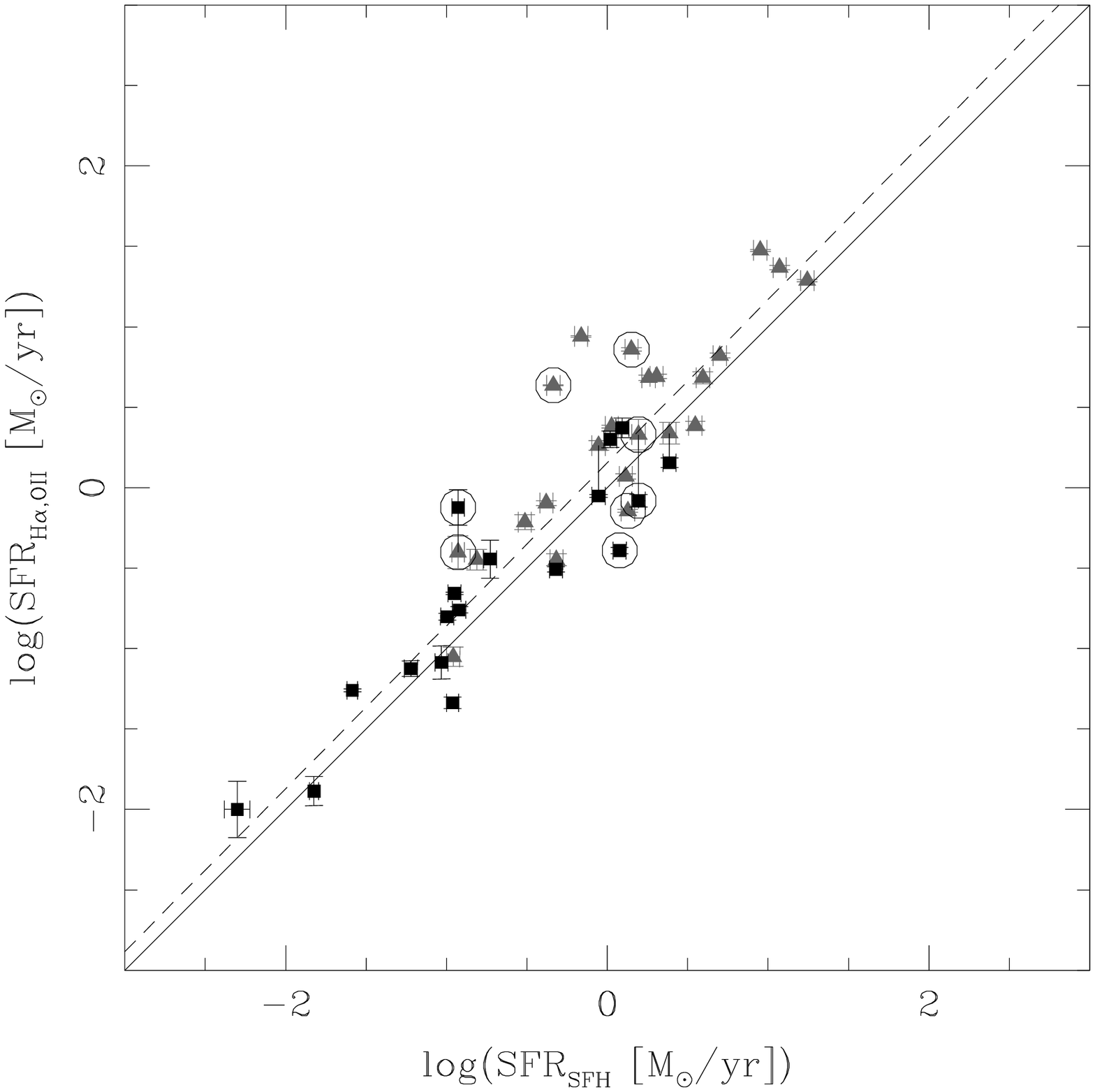}
\caption{\footnotesize{Left panel: SFR$_{\text{H}\alpha,\text{[OII]}}$ vs SFR$_{\text{UV}}$. Line fluxes were corrected using the $E(B-V)$ calculated through the Balmer decrement or the Kewley, Geller and Jansen \citeyearpar{Kewley2} recipe, depending on the availability of the necessary emission  lines. The UV fluxes were corrected using the $E(B-V)$ estimated from the SED fitting.$-$ Right panel: SFR$_{\text{H}\alpha,\text{[OII]}}$ vs SFR$_{\text{SFH}}$, obtained using exponentially declining SFHs (SFR$\sim$ $e^{-t/\tau}$). In both panels, points connected with a solid line represent the SFR calculated from H$\alpha$ (black squares) or [OII] (gray triangles) for the same galaxy. The solid line represents a 1:1 correlation, while the dashed line shows the linear fit applied to the sample. Empty circles represent our AGN candidates.}}\label{sfr}
\end{center}
\end{figure}

\begin{figure}
\begin{center}
\includegraphics[width=0.49\textwidth]{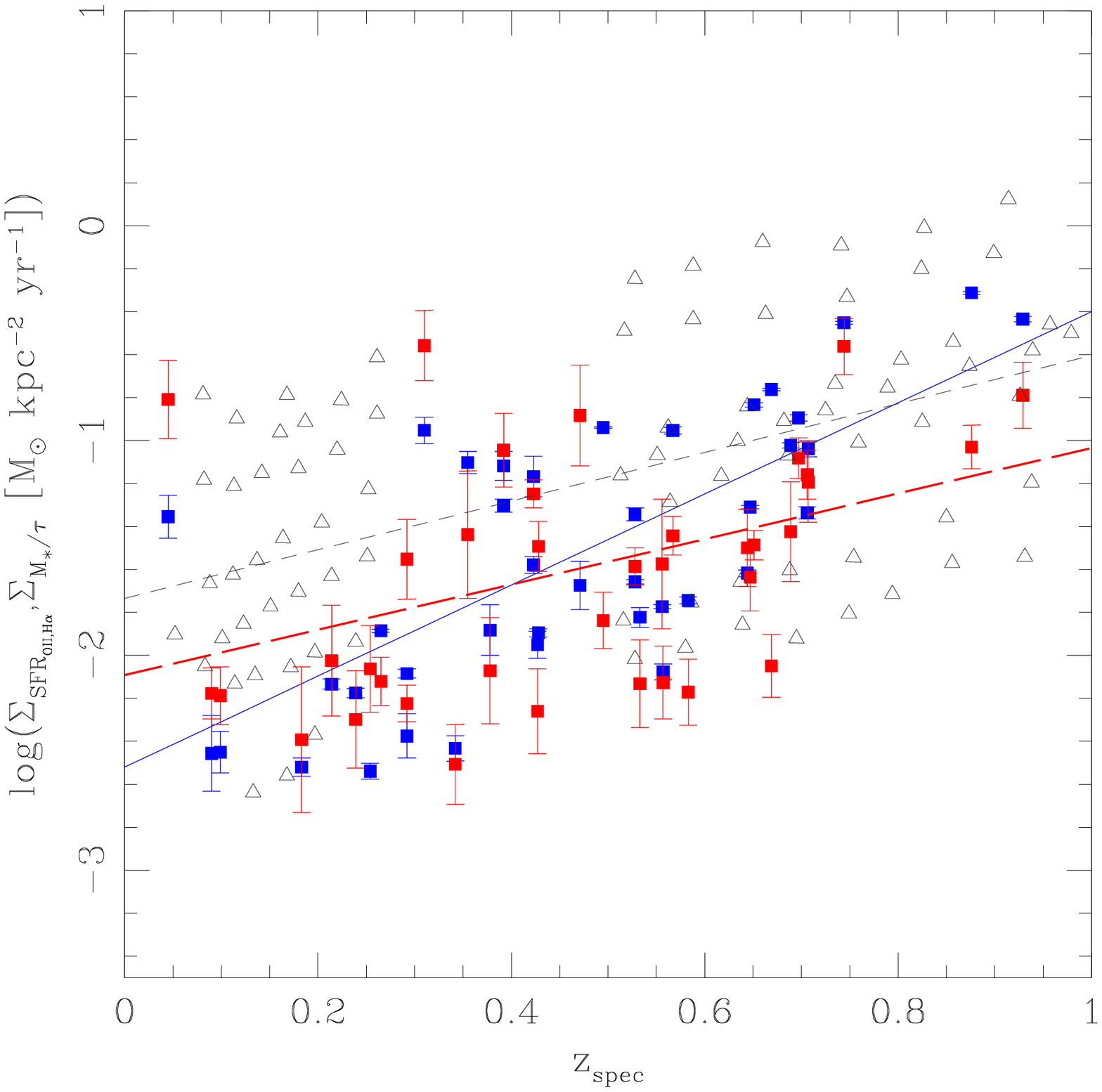}
\includegraphics[width=0.49\textwidth]{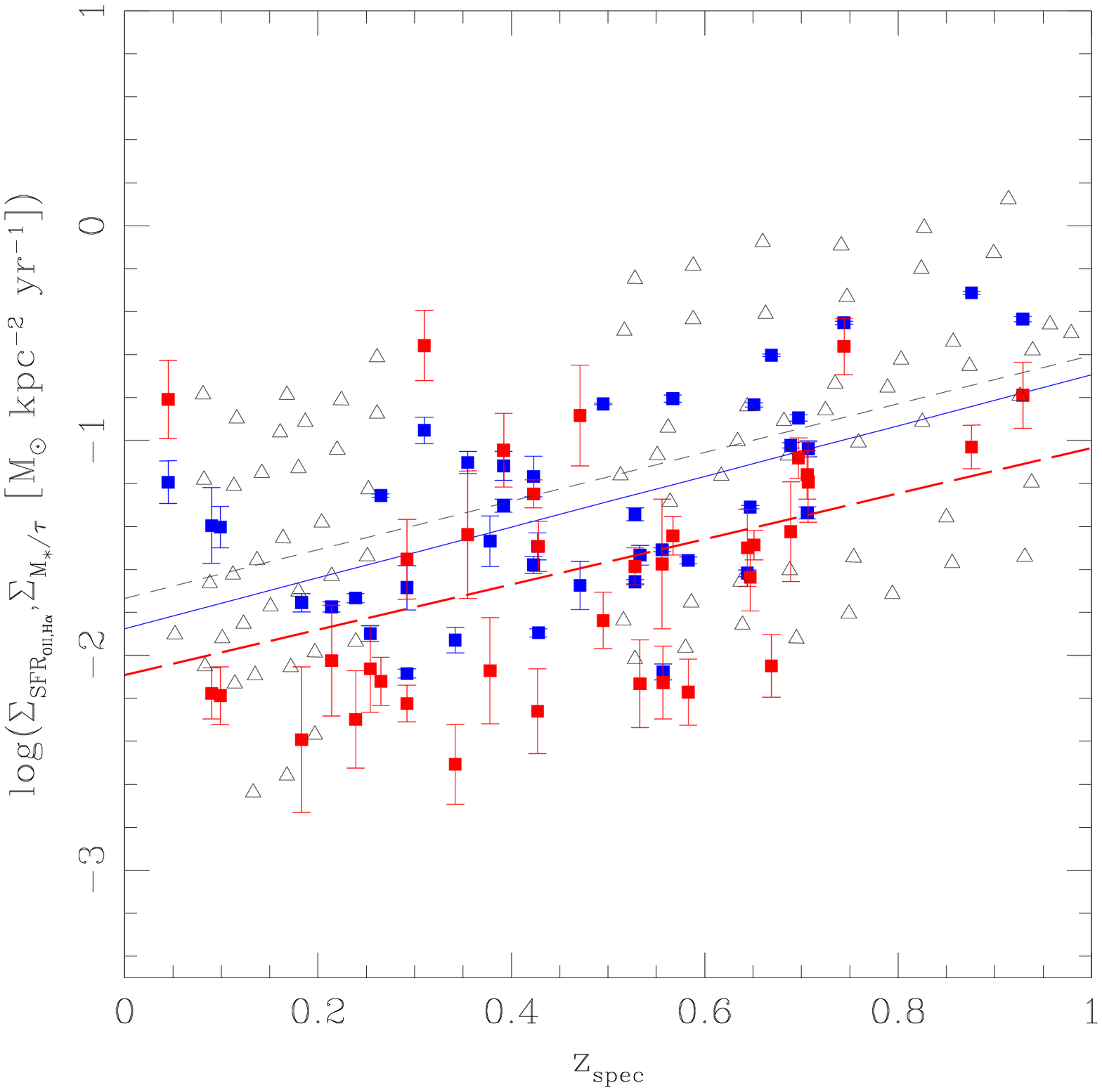}
\caption{\footnotesize{$\Sigma_{\text{SFR}_{\text{H}\alpha,\text{[OII]}}}$ (blue squares) and $\Sigma_{M_{\ast}/\tau}$ (red squares) as a function of redshift. Left panel shows $\Sigma_{\text{SFR}_{\text{H}\alpha,\text{[OII]}}}$ values prior to stellar mass normalization, where a systematic bias can be observed. In the right panel $\Sigma_{\text{SFR}_{\text{H}\alpha,\text{[OII]}}}$ values for galaxies with $M_{\ast}<$ 10$^{9}M_{\odot}$ were normalized to 10$^{9}M_{\odot}$ (see text for details). In both panels the long-dashed red and solid blue lines show the linear fits for $\Sigma_{M_{\ast}/\tau}$ and $\Sigma_{\text{SFR}_{\text{H}\alpha,\text{[OII]}}}$, respectively. As a comparison, we also plotted the $\Sigma_{\text{SFR}_{\text{[OII]}}}$ values calculated for zCOSMOS star-forming galaxies (\citeauthor{Silverman}) with redshifts spanning 0 $<$ $z$ $<$ 1.0 (empty triangles). The short-dashed line shows the linear fit for this sample.}}\label{sfrevol}
\end{center}
\end{figure}
\clearpage

\begin{figure}
\begin{center}
\includegraphics[width=0.48\textwidth]{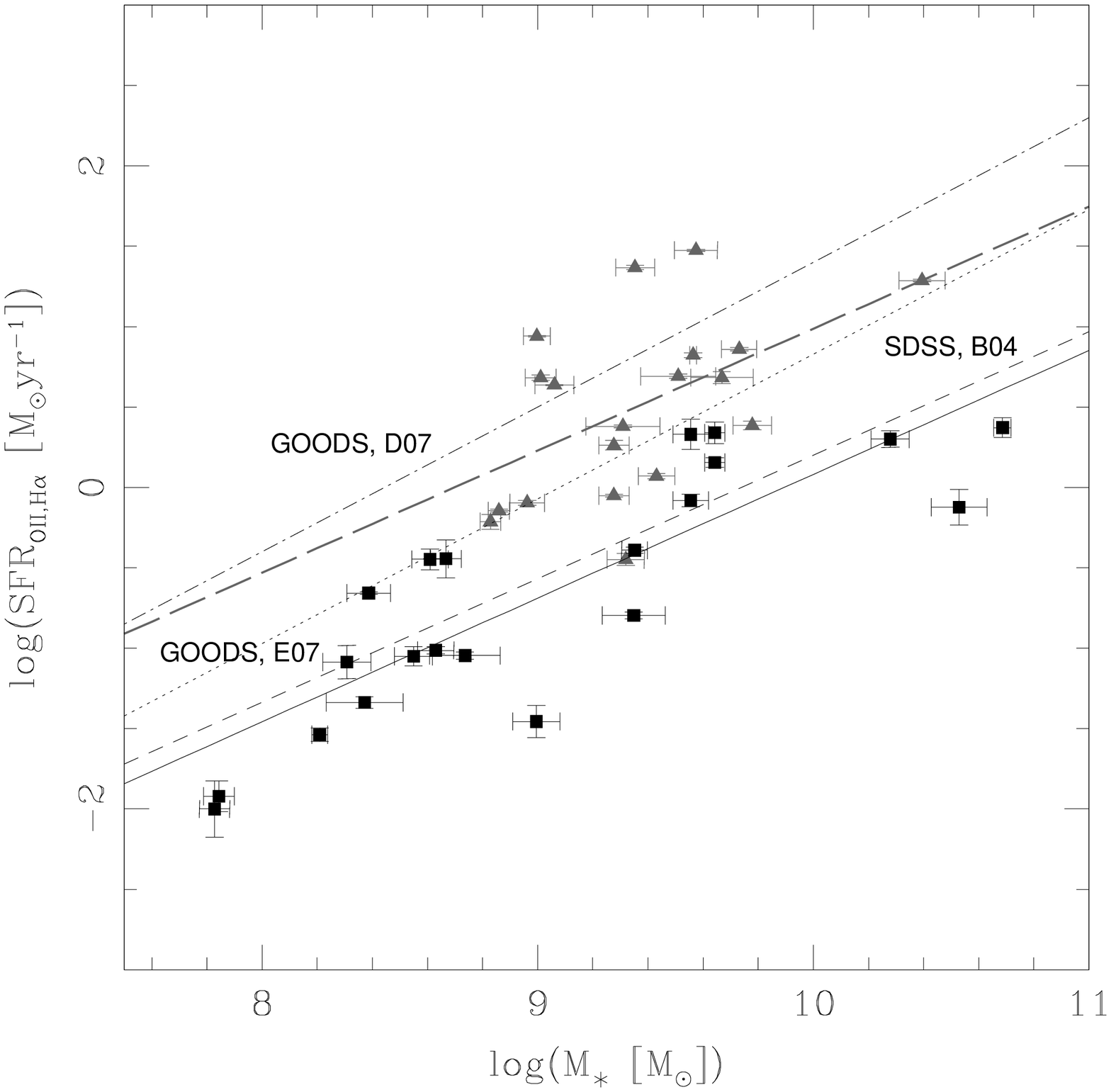}
\caption{\footnotesize{SFR$_{\text{[OII],H}\alpha}$ vs total stellar mass. Black squares represent our low redshift galaxies (0 $<$ $z$ $<$ 0.5), while gray triangles are our intermediate redshift galaxies (0.5 $<$ $z$ $<$ 1.0). The solid and long-dashed lines show the linear fits obtained for each group. The short-dashed line shows the trend observed in SDSS at redshift $z$ $<$ 0.2 (\citeauthor{Brinchmann}), the dotted line shows the trend observed in GOODS by Elbaz et al. \citeyearpar{Elbaz} at redshift 0.8 $<$ $z$ $<$ 1.2, while the dot-dashed line shows the trend found by Daddi et al. \citeyearpar{Daddi2} in GOODS at 1.4 $<$ $z$ $<$ 2.5.}}\label{massfr}
\end{center}
\end{figure}

\begin{figure}
\begin{center}
\includegraphics[width=0.48\textwidth]{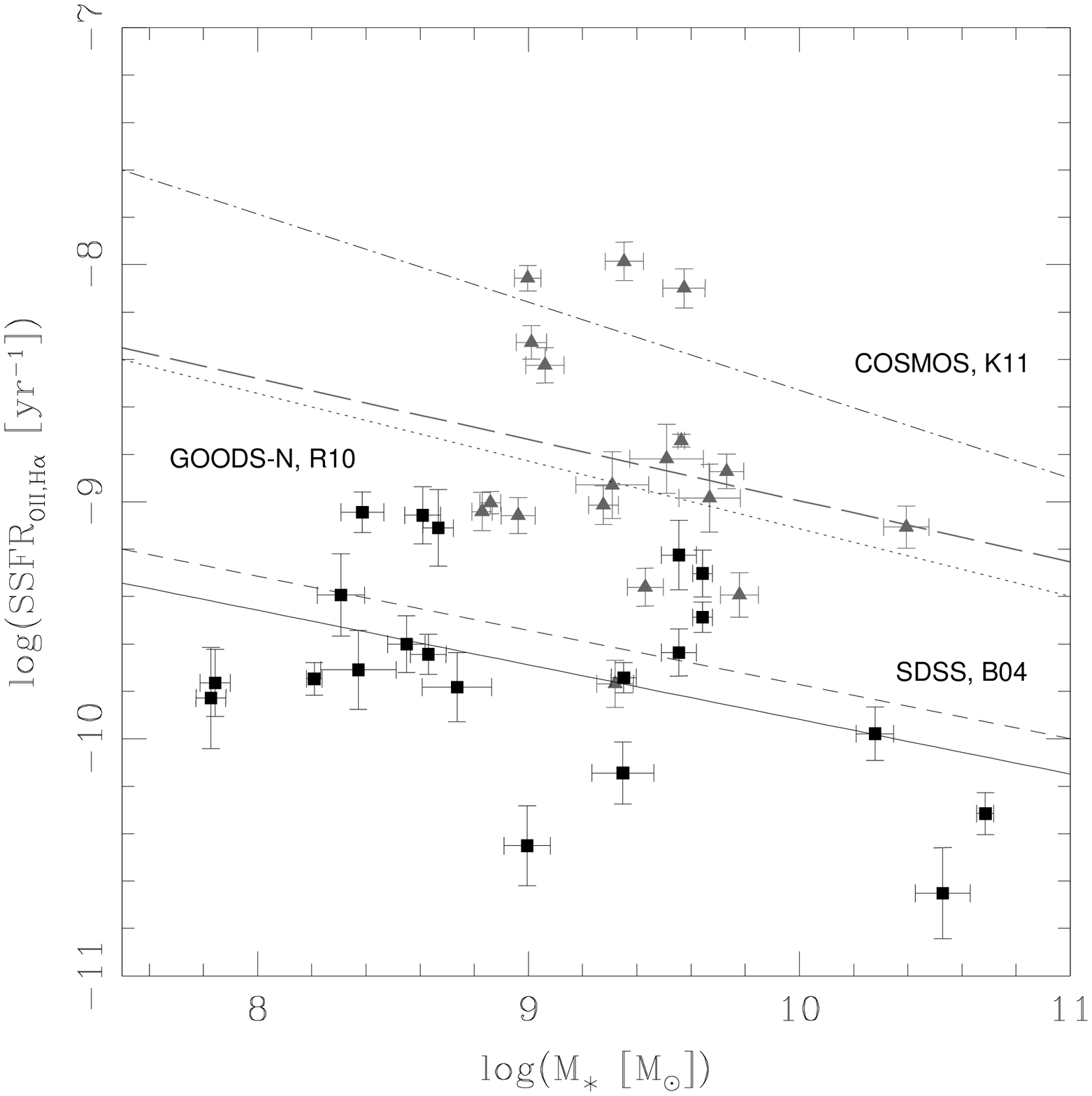}
\caption{\footnotesize{SSFR$_{\text{[OII],H}\alpha}$ vs total stellar mass. Black squares represent our low redshift galaxies (0 $<$ $z$ $<$ 0.5), while gray triangles are our intermediate redshift galaxies (0.5 $<$ $z$ $<$ 1.0), with the solid and long-dashed lines showing the linear fits obtained for each group, respectively. For comparison, the dashed line shows the trend observed in the SDSS sample at redshift $z$ $<$ 0.2 (\citeauthor{Brinchmann}), the dotted line shows the trend found by Rodighiero et al. \citeyearpar{Rodighiero} for GOODS-N galaxies at redshift 0.5 $<$ $z$ $<$ 1.0, and the dot-dashed line shows the trend found by Karim et al. \citeyearpar{Karim} in COSMOS at 1.6 $<$ $z$ $<$ 2.0.}}\label{ssfr}
\end{center}
\end{figure}
\clearpage

\begin{figure}[t]
\begin{center}
\includegraphics[width=0.6\textwidth]{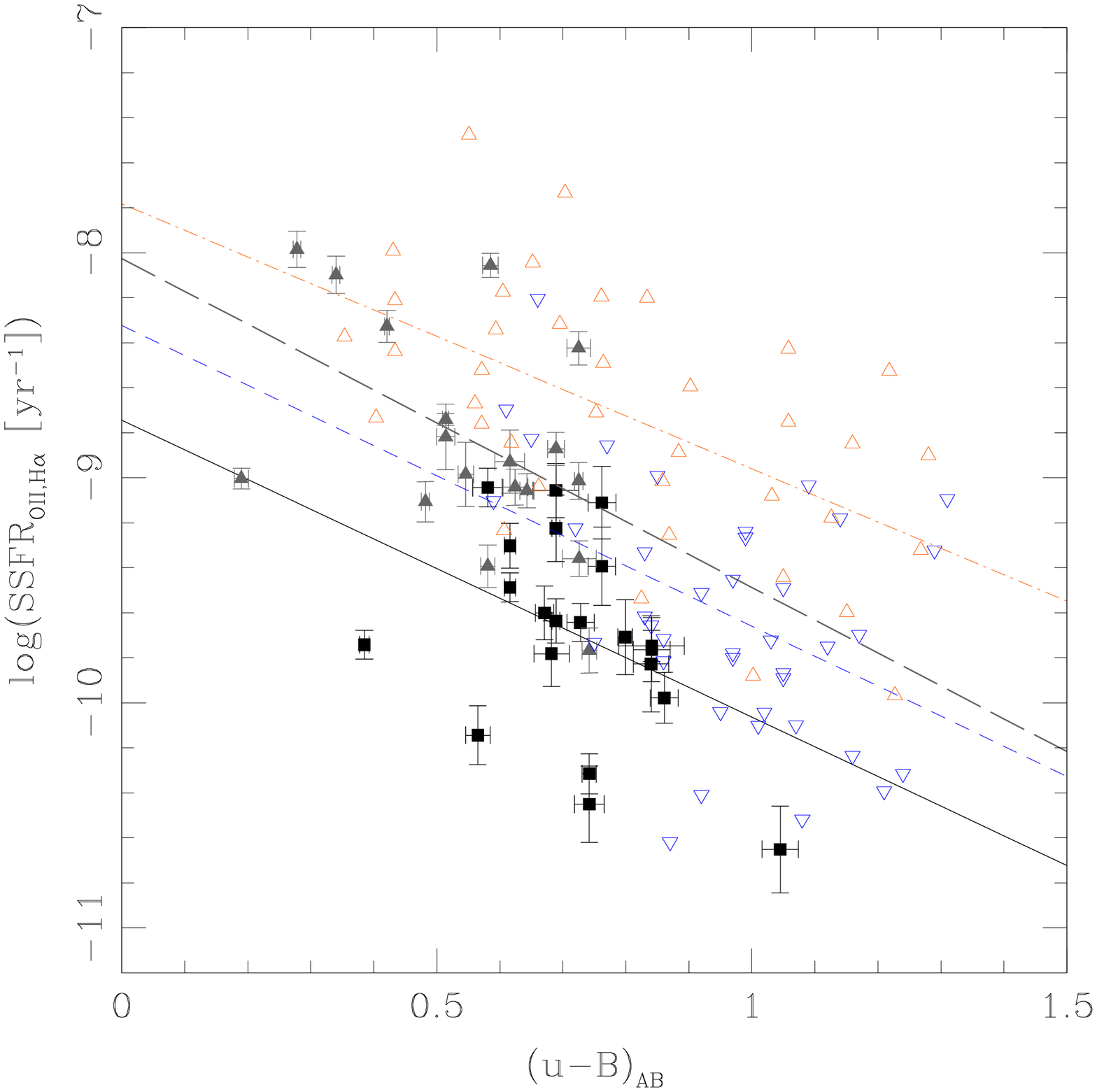}
\caption{\footnotesize{SSFR$_{\text{[OII],H}\alpha}$ vs rest frame color $(u-B)_{AB}$. Black solid squares are our low redshift galaxies at 0 $<$ $z$ $<$ 0.5, while gray solid triangles are our intermediate redshift galaxies at 0.5 $<$ $z$ $<$ 1.0, with the solid and long-dashed lines showing the linear fit for each group, respectively. For comparison, the blue empty inverted triangles are galaxies at redshift 0.55 $<$ $z$ $<$ 1.23 from the Twite et al. \citeyearpar{Twite} sample, and the red empty triangles are galaxies from the GOODS-NICMOS sample of Bauer et al. \citeyearpar{Bauer2} at 1.5 $<$ $z$ $<$ 2. The blue dashed line and red dot-dashed lines show the linear fit estimated for the Twite et al. \citeyearpar{Twite} and Bauer et al. \citeyearpar{Bauer2} galaxies, respectively.}}\label{cssfr}
\end{center}
\end{figure}
\clearpage


\begin{deluxetable}{c c}
\tabletypesize{\small}
\tablewidth{0pt}
\tablecaption{MosaicII Observations\label{uphot}}
\tablehead{\colhead{Date} & \colhead{Exp. Time (s)}}
\startdata
2006 Sep 16 & 6 $\times$ 1000\\
2006 Sep 17 & 5 $\times$ 1000\\
 & 5 $\times$ 600\\
2006 Sep 19 & 10 $\times$ 1000\\
2006 Sep 20 & 12 $\times$ 1000\\
\enddata
\end{deluxetable} 

\begin{deluxetable}{c c c}
\tabletypesize{\footnotesize}
\tablewidth{0pt}
\tablecaption{GMOS Observations.\label{gmos}}
\tablehead{\colhead{Date} & \colhead{Field} & \colhead{Exp. Time (s)}}
\startdata
2008 Nov 1 & SXDF3 & 2 $\times$ 1830\\
2008 Nov 3 & SXDF3 & 2 $\times$ 1830\\
& & 1 $\times$ 1037\\
2008 Nov 4 & SDXF3 & 4 $\times$ 1830\\
2008 Nov 5 & SXDF3 & 4 $\times$ 1830\\
2008 Nov 6 & SXDF3 & 2 $\times$ 1830\\
2008 Nov 20 & SXDF1 & 4 $\times$ 1830\\
2008 Nov 21 & SXDF1 & 4 $\times$ 1830\\
2008 Nov 22 & SXDF1 & 2 $\times$ 1830\\
2008 Nov 25 & SXDF1 & 3 $\times$ 1830\\
2008 Nov 26 & SXDF1 & 2 $\times$ 1830\\
2008 Dec 24 & SXDF2 & 5 $\times$ 1830\\
2008 Dec 25 & SXDF2 & 5 $\times$ 1830\\
2008 Dec 26 & SXDF2 & 2 $\times$ 1830\\
2008 Dec 27 & SXDF2 & 2 $\times$ 1830\\
\enddata
\end{deluxetable} 

\begin{deluxetable}{c c c c c c c c}
\tabletypesize{\scriptsize}
\tablewidth{0pt}
\tablecaption{Spectroscopic sample obtained with IMACS and GMOS.\label{spectra}}
\tablehead{\colhead{Catalogue ID} & \colhead{Instrument} & \colhead{R.A.} & \colhead{Dec.} & \colhead{${z}_{spec}$} & \colhead{$u_{AB}$\tablenotemark{a}} & \colhead{err $u$} & \colhead{Comments\tablenotemark{b}}}
\startdata
SXDF021811.7-050353 & GMOS & 02:18:11.735 & -05:03:53.84 & 0.183 & 24.34 & 0.07 & \\
SXDF021804.3-050022 & GMOS & 02:18:04.308 & -05:00:22.77 & 0.292 & 24.53 & 0.07 & \\
SXDF021758.7-050035 & GMOS & 02:17:58.754 & -05:00:35.44 & 0.423 & 23.22 & 0.03 & \\
SXDF021759.4-050523 & GMOS & 02:17:59.405 & -05:05:23.55 & 0.265 & 23.84 & 0.05 & \\
SXDF021826.8-050352 & GMOS & 02:18:26.889 & -05:03:52.36 & 0.528 & 24.21 & 0.06 & \\
SXDF021826.8-050449 & GMOS & 02:18:26.896 & -05:04:49.86 & 0.378 & 23.95 & 0.06 & \\
SXDF021834.7-050432 & GMOS & 02:18:34.780 & -05:04:32.90 & 0.471 & 24.57 & 0.07 & 24$\mu$m source \\
SXDF021838.3-050410 & GMOS & 02:18:38.359 & -05:04:10.69 & 0.099 & 24.39 & 0.06 & \\
SXDF021836.4-050423 & GMOS & 02:18:36.415 & -05:04:23.04 & 0.310 & 21.83 & 0.02 & 24$\mu$m, 70$\mu$m source\\
SXDF021839.0-050423 & GMOS & 02:18:39.021 & -05:04:23.58 & 0.428 & 24.21 & 0.06 & \\
SXDF021835.5-050145 & GMOS & 02:18:35.522 & -05:01:45.32 & 0.254 & 24.13 & 0.06 & \\
SXDF021730.0-045844 & GMOS & 02:17:30.043 & -04:58:44.79 & 0.392 & 22.76 & 0.03 & \\
SXDF021739.4-050305 & GMOS & 02:17:39.428 & -05:03:05.90 & 0.239 & 23.61 & 0.04 & \\
SXDF021737.5-050437 & GMOS & 02:17:37.505 & -05:04:37.52 & 0.355 & 21.83 & 0.02 & 24$\mu$m source \\
SXDF021742.5-050424 & IMACS & 02:17:42.506 & -05:04:24.82 & 0.045 & 17.26 & 0.01 & 24$\mu$m-850$\mu$m source \\
SXDF021801.8-050930 & IMACS & 02.18:01.860 & -05.09:30.75 & 0.214 & 22.78 & 0.03 & \\
SXDF021833.2-050614 & IMACS & 02:18:33.250 & -05:06:14.81 & 0.090 & 24.24 & 0.06 & \\
SXDF021845.2-045640 & IMACS & 02:18:45.231 & -04:56:40.21 & 0.292 & 22.25 & 0.02 & \\
SXDF021814.7-050254 & GMOS & 02:18:14.763 & -05:02:54.69 & 0.644 & 23.93 & 0.07 & \\
SXDF021806.4-045944 & GMOS & 02:18:06.425 & -04:59:44.79 & 0.557 & 24.53 & 0.07 & \\
SXDF021804.7-045923 & GMOS & 02:18:04.776 & -04:59:23.15 & 0.707 & 24.14 & 0.06 & \\
SXDF021809.3-050322 & GMOS & 02:18:09.363 & -05:03:22.16 & 0.929 & 23.78 & 0.04 & \\
SXDF021801.3-050441 & GMOS & 02:18:01.328 & -05:04:41.28 & 0.697 & 23.74 & 0.04 & \\
SXDF021827.8-050518 & GMOS & 02:18:27.894 & -05:05:18.22 & 0.706 & 24.21 & 0.08 & \\
SXDF021830.0-050514 & GMOS & 02:18:30.050 & -05:05:14.41 & 0.533 & 25.16 & 0.10 & \\
SXDF021827.0-050031 & GMOS & 02:18:27.040 & -05:00:31.74 & 0.689 & 24.44 & 0.07 & \\
SXDF021834.1-050012 & GMOS & 02:18:34.168 & -05:00:12.19 & 0.647 & 23.85 & 0.05 & \\
SXDF021740.4-050223 & GMOS & 02:17:40.450 & -05:02:23.27 & 0.876 & 23.23 & 0.04 & \\
SXDF021721.6-050245 & GMOS & 02:17:21.626 & -05:02:45.81 & 0.495 & 24.17 & 0.06 & \\
SXDF021732.8-050303 & GMOS & 02:17:32.872 & -05:03:03.20 & 0.567 & 23.52 & 0.04 & \\
SXDF021721.3-050220 & GMOS & 02:17:21.345 & -05:02:20.46 & 0.583 & 25.22 & 0.11 & \\
SXDF021750.0-050342 & IMACS & 02.17:50.030 & -05:03:42.63 & 0.669 & 24.61 & 0.08 & \\
SXDF021803.9-050749 & IMACS & 02.18:03.930 & -05:07:49.70 & 0.427 & 24.92 & 0.10 & \\
SXDF021741.0-045356 & IMACS & 02:17:41.013 & -04:53:56.45 & 0.744 & 22.79 & 0.03 & 24$\mu$m source \\
SXDF021758.1-045055 & IMACS & 02:17:58.100 & -04.50:55.50 & 0.342 & 24.54 & 0.08 & \\
SXDF021821.7-044659 & IMACS & 02:18:21.723 & -04:46:59.21 & 0.651 & 23.73 & 0.04 & \\
SXDF021836.9-045046 & IMACS & 02:18:36.970 & -04.50:46.19 & 0.556 & 24.96 & 0.09 & \\
\enddata
\tablenotetext{a}{Measured observer frame $u$-band photometry in AB magnitudes.}
\tablenotetext{b}{Detection of a FIR and/or submillimeter counterpart.}
\end{deluxetable}

\end{document}